\documentclass{aa}  

\usepackage{graphicx}
\usepackage{txfonts}
\usepackage{amsmath}
\usepackage{amssymb}
\usepackage{mathabx}
\usepackage{xcolor}
\usepackage{commath}
\usepackage{relsize}
\usepackage{mathtools}
\usepackage{tabularx}
\usepackage{booktabs}
\usepackage[thinc]{esdiff}
\usepackage{nicefrac}
\usepackage{cancel}
\usepackage{csquotes}
\usepackage[justification=centering]{subcaption}
\usepackage{enumitem}
\usepackage{afterpage}
\usepackage{multirow}
\usepackage{url}
\usepackage[colorlinks=true, allcolors=blue]{hyperref}
\usepackage[noabbrev, capitalise]{cleveref}

\makeatletter
\renewcommand*\aa@pageof{, page \thepage{} of \pageref*{LastPage}}
\makeatother

\newcolumntype{L}{>{\raggedright\arraybackslash}X}
\newcolumntype{R}{>{\raggedleft\arraybackslash}X}

\begin{document}

   \title{Populating the Milky Way:}
   \subtitle{Characterising planet demographics by combining galaxy formation simulations and planet population synthesis models}

   \author{C. Boettner
          \inst{1}
          \fnmsep
          \inst{2}
          \and
          P. Dayal
          \inst{1}
          \and
          M. Trebitsch
          \inst{1}
          \and
          N. Libeskind
          \inst{3}
          \and
          K. Rice
          \inst{4,5}
          \and
          C. Cockell
          \inst{6}
          \and
          B. I. Tieleman
          \inst{2}
          }

   \institute{Kapteyn Astronomical Institute, University of Groningen,
              Landleven 12 (Kapteynborg, 5419) 9747 AD Groningen\\
              \email{boettner@astro.rug.nl}
              \and
              GELIFES Institute, University of Groningen,
              Nijenborgh 7 9747 AG Groningen\\
              \and
              Leibniz-Institut für Astrophysik Potsdam (AIP), An der Sternwarte 16, 14482 Potsdam\\
              \and
              SUPA, Institute for Astronomy, University of Edinburgh, Royal Observatory, Blackford Hill, Edinburgh, EH9 3HJ\\
              \and
              Centre for Exoplanet Science, University of Edinburgh, Edinburgh, EH9 3FD \\
              \and
              School of Physics and Astronomy, The University of Edinburgh, Peter Guthrie Tait Road, Edinburgh, EH9 3FD\\
             }

   \date{Received xxx; accepted xxx}
 
  \abstract
   {Stellar populations and their distribution differ widely across the Galaxy,  which is likely to affect planet demographics. Our local neighbourhood is dominated by young, metal-rich stars in the galactic thin disc, while the stellar halo and galactic bulge host a large fraction of older, metal-poor stars.}
   {We study the impact of these variations on planet populations in different regions of the Galaxy by combining a high-resolution galaxy formation simulation with state-of-the-art planet population synthesis models. }
   {We constructed a population model to estimate occurrence rates of different planet types, based on the New Generation Planet Population Synthesis (NGPPS). We applied this model to a simulated Milky Way (MW)\ analogue in the HESTIA galaxy formation simulation. We studied the planet occurrence rate in the metal-rich regions of the inner Galaxy, namely, in the galactic bulge and thin disc. We compared these result with the frequencies in the more distant, metal-poor region such as the thick disc and stellar halo.}
   {We find that the planet demographics in the central, metal-rich regions of the MW analogue differ strongly from the planet populations in the more distant, metal-poor regions. The occurrence rate of giant planets ($>300 M_\Earth$) is 10 to 20 times larger in the thin disc compared to the thick disc, driven by the low amounts of solid material available for planet formation around metal-poor stars. Similarly, low-mass Earth-like planets around Sun-like stars are most abundant in the thick disc, being 1.5 times more frequent than in the thin disc. Moreover, low-mass planets are expected to be abundant throughout the galaxy, from the central regions to the outer halo, due to their formation processes being less dependent on stellar metallicity. The planet populations differ more strongly around Sun-like stars compared to dwarfs with masses 0.3 -- 0.5 $M_\Sun$, caused by a weaker correlation between [Fe/H] metallicity and planet mass. However, it is important to note that the occurrence rates of low-mass planets are still uncertain, making our findings strongly model-dependent. Massive planets are more comprehensively understood and our findings are more robust. Nonetheless, other systematic effects have the potential to alter the giant planet population that we have not addressed in this study. We discuss some of these limitations and offer further directions for future research.}
   {}

   \maketitle
%
%-------------------------------------------------------------------
\section{Introduction}
\label{sec:introduction}
The architectures of planetary systems are tightly connected with the properties of their host stars, including their effective temperature, age, and metallicity. This relationship is most pronounced in the established correlation between stellar metallicity [Fe/H] and the occurrence rates of giant planets. Multiple observational studies have shown that close-in giant planets are more commonly found around metal-rich stars \citep{Gonzalez1997, Fischer2005, Johnson2010}. Additionally, observations suggest a decline in the fraction of stars hosting planets and a decrease in planet multiplicity in line with the increasing mass of the host star \citep{Yang2020, He2021}. Across the galaxy, stellar populations exhibit significant variations in both age and chemistry \citep{Gilmore1983, Prochaska2000, Reddy2006, Schuster2006, Bensby2014, Masseron2015}, implying that planet demographics could strongly depend on the local galactic environment.

As galaxies grow and evolve, their local chemical profiles undergo changes due to stellar evolution \citep{McWilliam1997}. Throughout their life cycles, stars enrich the  interstellar medium (ISM)\ with heavy elements. This process, along with the hierarchical growth of galaxies, results in a metallicity gradient across the galaxy \citep{Lemasle2007, Luck2011, Genovali2014}, as well as the formation of distinct stellar populations \citep{Gilmore1983, Bland-Hawthorn2016}. The Milky Way (MW), for instance, is typically divided into four components: the galactic bulge, thin disc, thick disc, and halo, each characterized by unique stellar populations with varied chemical compositions. The halo is particularly known for containing some of the oldest and most metal-poor stars in the galaxy, with metallicities ranging from -7 < [Fe/H] < -0.5 \citep{Beers2005, Frebel2015}. In contrast, the thin disc hosts a younger stellar population with metallicities near solar levels, [Fe/H] $\approx$ 0 \citep{Kilic2017}. The thick disc, though not as metal-rich as the thin disc, is distinguished by its $\alpha$-enhanced stars \citep{Wallerstein1962}. The bulge, representing the most heterogeneous of these populations, spans a broad range of metallicities, including both extremely metal-poor and metal-rich stars \citep{McWilliam2016}.

The majority of exoplanet studies to date have concentrated on stars in the nearby thin disc \citep{Bashi2019}. This focus is primarily due to observational constraints, as planets orbiting nearby stars are more readily detected via transit and radial velocity (RV) methods. Despite these limitations, a comprehensive picture of planet formation is emerging \citep[e.g.][]{Pollack1996, Bitsch2019, Johansen2019, Drazkowska2023}. Planetary systems are understood to originate from the protoplanetary discs that encircle young host stars. The prevailing theory is that most planets form through the core-accretion model, where large planet embryos grow by accumulating planetesimals and pebbles. Considering that the chemical composition of these protoplanetary discs reflects that of the host star, it is reasonable to expect a relationship between the architecture of planetary systems and the metallicity of their host stars, and such correlations have indeed been observed \citep{Gonzalez1997, Santos2001, Mulders2015, Petigura2018}.

Planet formation is an intricate process that involves a wide range of physical phenomena, and considerable effort has been dedicated to modeling and simulation work. Key aspects of this process include the formation of planetesimals and planetary embryos, the evolution and eventual dissipation of the protoplanetary disc, and the growth of planets through mechanisms such as accretion, migration, and the development of planetary interiors and atmospheres \citep[e.g.][]{Pollack1996, Ida2004, Ida2004a, Bitsch2019, Johansen2019, Alessi2018a, Alessi2020}. To align these models with observational data, planet population synthesis has become a crucial methodology \citep{Ida2004, Mordasini2009a, Mordasini2009}. This approach involves generating representative samples of synthetic exoplanets for specific stellar populations, which can then be systematically compared to the actual observed exoplanet data.

The heterogeneity of stellar populations, and their connection to galaxy evolution is an area of recent research that has has benefitted greatly from the advancements in simulation techniques \citep{Crain2023}. Hydrodynamical simulations of galaxy evolution, in particular, have had great success in reproducing the properties of galaxies in the local universe \citep[e.g.][]{Navarro1994, Vogelsberger2014, Schaye2015, Dubois2016}, including MW-like galaxies. Amongst the MW-reproducing simulations, one successful approach has been to model the local (galaxy) environment through the use of constrained simulations. These simulations are able to generate MW analogues within environments that closely resemble the Local Group \citep{Hoffman2009, Gottloeber2010, Libeskind2020}.

In this work, we aim to combine the recent advances in both galactic modelling and planetary population synthesis. We employ the HESTIA suite of constrained hydrodynamical galaxy formation simulations to study planet populations throughout the MW \citep{Libeskind2020}. HESTIA reproduces the Local Group, including the MW, Andromeda (M31), and surrounding satellite galaxies within the correct cosmographic context. As a zoom-in simulation, HESTIA creates a high-resolution MW-like galaxy with a smoothing length of approximately 220 pc and a star particle mass resolution of $10^4 M_\Sun$. The simulated galaxy closely matches the MW in terms of total mass, stellar mass, metallicity distribution, and size. We combine one of these simulated galaxies with the state-of-the-art New Generation Planet Population Synthesis (NGPPS) data, obtained from the Bern model for planet formation and evolution, which generates realistic planet populations for a wide range of host star masses and metallicities \citep{Emsenhuber2021}. Its goal is to study the demographics of exoplanet throughout the MW.

This paper is organized as follows. \cref{sec:bern_model} details the Bern model and NGPPS planet population, including the construction of a planet assignment model to associate planet occurrence rates with the mass and metallicity of the host stars. \cref{sec:HESTIA} describes the simulated MW analogue in HESTIA, comparing  the key properties with those of the actual MW and presenting a decomposition of the simulated galaxy into bulge, thin disc, thick disc, and halo. This section also details the integration of the planet assignment model from \cref{sec:bern_model} into the MW analogue. In \cref{sec:results}, we present the resulting planet populations in different parts of the galaxy, focussing on Sun-like and low-mass stars. These results are discussed in \cref{sec:discussion}, where we also highlight the limitations of our study. Finally, we summarise our findings in \cref{sec:summary}.

\section{Bern planet model and planet populations }
\label{sec:bern_model}
\begin{figure*}
    \centering
    \includegraphics[width=\textwidth]{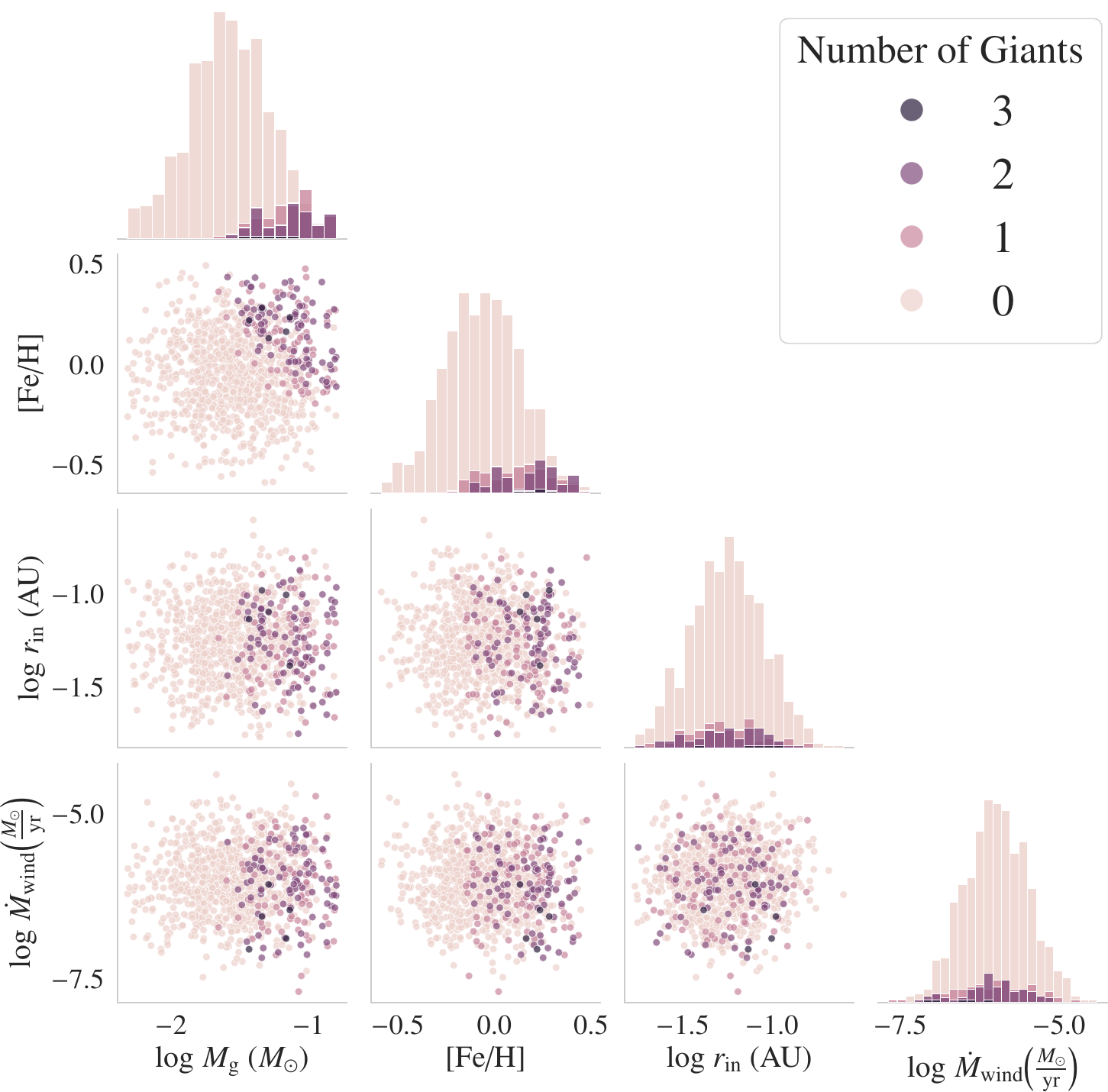}
    \caption{Distribution of Monte Carlo variables for original sample from \citet{Emsenhuber2021a}, and effect on number of giant planets for the $M_\star = 1$ and $N_\mathrm{Embryos} = 50$ model run. The colour marks the number of planets in the system after 20 Myr of evolution. Shown are the correlation between the gas mass, $M_\mathrm{g}$, metallicity, [Fe/H], inner protoplanetary disc radius, $r_\mathrm{in}$, and external photo-evaporation rate, $\dot{M}_\mathrm{wind}$.}
    \label{fig:original_sample_pairplot}
\end{figure*}
The third
generation of the Bern global model for planet formation and evolution, NGPPS, is based on  a semi-numerical approach to study the formation of planets and evolution of planetary systems \citep{Emsenhuber2021}. The Bern model is based on the core-accretion paradigm \citep{Pollack1996}, which is thought to be the primary formation pathway for low-mass planets, and a majority of massive planets. Specifically, it does not include giant planet formation through gravitational instability \citep{Boss1997}, which is thought to be a potential origin for very wide orbits \citep[$> 10$ AU,][]{Rafikov2005, Schib2021} and very massive planets \citep[$> 4 M_\mathrm{Jupiter}$,][]{Schlaufman2018}. The third generation of the model encompasses the growth of planets from (1 -- 100) planet embryos with a mass of about $0.01 M_\Earth$ into planets with masses ranging from Earth-like planets to super-Jupiter giants. The planet embryos are immersed into fluid-like protoplanetary discs, which are described by gas and planetesimal fields from which the planets grow. These fields are described by continuous density fields, coupled to the planet embryos. In the current generation of the model, direct N-body interaction between the growing planets is also included, and \citet{Emsenhuber2021} have found that embryo-embryo interactions increase the number of massive planets compared to isolated evolution.

The model is initialised at the stage when planet embryos have formed; the growth stages from dust to pebbles and planetesimals are not included. The initial number and distribution of the embryos are therefore free parameters. The simulation starts with a formation phase in which the planet embryos interact with the gas and planetesimal fields. This stage lasts for a fixed period of 20Myr, after which the disc is considered dispersed. From this point on, the planets evolve isolated, affected by tidal migration and atmospheric escape. The assumed formation phase of 20Myr is longer than typical protoplanetary disc lifetimes and therefore encompasses most of the growth stage. However, the late-stage giant impact phase that can affect low-mass planets is not fully simulated, which can lead to an underestimation of the masses for rocky planets. \citet{Emsenhuber2021} found that for a minimum-mass solar nebula-like surface density of solids, planet formation is mostly complete within $\sim 1$ AU after 20Myr, while more distant planets have not yet reached their final state. The model provides a wealth of data about the planet populations in the systems, including their mass, distance to the host star, radii, luminosities, and evolution tracks.

A wide range of physical processes are included in the model and, correspondingly, a large number of parameters need to be set. Some of these parameters are fixed to specific values \citep[][see Table 1]{Emsenhuber2021a}; however, some parameters are considered to be variable and a suite of simulations has been run varying these values. Four quantities are varied randomly and dubbed Monte Carlo variables.

\textbf{Initial gas disc mass}: The initial mass of the gas disc around the star, denoted as $M_\mathrm{g}$, provides the building material for planet formation, with a more massive disc potentially leading to more numerous and larger planets. \citet{Emsenhuber2021a} opted for log-normal distributions (with median $\mu = -1.49$, and standard deviation $\sigma=0.35$) based on Very Large Array (VLA) observations reported by \citet{Tychoniec2018}, which focussed on Class I discs in the Perseus star-forming region. They restricted the range of of masses to 0.004 -- 0.16 $M_\Sun$ to guarantee that the discs have non-negligible mass and are self-gravitationally stable.

\textbf{External photo-evaporation rate}: The external photo-evaporation $\dot{M}_\mathrm{wind}$ measurement is the main parameter determining the lifetime of the protoplanetary disc. While the inner part of the disc is evaporated by the internal radiation of the host star, in the outer regions (where most of the mass resides) evaporation is initiated primarily by the far-ultraviolet (FUV) radiation of nearby massive stars \citep{Matsuyama2003a}. \citet{Emsenhuber2021a} assumed a fixed disc viscosity and a normal distribution ($\mu = -6$, $\sigma=0.5$) for $ \log \dot{M}_\mathrm{wind}$ to reproduce observed protoplanetary lifetimes. With their parametrisation 50\% of discs will have a lifetime of $<4$Myr and virtually all discs are dispersed within 10Myr.
    
\textbf{Dust-to-gas ratio}: The dust-to gas ratio, $f_\mathrm{D/G} = M_\mathrm{s}/M_\mathrm{g}$, determines the amount of solid material in the disc, which is crucial for the growth of solid planets and the cores of gas giants. In the \citet{Emsenhuber2021} model, the dust-to-gas ratio is assumed to be a function of stellar metallicity given by $\frac{f_\mathrm{D/G}}{f_\mathrm{D/G,\Sun}} = 10^\mathrm{[Fe/H]}$, where [Fe/H] is the iron abundance and $f_\mathrm{D/G, \Sun} = 0.0149$ is the dust-to-gas ratio of the sun. \citet{Emsenhuber2021a} chose a normal distribution ($\mu = -0.02$, $\sigma=0.22$) for the [Fe/H] distribution truncated to the range [-0.6, 0.5]. This encompasses the vast majority of stars in the solar neighbourhood. However, in the galactic centre and stellar halo metal-poor stars exhibit significantly lower [Fe/H] abundances. This needs to be considered for the HESTIA planet population model.
    
\textbf{Inner edge of the gas disc}: The position the inner edge of the protoplanetary (gas) disc, $r_\mathrm{in}$, is important for the final locations of close-in planets, since gas drag is a major driver of migration. \citet{Emsenhuber2021a} chose a log-normal distribution ($\mu = -1.26$, $\sigma=0.2$) for this quantity. 

The mass of the planetesimal (i.e. solid material) disc, $M_\mathrm{s}$, is not a Monte Carlo variable itself, but of central importance to planet formation since it provides the building blocks for rocky planets and giant cores. Within the Bern model, this is fully characterised by the mass of the gas disc and the dust-to-gas ratio, $M_\mathrm{s} = f_\mathrm{D/G} \cdot M_\mathrm{g}$. Furthermore, the embryos are initialised with a a mass of $10^{-2} M_\Earth$, and distributed uniformly in the logarithm of distance between $r_\mathrm{in}$ and 40 AU.
In addition to these quantities, two more parameters are varied systematically:
\begin{itemize}
\item \textbf{Initial number of embryos}: The initial number of lunar-mass planetary embryos $N_\mathrm{Embryo}$ is crucial for the final number and properties of the planetary systems. Since the Bern model does not include the initial growth stage of dust and pebbles, this quantity cannot be predicted and needs to be set manually. \citep{Emsenhuber2021a} study the evolution for 1, 10, 20, 50 and 100 embryos. We will however only consider the range of 10 -- 100 embryos, since they were created with a consistent sample of Monte Carlo variable and are thus more easily comparable.

\item \textbf{Host star mass}: The host star mass ($M_\star$) is fixed to $1 M_\Sun$ in the work by \citet{Emsenhuber2021a}. However \citet{Burn2021} extend the analysis to low mass stars with 10\%, 30\%, 50\% and 70\% of the sun's mass. The host star's mass affects the planet evolution in direct ways, for instance, through radiation, but is also correlated with other relevant quantities, such as the mass of the protoplanetary disc. In the work by \citet{Burn2021}, only a $N_\mathrm{Embryo}=50$ run is considered, the Monte Carlo variables follow the same distribution as the for the Sun-like star but re-scaled through some fixed prescriptions, with $M_\mathrm{g} \propto M_\star$ and $r_\mathrm{in} \propto M_\star^{\nicefrac{1}{3}}$. The metallicity and external photo-evaporation-rate distributions are assumed to be identical to the solar-mass case. Stars more massive than the sun were not studied, due to their relative rarity and short lifespan.
\end{itemize}

\subsection{Original NGPPS sample and planet types}
In the original set of publications, \citet{Emsenhuber2021} have performed and analysed a set of population runs based on the parameter distributions described before. In their work, they fix the host star mass to $M_\star = 1$ \citep{Emsenhuber2021a}, and run simulations for the different assumed initial embryo numbers. For each fixed value of $N_\mathrm{Embryo}$, they draw 1000 random samples from the Monte Carlo variable distributions and run their simulations with these parameters. In a subsequent paper, \citet{Burn2021} perform a similar analysis for a fixed number of embryos $N_\mathrm{Embryo}=50$ and varying the host star mass between 0.1 and 1 $M_\Sun$, using the same sample of Monte Carlo variables but re-scaled based on the host star mass. 

Both \citet{Emsenhuber2021} and \citet{Burn2021} classify the planets into categories based on their mass at the end of the formation phase. These encompass Earth-like, Super-Earth, Neptunian, Sub-giant, and giant, as detailed in \cref{tab:planet_categories}.
\cref{fig:original_sample_pairplot} shows the distribution of the Monte Carlo variables and number of giant planets found in this system for the $M_\star = 1$ and $N_\mathrm{Embryo} = 50$ run.
\begin{table}[!ht]
    \centering
    \caption{Planet types as defined in \citet{Emsenhuber2021a}.}
    \begin{tabularx}{\columnwidth}{l|R|R}
         & Min. Mass [$M_\Earth$] & Max. Mass [$M_\Earth$] \\
        \addlinespace
        \hline\hline
        \addlinespace
        Earth & 0.5 & 2 \\ 
        Super-Earth & 2 & 10 \\ 
        Neptunian & 10 & 30 \\ 
        Sub-giant & 30 & 300 \\
        Giant & 300 &  \\ 
    \end{tabularx}
    \label{tab:planet_categories}
\end{table}

\begin{figure}
    \includegraphics[width=\columnwidth]{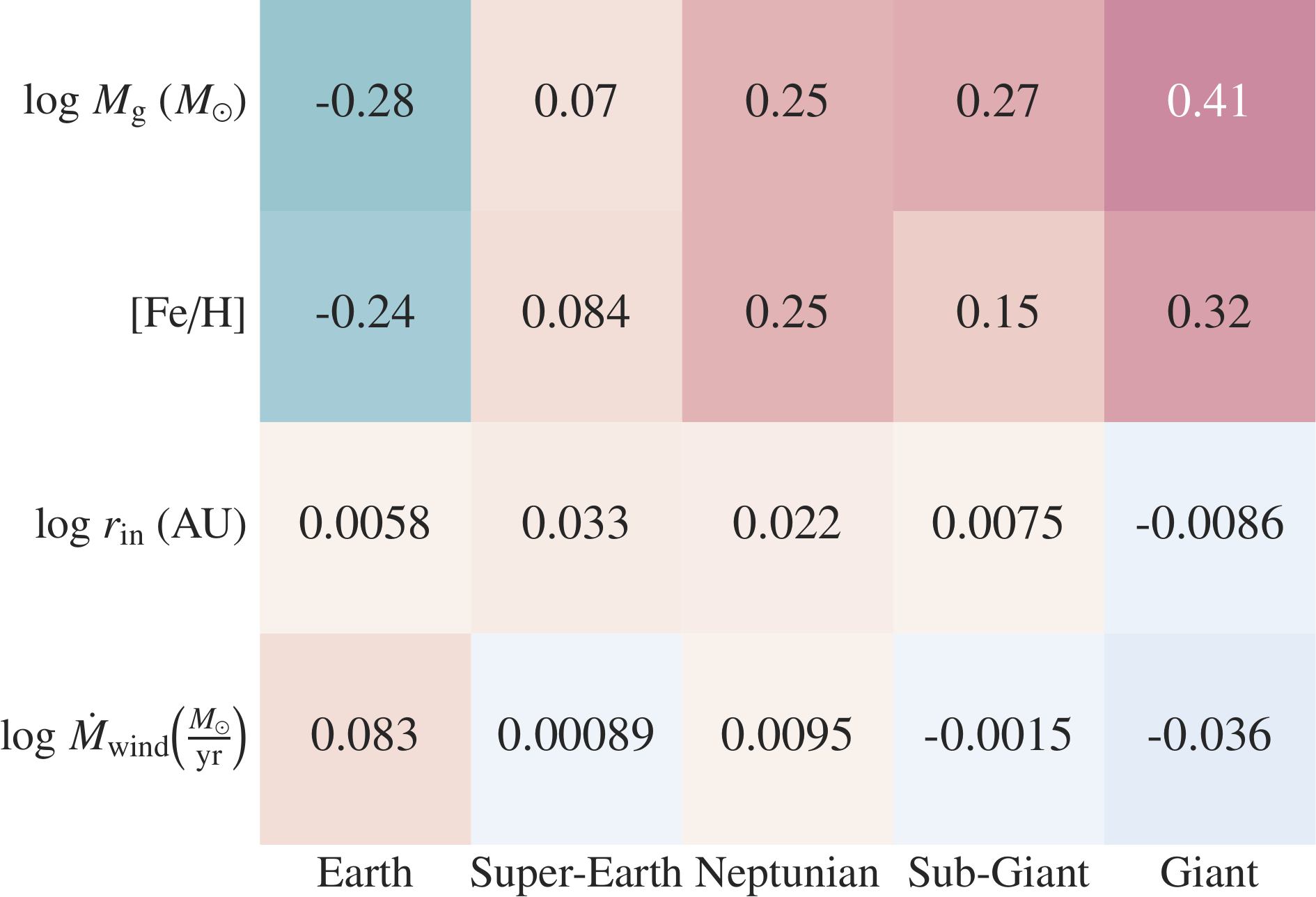}
    \caption{Kendall correlation matrix between Monte Carlo variables and planet types for the $M_\star = 1$ and $N_\mathrm{Embryos} = 50$ run \citep{Emsenhuber2021a}.}
    \label{fig:original_sample_correlation_matrix}
\end{figure}
\subsection{Correlations between Monte Carlo variables and planet type abundances}
Of primary concern for the study of planet populations throughout the galaxy is the relation between the number of planets in a given category and the chosen parameter, since those expected to vary systematically as a function of galactic environment. For example, \cref{fig:original_sample_pairplot} indicates that higher values of [Fe/H] and $M_\mathrm{g}$ lead to a larger number of giant planets. To get a feeling for the general relations, the correlation coefficients between the planet types and Monte Carlo variables are shown in \cref{fig:original_sample_correlation_matrix} for the $M_\star = 1$ and $N_\mathrm{Embryos} = 50$ population. It is easy to see that the more massive planet types are positively correlated with a larger initial disc mass, while low-mass planets have a negative correlation. The same holds true for the metallicity. In contrast, correlations with the external photo-evaporation rate and inner disc edge are much weaker for the considered range of values. These results persist for initial values for $N_\mathrm{Embryos}$ between 10 and 100, and similarly for lower host star masses, although the correlations become weaker with decreasing host mass.

\begin{table}[!ht]
    \centering
    \caption{Comparison of the occurrence rates for various planet types under the $M_\star = 1$ and $N_\mathrm{Embryos} = 50$ model run, showcasing mean values (upper value) and interquartile range (95\% percentile - 5\% percentile, lower value). The reduced model tends to slightly underestimate the frequency of Earth-like planets and overestimate that of giants relative to the full model, with a consistently higher variance observed across all planet categories.}
    \begin{tabularx}{\columnwidth}{l|R|R}
         & Full Model & Reduced Model \\
        \addlinespace
        \hline\hline
        \addlinespace
        Earth & 4.1 & 3.9 \\ 
        & 6.1 & 6.4\\
        \addlinespace
        Super-Earth & 3.4 & 3.0 \\ 
        & 6.4 & 7.1 \\
        \addlinespace
        Neptunian & 0.43 & 0.42 \\
        & 1.2 & 1.6 \\
        \addlinespace
        Sub-giant & 0.072 & 0.066 \\
        & 0.43 & 0.53 \\
        \addlinespace
        Giant & 0.31 & 0.35  \\ 
        & 1.3 & 1.5 \\
    \end{tabularx}
    \label{tab:planet_model_comparison}
\end{table}

\subsection{Planet assignment model}
\label{sec:assignment_model} 
To populate the star particles in the HESTIA galaxy simulation with planets, we need a model that connects the properties of the host star (or stellar population) with the properties of the planets forming around the star. We constructed a simple assignment model that connects the occurrence rate of planets in each category with the Monte Carlo variable, where the occurrence rate is defined as: 
\begin{equation}
    \text{Occurrence rate} = \frac{\text{Total number of planets}}{\text{Total number of stars}}.
    \label{eq:occurrencerate}
\end{equation}
We note that this quantity differs from the fraction of planet-hosting stars, which compares the total number with systems with a planet of a given type compared to the total number of stars.

We determine the number of planets per category (as defined in \cref{tab:planet_categories}) across the 1000 Monte Carlo runs (for a fixed $N_\mathrm{Embryo}$ and $M_\star$). The results are smoothed and interpolated using a $k$-nearest neighbours algorithm with $k=30$ (chosen via a ten-fold cross-validation on the original sample). 

Considering that the photo-evaporation rate and inner disc radius exert minimal influence on planet classifications, we chose to further reduce the model by incorporating only the initial gas disc mass and [Fe/H] as occurrence rate predictors. This increases interpretability of the model at the cost of slightly reduced predictive power. The predicted occurrence rate of planets for the full model (incorporating all four Monte Carlo variables) and the reduced model (using only $M_\mathrm{g}$ and [Fe/H]) are shown in \cref{tab:planet_model_comparison} for the $M_\star = 1$ and $N_\mathrm{Embryos} = 50$ population. The reduced model yields slightly fewer low-mass planets and more giant ones, as elevated photo-evaporation rates may shorten disc lifetimes and thus stop gas accretion early. These effects are however well within the variability of the model.

In \cref{fig:contourplots_planet_model}, the occurrence rates of planets are shown as a function of $M_\mathrm{g}$ and [Fe/H], for the $M_\star = 1$ and $N_\mathrm{Embryos} = 50$ population. Within the Bern model, occurrence rates of the different planet types are strongly dependent on the initial planetesimal mass in the disc (diagonal lines in \cref{fig:contourplots_planet_model}, since $M_\mathrm{s} = f_\mathrm{D/G} \cdot M_\mathrm{g}$), tend to peak around some value of $M_\mathrm{s}$ (dependent on the planet mass), and taper off to either side. The peak value increases as the characteristic mass of the planet type increases, suggesting an optimal range of available solid material to form planets within specific mass ranges. For Earth-like planets, the model indicates a value $\sim 0.3 M_\mathrm{Jup}$, while Super-Earths are preferably formed in discs with $\sim 0.7 M_\mathrm{Jup}$ of available solid material. Sub-giant and giant planets in particular almost exclusively occur if the solid disc mass $> 1 M_\mathrm{Jup}$. Moreover, \cref{fig:contourplots_planet_model} also shows the planet occurrence rates in relation to metallicity [Fe/H] alone. Notably, giant planets exhibit a much more pronounced metallicity-dependence in their occurrence rates compared to low-mass planets, in line with observations.

As part of the original publication series on NGPPS, \citet{Schlecker2021} performed a similar analysis on the predetermination of planet types from disc properties based on the Bern Model. Their approach differs from the one taken here in that they do not pre-define planet categories but use an unsupervised learning algorithm to categorise planets based on their mass, radius, and semi-major axis. Using a Gaussian Mixture Model, they identified four planet categories, which they call Neptunes, icy cores, giant planets, and (Super-)Earths. They found that the strongest predictor for a planet's type is the location of its initial embryo, and emphasised that consistent treatment of planetary embryo formation is a priority in the future development of the Bern model. Given a fixed starting location, they found that giant planet formation is strongly dependent on the mass of the available solid material, and that the photo-evaporation has no strong correlation with planet type for the considered range of values. They interpret the latter finding as an indication that planet formation, on average, concludes on shorter time scales than the lifetime of the disc within the NGPPS population. Our results are therefore consistent with their findings despite the differences in the approach.

\begin{figure}
    \centering
    \includegraphics[width=\columnwidth]{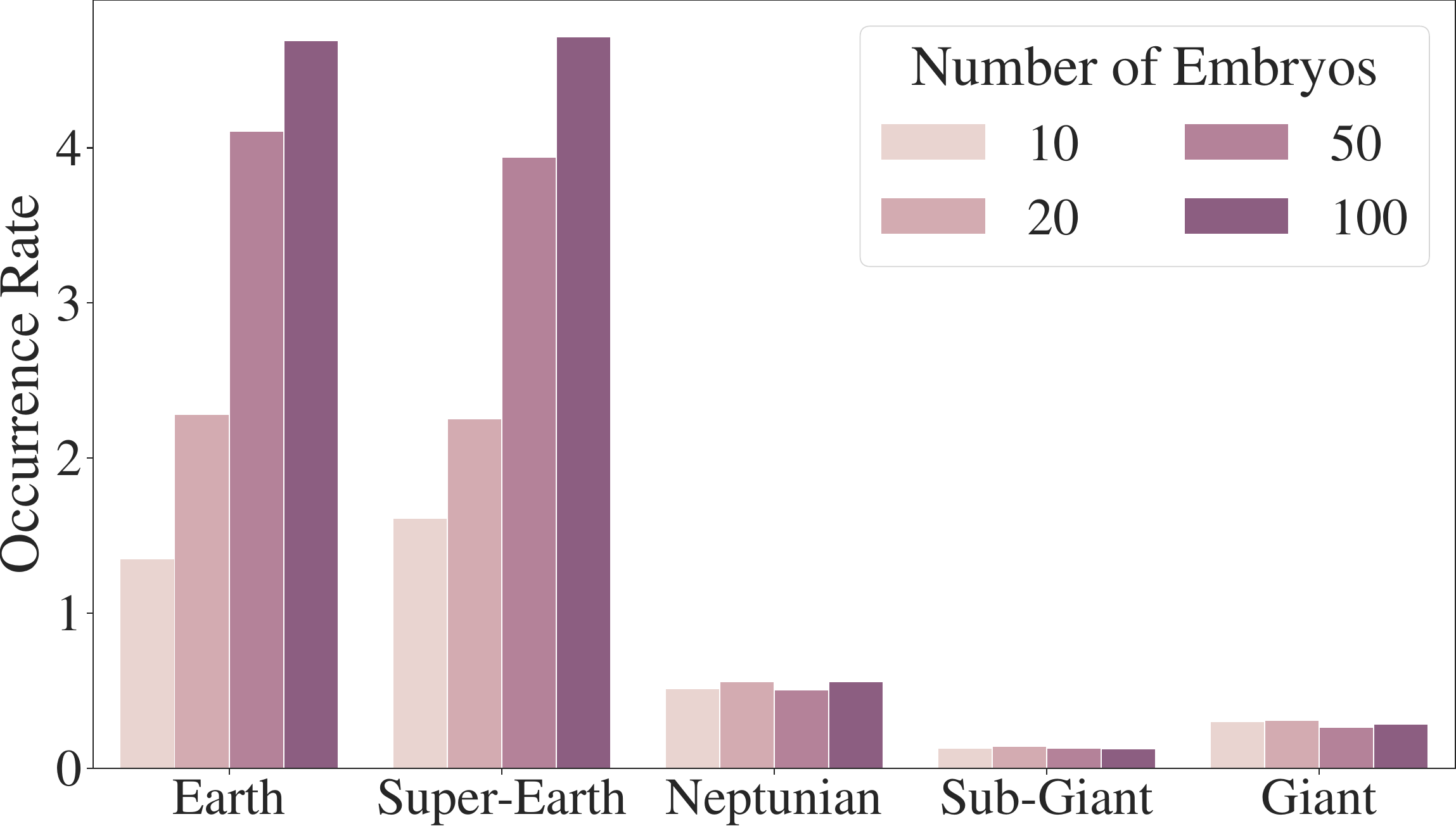}
    \caption{Variations of planet occurrence rates per category given different numbers of initial embryos, $M_\star = 1$ \citep{Emsenhuber2021a}.}
    \label{fig:original_sample_number_of_embryos}
\end{figure}

\begin{figure}
    \centering
    \includegraphics[width=\columnwidth]{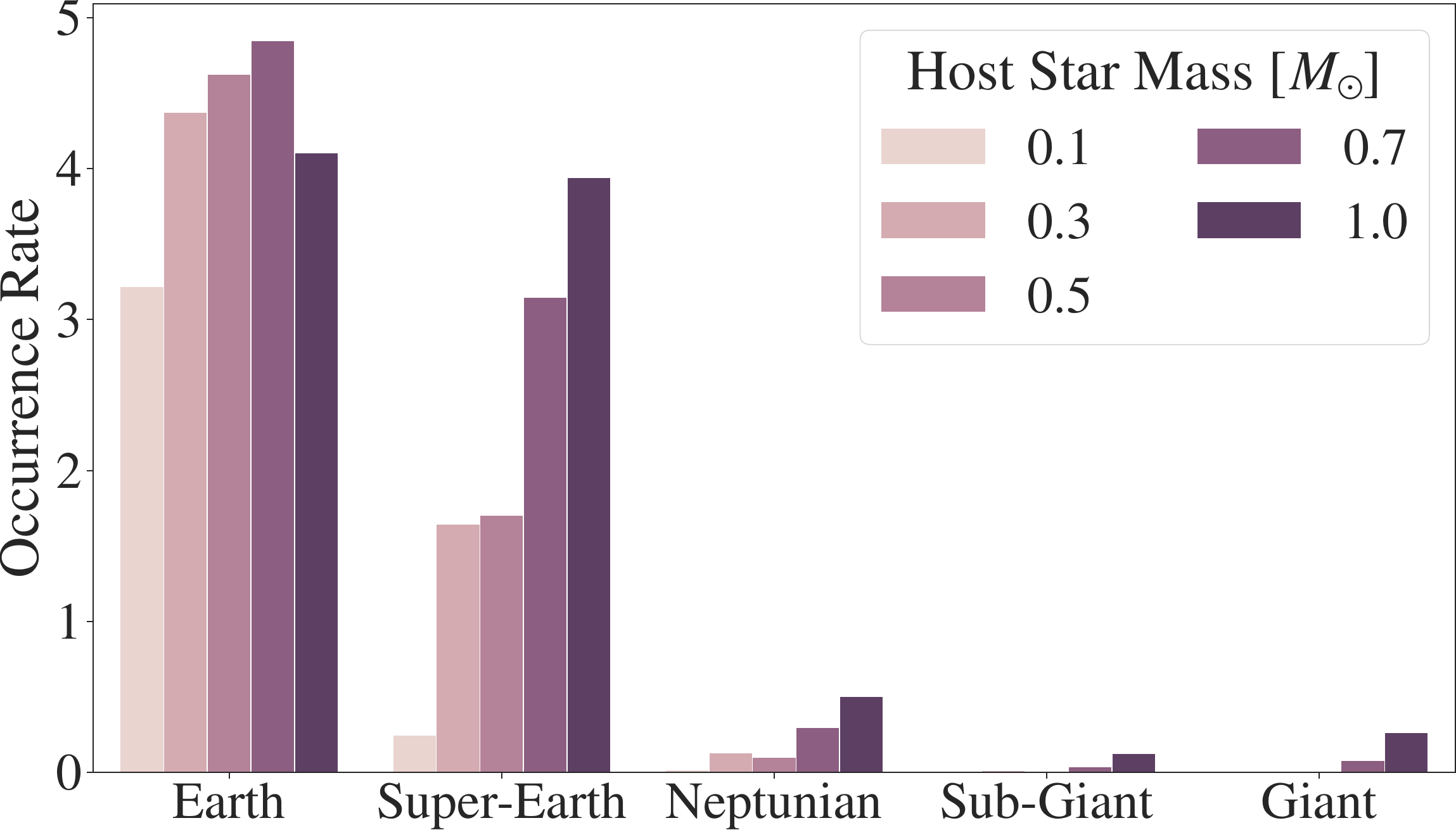}
    \caption{Variations of planet occurrence rates per category given different host star masses, $N_\mathrm{Embryos} = 50$ \citep{Burn2021}.}
    \label{fig:original_sample_host_star_mass}
\end{figure}

\subsection{Influence of the number of embryos and stellar mass}
One asepct that merits further interest is how the relation of planet populations to the systematically varied quantities. In \cref{fig:original_sample_number_of_embryos}, we show the total number of planet per category in all the simulation runs with a varying number of initial embryos (and fixed host star mass, $M_\star = 1 M_\Sun$). The number of massive planets (Neptunians, Sub-giants, giants) are nearly constant for all runs with $N_\mathrm{Embryo} \geq 10$. The number of Earth-like planets and Super-Earths, on the other hand, increases significantly with the number of embryos. \cref{fig:original_sample_host_star_mass} shows a similar figure for a fixed number of initial embryos ($N_\mathrm{Embryo} = 50$) and varying host star mass. Here, the number of Earth-like planets does not depend strongly on the host star mass for masses between 0.1 and 1 $M_\Sun$. The more massive planets however become more frequent for massive stars. Giant planets, in particular, only occur for host star masses $> 0.3 M_\Sun$. This agrees with our previous findings that occurrence rates depend primarily on the available solid mass. \citet{Burn2021} assume that the disc mass scales linearly with the mass of the host star. Low-mass stars therefore are less likely to possess the $\sim 1 M_\mathrm{Jup}$ solid material needed to form giant planets within the Bern model.

\begin{table*}[!ht]
    \centering
    \caption{Comparison of key properties between the simulated MW analogue and observational constrains. The columns correspond to the virial mass, stellar mass, effective radius of the Sérsic profile, scale length of the disc, Sérsic index, and the [Fe/H] gradient. References:
    [1] \citet{Posti2019a},
    [2] \citet{Hattori2018},
    [3] \citet{Monari2018},
    [4] \citet{Watkins2019},
    [5] \citet{Zaritsky2017},
    [6] \citet{Licquia2015a}, 
    [7] \citet{McMillan2017},
    [8] \citet{Kafle2014},
    [9] \citet{Libeskind2020}, 
    [10] \citet{Bland-Hawthorn2016a}, 
    [11] \citet{Lemasle2007}, 
    [12] \citet{Luck2011},
    [13] \citet{Genovali2014},
    [14] \citet{Lemasle2018}.}
    \begin{tabular}{l|rrrrrr}
         & $M_\mathrm{vir}$ & $M_\mathrm{stars}$ & $R_\mathrm{eff}$ & $R_\mathrm{disk}$ & 
         $n_\mathrm{S\acute{e}rsic}$ 
         & $\nabla$ [Fe/H] \\
         & $10^{12} M_\Sun$ & $10^{10} M_\Sun$ & kpc & kpc & & $10^{-2}$ dex/kpc \\
        \addlinespace
        \hline\hline
        \addlinespace
        HESTIA analogue & 1.04 & 5.87  & 0.71 & 2.93  & 1.80 & -2.7\\ 
        \addlinespace
        Milky Way & 1.0--2.1 & 5--10 & $\sim 1$ & $2.5 \pm 0.5$ & -- & - (3.8 -- 8.0) \\ 
        References & [1, 2, 3, 4, 5] & [6, 7, 8] & [9] & [10] &  & [11, 12, 13, 14]\\
    \end{tabular}
    \label{tab:milky_way_comparison}
\end{table*}
\subsection{Comparison with observations}
\label{sec:model_observation_comparion}
The relationship between stellar metallicity [Fe/H] and occurrence rates of giant planets has been well-established over the years \citep{Gonzalez1997, Santos2001, Fischer2005, Johnson2010}. Similarly, close-in ($P < 10$ days) small ($< 2~M_\Earth$) planets are also more likely to be found around metal-rich stars \citep{Mulders2016, Lu2020}, hinting at a metallicity-dependence for rocky planets as well, albeit much weaker than for giant planets. \citet{Petigura2018} find that warm Super-Earths with a period between 10-100 days have a nearly constant occurrence rate over metallicities in the range -0.4 < [Fe/H] < 0.4, while warm Sub-Neptune occurrence rates double over the same range. The radii of planets have also been found to increase with host star metallicity \citep{Narang2018, Swastik2022}. This agrees well with our assignment model, where the metallicity-dependence is similarly weakened for low-mass planets (\cref{fig:contourplots_planet_model}). Giant planets are generally found to be rare around metal-poor stars \citep{Buchhave2012, Thorngren2016}, which is also reproduced in the assignment model.

\citet{Howard2012} and \citet{Mulders2015} find that the occurrence rate of Earth- to Neptune-sized close-in planets decreases for increasing stellar mass and temperature. \citet{He2021} and \citet{Yang2020} suggest that the fraction of stars with planets changes from $\sim 0.3$ for F-type stars to $\sim 0.8$ for K-types. Our assignment model, on the other hand, predicts an increase in both Super-Earth and Neptunian occurrence rates with increasing stellar mass (although the occurrence rate for Earth-like planets slightly decreases for host star masses between 0.7 and 1 $M_\Sun$), therefore contradicting the observations. \citet{vanderMarel2021} suggest that gap formation in the protoplanetary disc caused by giant planet formation could impede inward migration of lower mass planets, leading to the observed anti-correlation. In the subsequent analysis, we therefore only compare planet population at fixed host star mass $M_\star$, rather than across masses in order to avoid biases based on this spurious behaviour of the model. 

\section{Milky Way analogue in HESTIA}
\label{sec:HESTIA}
High-Resolution Environmental Simulations of The Immediate Area (HESTIA) is a suite of hydrodynamical cosmological simulations designed to replicate the Local Group in its correct cosmographic context \citep{Libeskind2020}. It is based on the \texttt{Arepo} code for 
gravitational N-body systems and magnetohydrodynamics, while the baryonic physics is handled 
by the \texttt{Auriga} \citep{Grand2017} model of galaxy formation.

The initial conditions are derived from cosmographic observations to faithfully reconstruct the Local Group, both in terms of its internal structure and nearby large-scale structures. It accurately recreates prominent features of the local Universe such as the Virgo Cluster and Local Void, including their appropriate masses and relative locations to the Local Group \citep{Tully2009, Karachentsev2010}.
Centered on the Local Group midpoint, the simulation contains a high-resolution region with a radius of 3-5 Mpc. An ensemble of dark matter-only and low-resolution hydrodynamical simulations were run in order to construct a Local Group-like pair of MW  and Andromeda analogues with accurate mass ranges and separations. The three most promising simulations have been re-run at a higher resolution.
These analogues closely resemble their observational counterparts in terms of total mass, mass ratio, stellar disc mass, morphology separation, relative velocity, rotation curves, bulge-disc morphology, satellite galaxy stellar mass function, and satellite radial distribution. In terms of their formation history, 
the MW and Andromeda analogues grow slower than than comparable halos in unconstrained 
\citep{McBride2009, Libeskind2020}.
In the highest-resolution runs, the attained mass resolutions of the individual particles are 
approximately $m_\mathrm{DM} \sim 10^5$, $m_\mathrm{gas} \sim 10^4$, and $m_\mathrm{stars} \sim 
10^4$, while the spatial resolution is around $220$ pc. It ought to be highlighted that the resolution is not sufficient to resolve individual stars, rather the star particles are treated as single stellar populations with uniform age and metallicity. 

In this study, we focus on the MW analogue in the highest-resolution \texttt{37\_11} 
run, which will be described in more detail in the following sections.

\begin{figure*}
     \centering
    \begin{subfigure}[b]{0.245\textwidth}
        \centering
        \includegraphics[width=\textwidth]{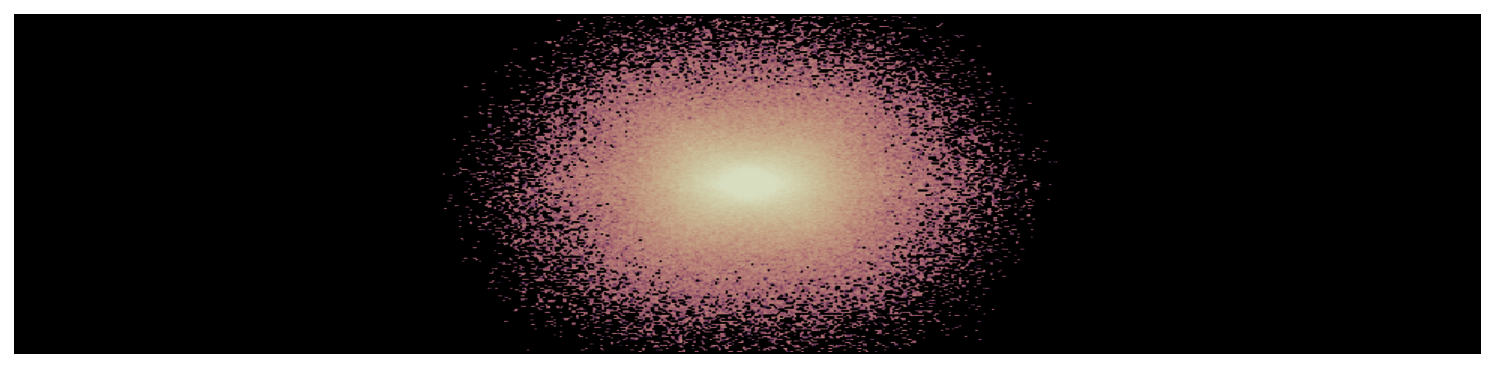}
        \label{fig:bulge_map_x}
     \end{subfigure}
     \begin{subfigure}[b]{0.245\textwidth}
        \centering
        \includegraphics[width=\textwidth]{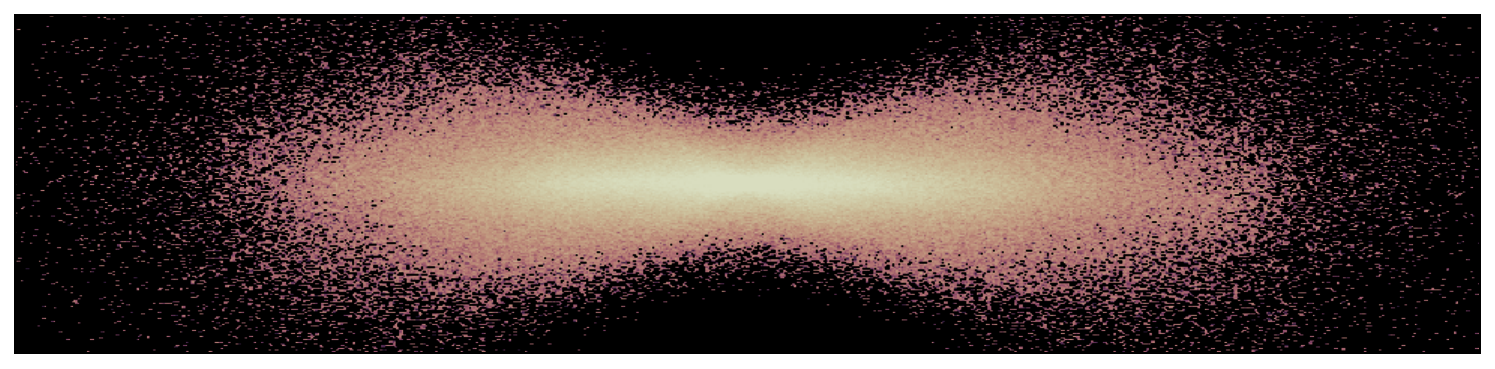}
        \label{fig:thin_disk_map_x}
     \end{subfigure}
     \begin{subfigure}[b]{0.245\textwidth}
        \centering
        \includegraphics[width=\textwidth]{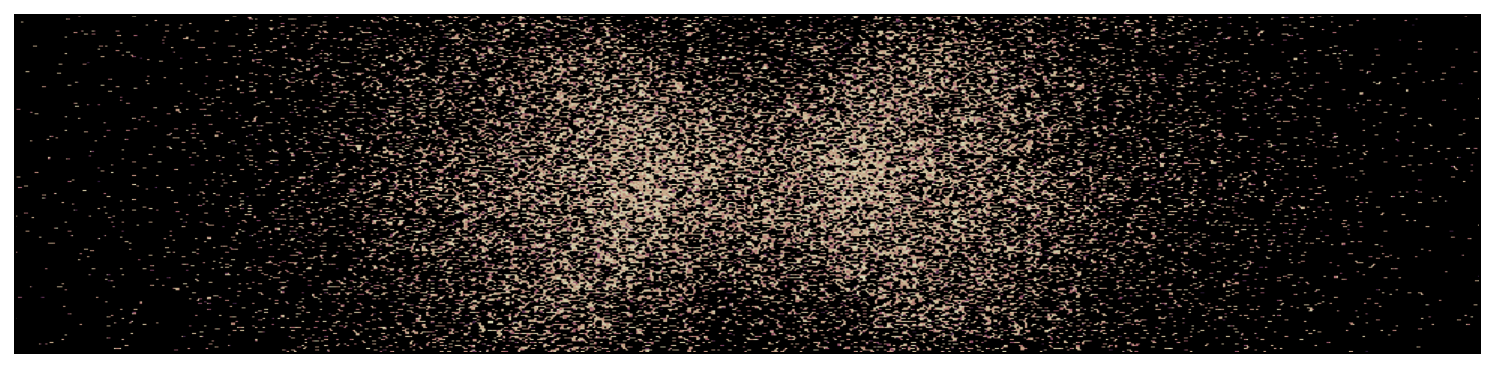}
        \label{fig:thick_disk_map_x}
     \end{subfigure} 
     \begin{subfigure}[b]{0.245\textwidth}
        \centering
        \includegraphics[width=\textwidth]{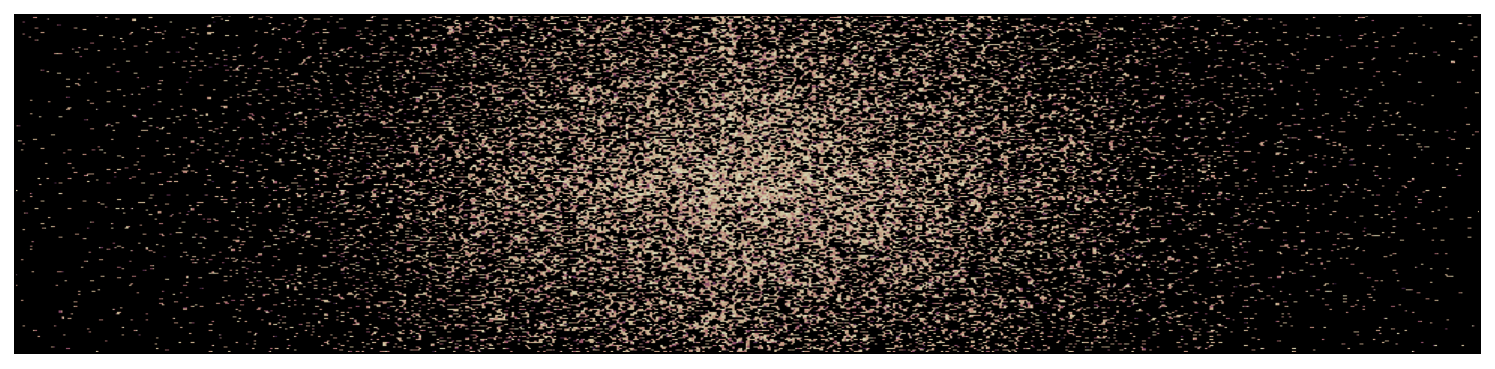}
        \label{fig:halo_map_x}
     \end{subfigure}
     \\[-3ex]
     
    \begin{subfigure}[b]{0.245\textwidth}
        \centering
        \includegraphics[width=\textwidth]{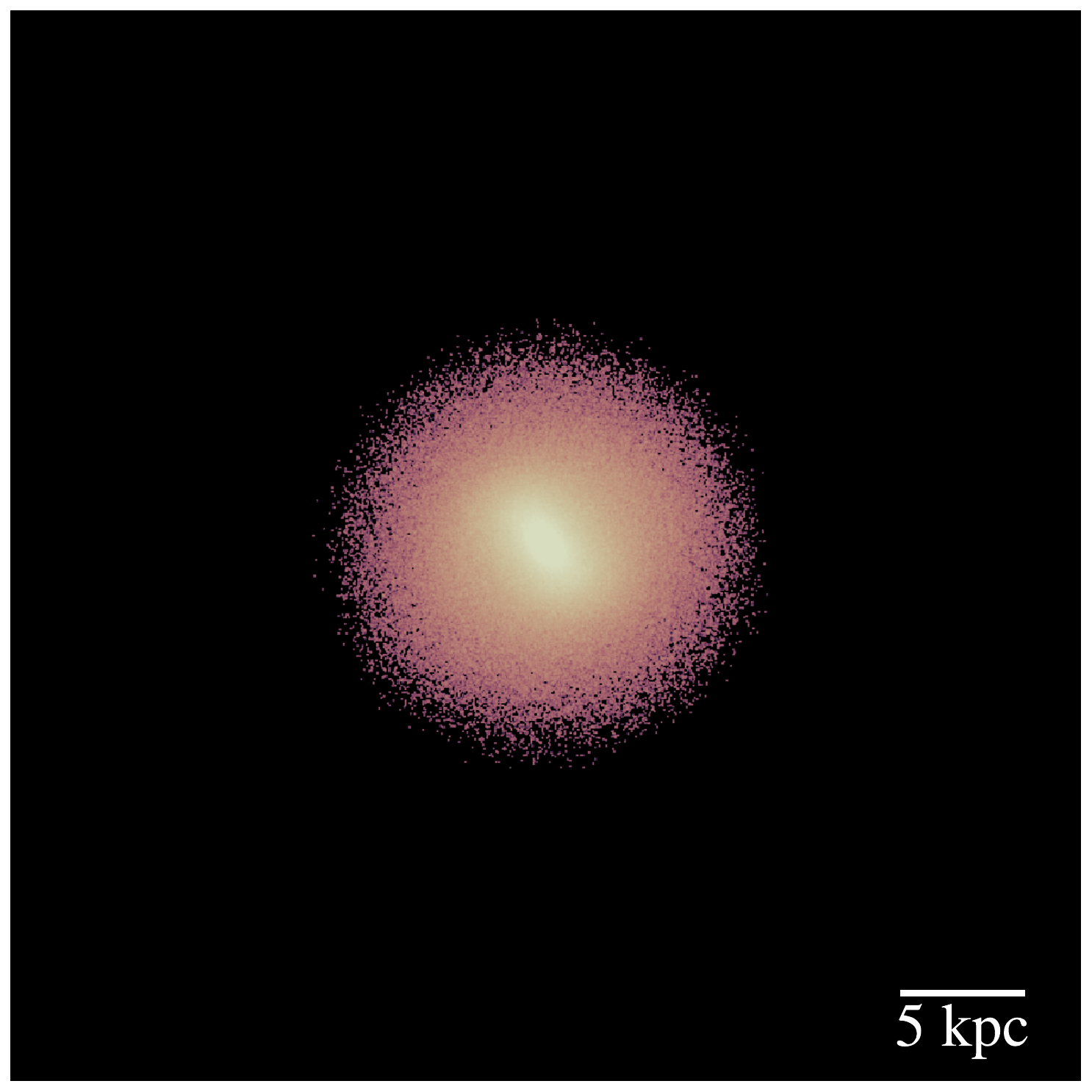}
        \caption{Bulge.}
        \label{fig:bulge_map_z}
     \end{subfigure}
     \begin{subfigure}[b]{0.245\textwidth}
        \centering
        \includegraphics[width=\textwidth]{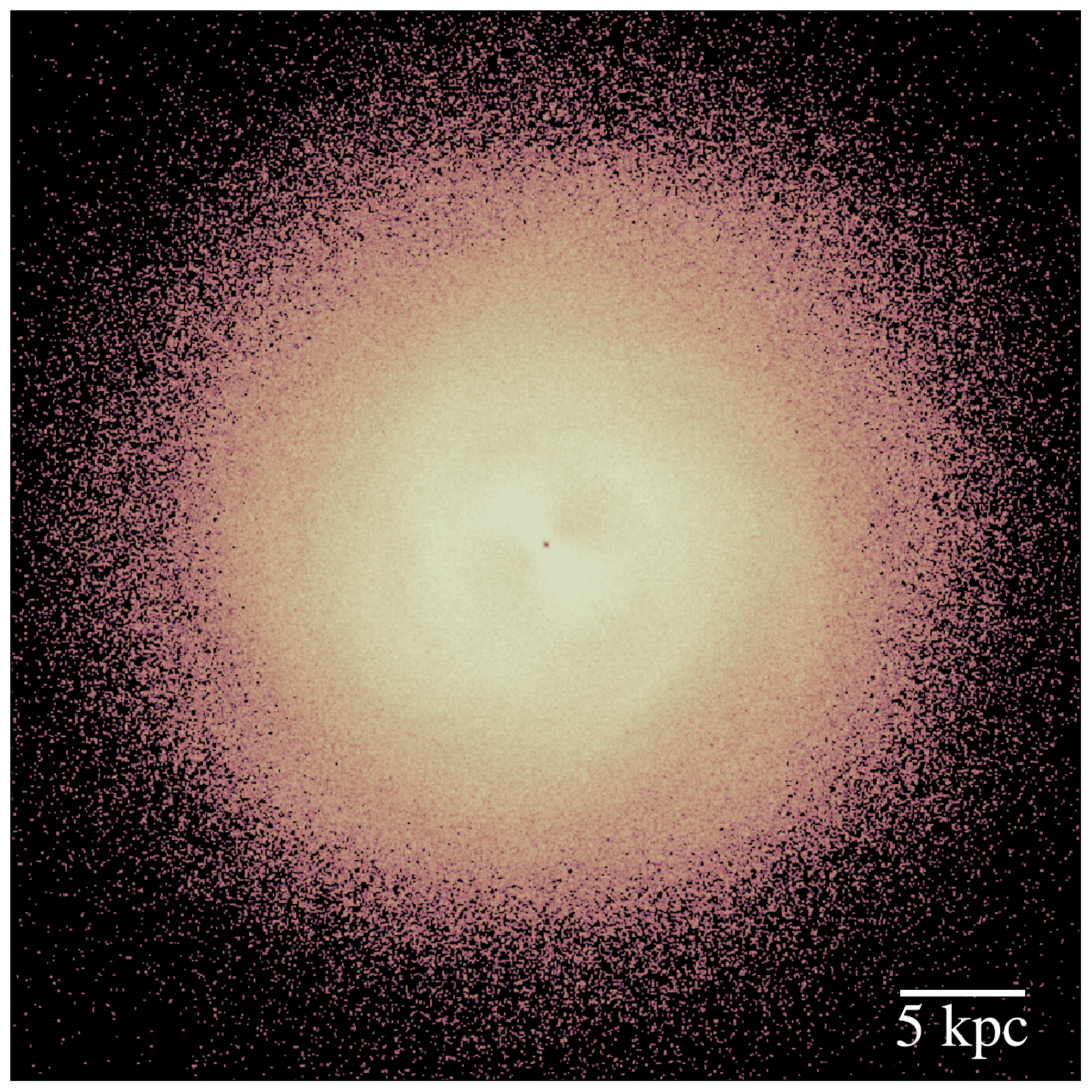}
        \caption{Thin disc.}
        \label{fig:thin_disk_map_z}
     \end{subfigure}
     \begin{subfigure}[b]{0.245\textwidth}
        \centering
        \includegraphics[width=\textwidth]{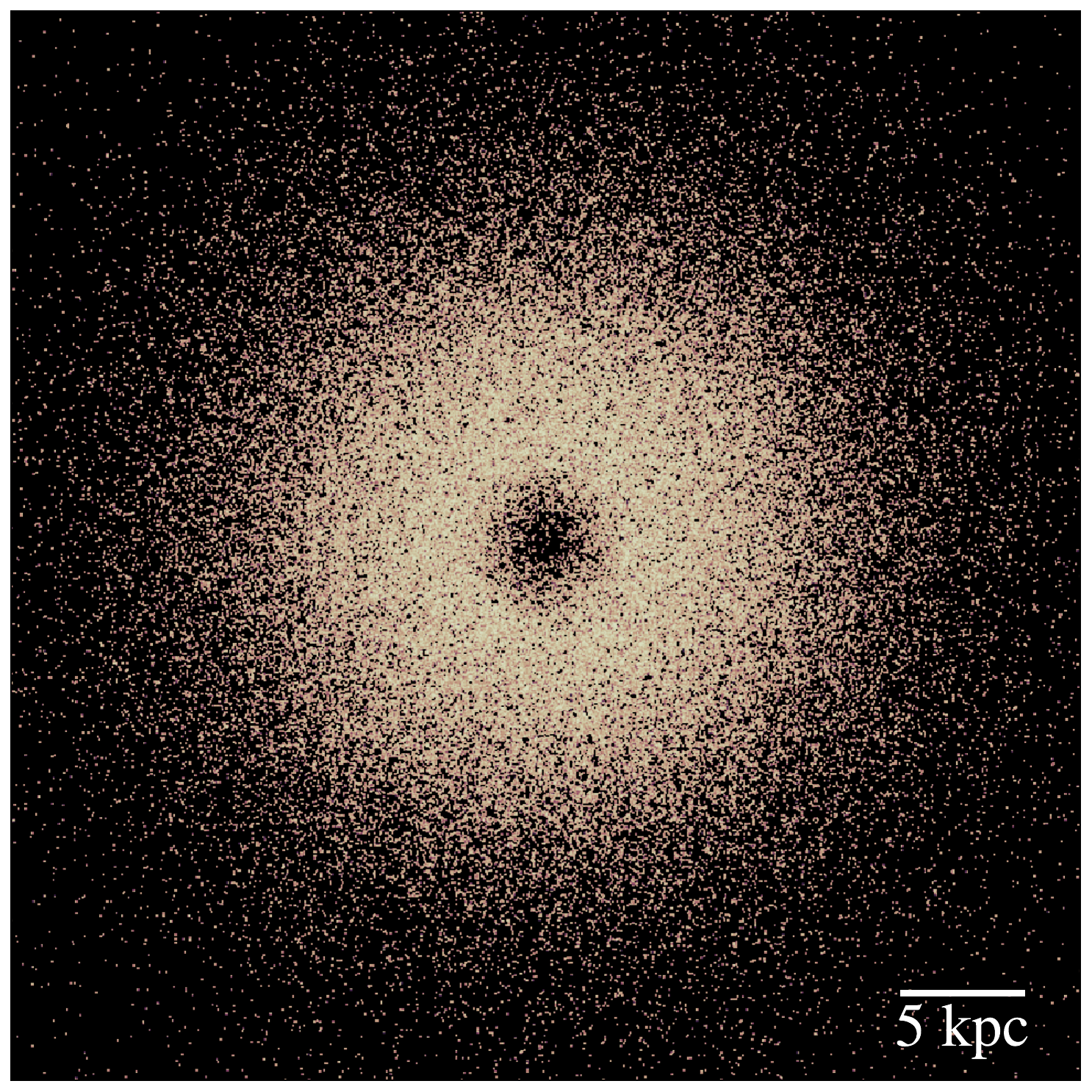}
        \caption{Thick disc.}
        \label{fig:thick_disk_map_z}
     \end{subfigure} 
     \begin{subfigure}[b]{0.245\textwidth}
        \centering
        \includegraphics[width=\textwidth]{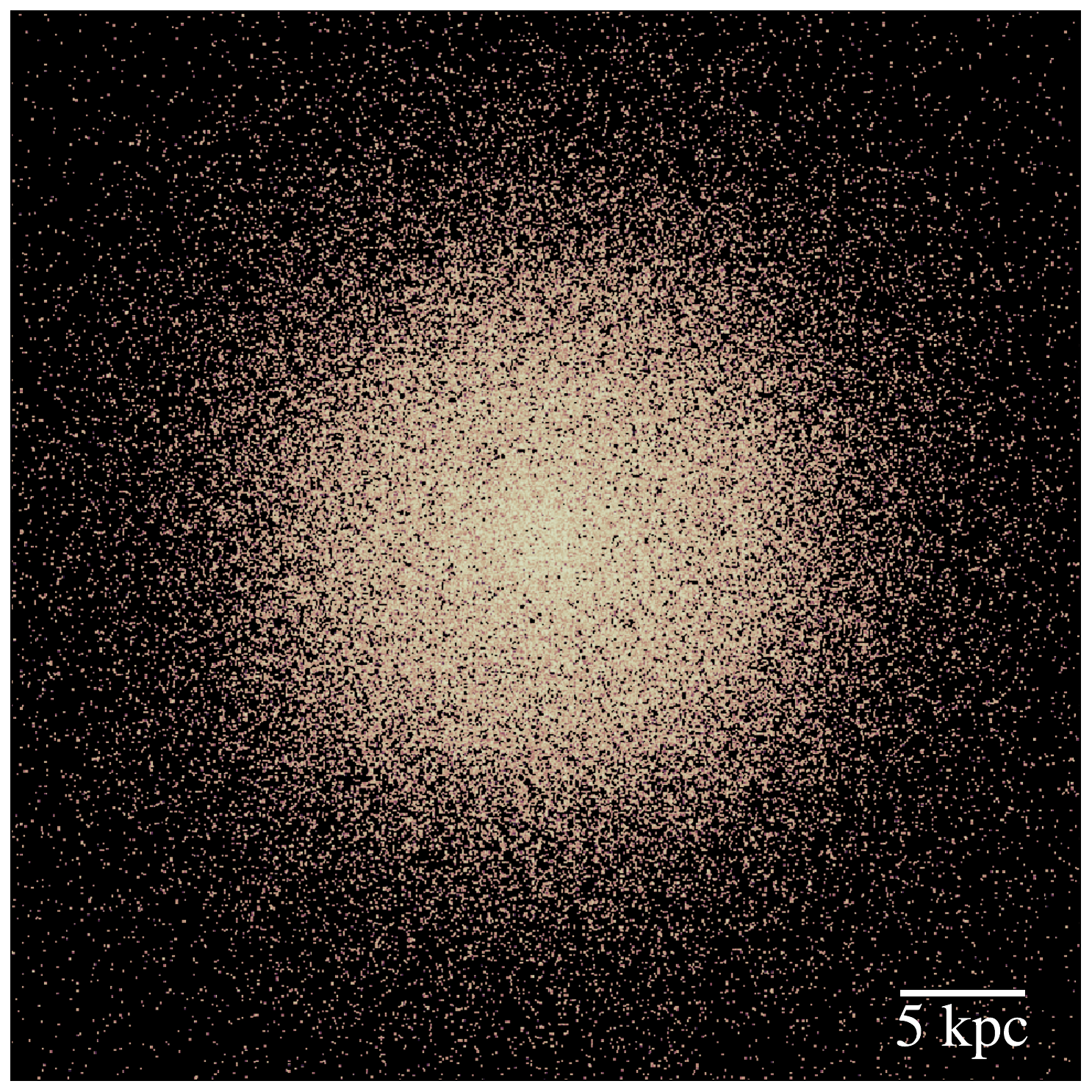}
        \caption{Halo.}
        \label{fig:halo_map_z}
     \end{subfigure}
     \caption{Side and face-on maps of the distribution of stars for the different galaxy components.}
     \label{fig:component_maps}
\end{figure*}
\subsection{General comparison}
As described in the previous section, the MW analogues in HESTIA have been purposefully crafted to mirror a galaxy that is MW-like in many properties relevant for galaxy formation and evolution. They were, however, not constructed to be a perfect copy of the MW, and some of the properties that are more relevant for planet formation were understandably not considered in the cosmological context they were built for. We therefore compare some of the more local and structural properties of our chosen MW analogue in more detail.

The virial mass and stellar masses of the analogue closely match those of the MW within observational constraints (see \cref{tab:milky_way_comparison}). This is unsurprising, given that the simulations were specifically designed to achieve this result. Assuming a \citet{Chabrier2003a} IMF, the stellar mass corresponds to roughly 135 billion stars in the analogue, of which 2.3 billion stars are main-sequence stars within 5\% of the mass of the sun. \citet{Libeskind2020} analyzed the structural parameters of the analogue by fitting a surface brightness profile, which they modelled as the sum of a Sérsic profile describing the bulge and an exponential profile for the disc. The obtained effective bulge and disc radii are are in good agreements with those of the MW (\cref{tab:milky_way_comparison}). While the Sérsic index of the MW bulge lacks precise constraints, the simulation value of 1.80 is comparable to the observational value of $2.2 \pm 0.3$ for the Andromeda galaxy \citep{Courteau2011}, the closest comparable spiral galaxy.

Furthermore, as discussed in \cref{sec:assignment_model}, a major factor in the planet assignment model is the metallicity of the host star. Within HESTIA, 9 elemental species are tracked throughout the simulation: H, He, C, N, O, Ne, Mg, Si, and Fe. From these, we can calculate the iron abundance [Fe/H] for any individual star particle. We assume all stars within a star particle (a single stellar population) share the same metallicity. The metallicity gradient of the MW analogue between 2 and 10 kpc from the galactic centre is -0.027 dex/kpc. This is a shallower slope compared to the values determined for the MW in observational studies \citep{Lemasle2007, Luck2011, Genovali2014}. Although, a more recent study by \citet{Lemasle2018} found MW values of –$0.045 \pm 0.007$ dex/kpc and –$0.040 \pm 0.002$ dex/kpc using F/10 double-mode Cepheids, closer to that of the analogue. Nevertheless, within the HESTIA analogue, an average solar neighbourhood metallicity value of [Fe/H] $\sim$ -0.1 to 0 \citep{Haywood2001} is reached at a distance of $\sim 7$--$8$ kpc from the galactic centre, in good agreement with the distance of the solar neighbourhood from the MW centre \citep[$\sim 8.2$ kpc,][]{McMillan2017}.

\subsection{Galaxy components}
Within the solar neighbourhood, the stars in the MW are generally categorised into three groups: the thin disc, the thick disc, and the stellar halo \citep{Gilmore1983}. These populations are distinguished by their kinematics \citep{Piffl2014, Sanders2015}, stellar age \citep{Schuster2006}, and chemical composition \citep{Bensby2014, Masseron2015, Hawkins2015}. Thin disc stars have lower orbital velocity and higher velocity dispersion compared to stars in the thick disc. Chemically, thick disc stars are metal-poor and $\alpha$-enhanced relative to thin-disc stars \citep{Prochaska2000, Reddy2003, Reddy2006}. Halo stars, in contrast, have even larger velocity dispersion \citep{Bond2010} and are metal-poor. Near the galactic centre, there is also a population of bulge stars characterised by high velocity dispersion \citep{Portail2015} and a wide range of metallicities, spanning from [Fe/H] $\sim$ -1.5 to [Fe/H] $\sim$ +0.5 \citep{Zoccali2008}. These variations in formation history and chemistry are likely to influence the planet populations within these galactic components \citep{Bashi2022, Nielsen2023}. Therefore, we decompose the HESTIA analogue into four categories, broadly consistent with the four MW components.

While observational surveys benefit strongly from the separation of components based on combined kinematics and chemistry criteria \citep{Bashi2019}, we adopt a simpler classification approach in this work. We categorised objects based on the binding energy and circularity of the star particles, aligning with previous studies on galaxy formation simulations \citep{Abadi2003, Yu2023} by employing the morphological decomposition code \texttt{MORDOR} by \citet{Zana2022}. Face- and side-on projection maps of the components are shown in \cref{fig:component_maps}.

The determined galaxy components broadly align with anticipated properties and observations of the MW. The bulge, thin disc, and thick disc closely match those of the MW in terms of stellar mass, mean stellar age, and mean [Fe/H], as well as the minor-to-major axis ratio. Discrepancies are found for the stellar halo, which is more massive and metal-rich in the HESTIA analogue compared to the MW. This can likely be attributed to a misclassification of some bulge and disc stars based on the purely kinematic decomposition. The key attributes are summarised in \cref{tab:component_comparison}, and some distributions are shown in \cref{fig:ridgeplots_galaxy_components}.
\begin{table*}[!ht]
    \centering
    \caption{Comparison of mass, mean stellar age, mean metallicity, and major-to-minor axis ratio of the galaxy components in the HESTIA MW analogue (simulation) with observational values. The observational major-to-minor axis ratios for the thin and thick disc are roughly approximated as the ratio between scale height (at the location of the sun) and scale length. References: [1] \citet{Bland-Hawthorn2016}, [2] \citet{Deason2019}, [3] \citet{Kilic2017}, [4] \citet{Sit2020}, [5] \citet{McWilliam2016}.}
    \begin{tabular}{l|rr|rr|rr|rr}
         & \multicolumn{2}{c|}{$M_\mathrm{stars}$ ($10^{10} M_\Sun$)} & \multicolumn{2}{c|}{$\langle \mathrm{Age}\rangle$ (Gyr)} & \multicolumn{2}{c|}{$\langle \mathrm{[Fe/H]}\rangle$} & \multicolumn{2}{c}{$c/a$} \\
         & HESTIA & MW & HESTIA & MW & HESTIA & MW & HESTIA & MW \\ 
        \addlinespace
        \hline\hline
        \addlinespace
        Bulge & 1.58 & 1.4 -- 1.7 & 8.6 & $\sim$ 8  & -0.03 & $\sim$ 0.06 & 0.35 & 0.26 -- 0.32 \\ 
        Thin Disc & 3.50 & 2.5 -- 4.5 & 6.0 & $\sim$ 7 -- 8 & 0.02 & $\sim$ -0.2 & 0.06 & 0.08 -- 0.17 \\ 
        Thick Disc & 0.35 & 0.3 -- 0.9 & 8.5 & $\sim$ 8 -- 10 & -0.55 & $\sim$ -0.6 & 0.38 & 0.33 -- 0.66 \\ 
        Halo & 0.43 & 0.04 -- 0.14 & 9.6 & $\sim$ 9 -- 13 & -0.74 & $\sim$ -1.5 & 0.62 & 0.60 -- 0.90 \\ 
        \addlinespace
        References &  & [1, 2] &  & [3, 4] &  & [2, 5] & & [1]\\ 
        
    \end{tabular}
    \label{tab:component_comparison}
\end{table*}

\subsection{Integrating the planet model}
To integrate the planet assignment model (\cref{sec:assignment_model}) into the HESTIA analogue, we associate each star particle with a certain number of planets for each planet type, based on the properties of the star particle. With a mass of approximately $10^4 M_\Sun$, we associate a star particle with a single stellar populations of uniform age and metallicity \citep{Chevance2022b}.

The hydrogen and iron masses of the star particles are inherited from the gas cells from which the particles originate, meaning each particle is equipped with an [Fe/H] value at birth. This value tracks the chemical evolution of the galaxy and can be fed directly into the planet assignment model.

Since the resolution of the simulation is on the order of approximately 200 pc, processes that potentially influence the formation and evolution of the protoplanetary discs cannot be tracked. Instead, we associate every star particle with a random initial disc mass $M_\mathrm{g}$ sampled from the distribution provided by \citet{Emsenhuber2021a}, and scaled to the host star mass according to the prescription given by \citet{Burn2021}. This approach includes the assumption that the mass distribution of discs does not change significantly in different regions of the galaxy. The implications of this assumption are discussed in \cref{sec:limitations}.

As discussion in \cref{sec:model_observation_comparion}, the assignment model predicts an increase of low-mass planet frequency with stellar mass, which is a behaviour not seen in observations. Since low-mass stars are common than massive ones, this could bias our conclusions if not controlled for. We therefore chose to analyse the planet populations for different host star masses, $M_\star$, independently, and not across stellar masses. To calculate the number of eligible stars associated with the star particle, we integrate the IMF between $M_\star \pm 5\%$, where $M_\star$ is the chosen (fixed) host star mass. Since planet formation is modeled only for main-sequence stars, we exclude stars with a lifetime shorter than the particle age.

In short, the planet prescription at fixed host star mass $M_\star$ is as follows. First, we choose an NGPPS population by fixing $N_\mathrm{Embryos}$ and $M_\star$. Then, we obtain the age, [Fe/H], and mass of the star particle. Next, we sample the Monte Carlo variable $M_\mathrm{g}$ from the distribution given by \citet{Emsenhuber2021a}. We integrate the (normalised) \citet{Chabrier2003a} IMF, $\xi$, in the interval $M_\star \pm 5\%$, and multiply by the mass of the star particle to obtain the number of stars. Post-main sequence stars are removed based on the age of the star particle \footnote{We assume a lifetime of 10 Gyr for a star with $M_\star = 1 M_\Sun$, and scale the lifetimes of less massive stars by a simple power law with an index of 2.5.}. We then assign the number of planets $n$ per star using the planet assignment model for each category (\cref{sec:assignment_model}). Finally, we multiply the number of stars by the number of planets per star. \footnote{The code used for the analysis in this work has been compiled into a python module \texttt{skaro} and can be found at \url{github.com/ChrisBoettner/skaro}.}

\subsection{Metal-poor stars}
\label{sec:metalpoor_stars}
The [Fe/H] distribution in the HESTIA analogue covers a wide range of values, with a majority of particles residing between [Fe/H] = -2.5 and [Fe/H] = 1. This interval is considerably broader than the range considered in the NGPPS database, which is limited to $-0.6 < \mathrm{[Fe/H]} < 0.5$. This range was selected to cover the typical metallicities within the solar neighbourhood. In the HESTIA analogue, 80\% of the star particles conform to this range. However, around 48\% of the star particles in the thick disc and stellar halo fall beneath the NGPPS lower threshold.

Drawing from trends highlighted in the planet assignment model discussed in \cref{sec:assignment_model}, considerably larger disc masses would be needed at such low metallicities to provide sufficient solid material for planet formation, especially for planets more massive than $2 M_\Earth$. In addition, observations suggest that protoplanetary discs disperse faster in low-metallicity environments \citep{Yasui2009, Yasui2021}.
From an observational standpoint, only 26 planets have been confirmed orbiting stars with [Fe/H] < -0.6 at the time of writing, with a minimum metallicity of [Fe/H] = -1.0, providing very limited information on the effects of low-metallicity environments on planet formation. \footnote{\url{exoplanetarchive.ipac.caltech.edu}}

Therefore, in our analysis, we consider two extreme scenarios. Firstly, we presume the planet frequencies to be unchanging at extremely low metallicities, adopting the same values as in the [Fe/H] = -0.6 case. Secondly, we assume that short disc lifetimes and a lack of solid material render planet formation highly 
unlikely to impossible, and assign an occurrence rate of zero to those particles.

\section{Results}
\label{sec:results}

\subsection{Planets around Sun-like stars}
\label{sec:sunlike_stars}
\begin{figure}
    \centering
    \includegraphics[width=\columnwidth]{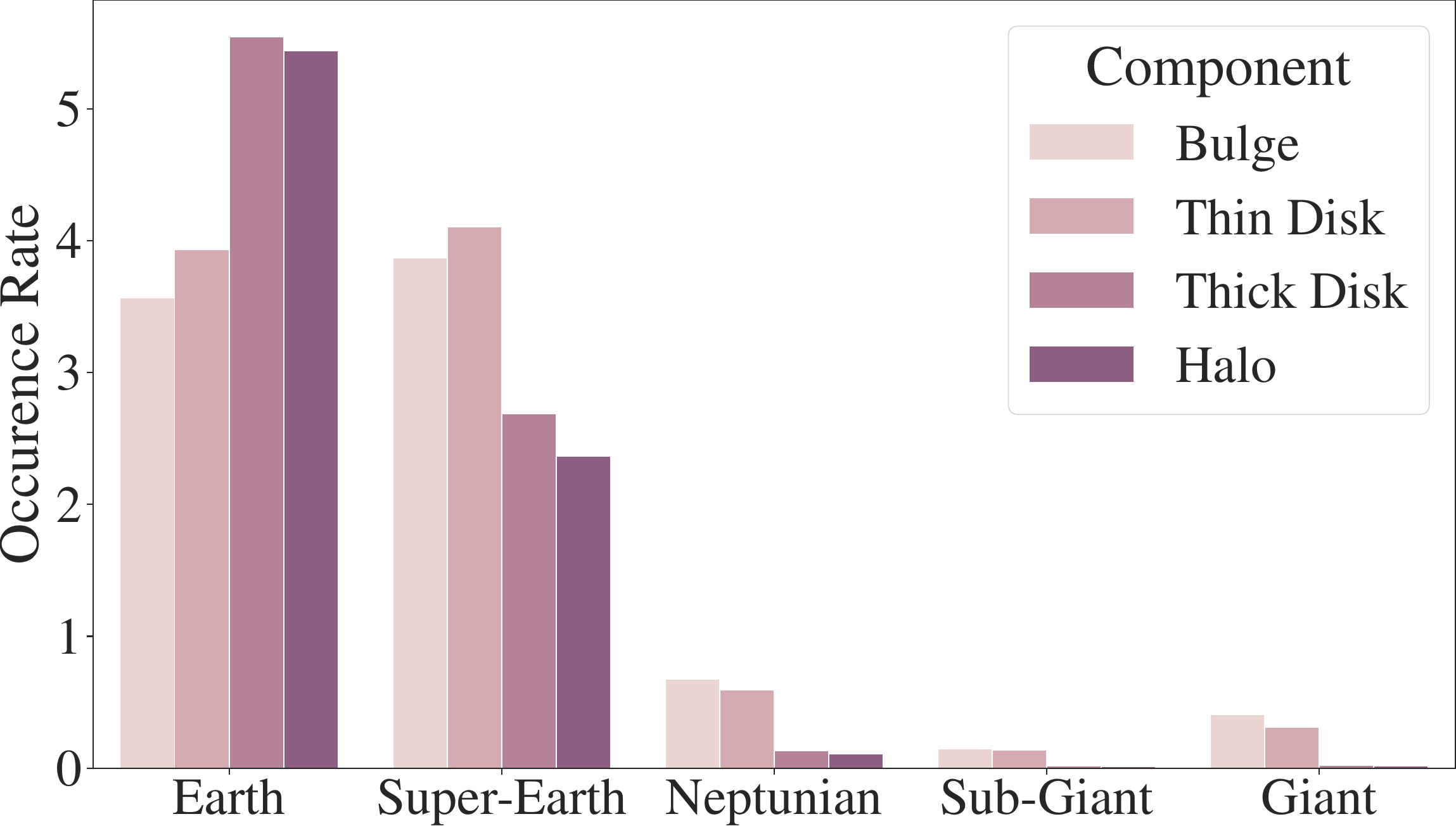}
    \caption{Occurrence rates of different planet types in the four galaxy components. The occurrence rates are calculated for the $M_\star=1 M_\Sun$ and $N_\mathrm{Embryo}=50$ run.}
    \label{fig:occurrence_rates}
\end{figure}

\begin{figure}
    \centering
    \includegraphics[width=\columnwidth]{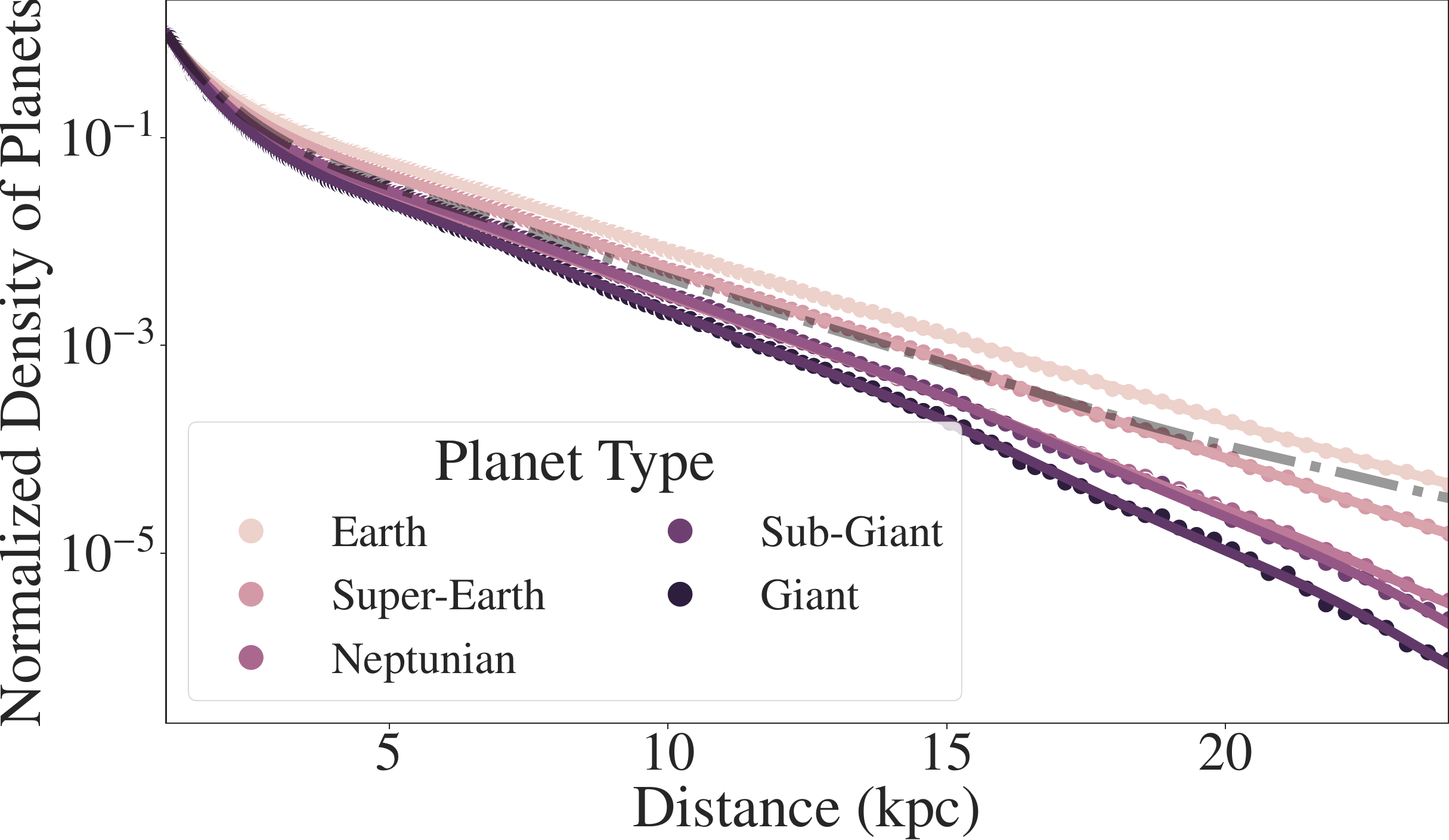}
    \caption{Spatial density of planets as a function of distance from the galactic centre (normalised to $R=0$). The points correspond to the estimated values per radial bin, while lines are smoothing splines to show trends. The grey, dash-dotted line corresponds to the normalized density of stars.}
    \label{fig:normalized_planet_density}
\end{figure}

\begin{figure*}
     \centering
     \begin{subfigure}[b]{0.97\textwidth}
        \centering
        \includegraphics[width=\textwidth]{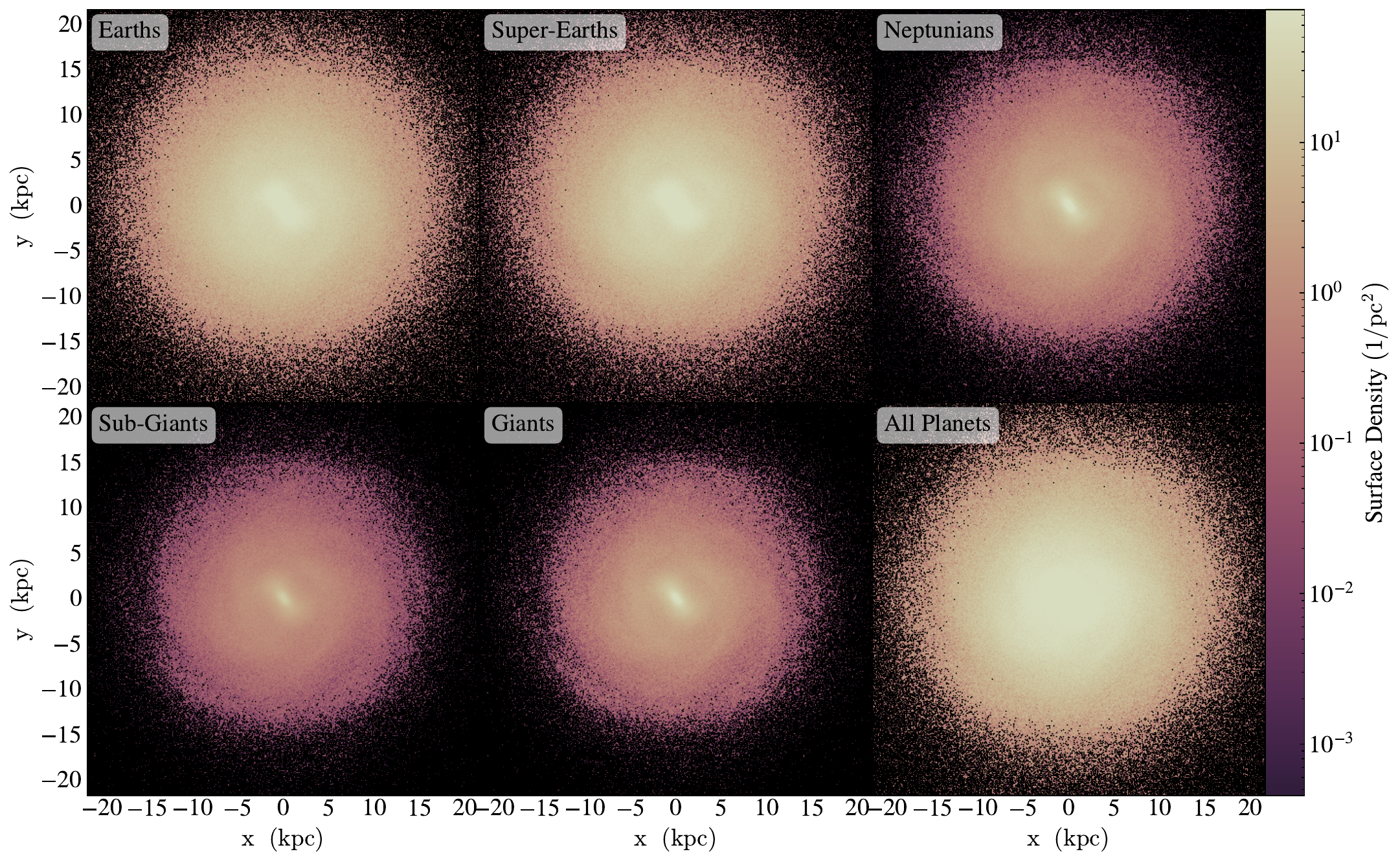}
        \caption{Face-on view.}
        \label{fig:face_map_sunlike}
     \end{subfigure}

     \begin{subfigure}[b]{0.97\textwidth}
        \centering
        \includegraphics[width=\textwidth]{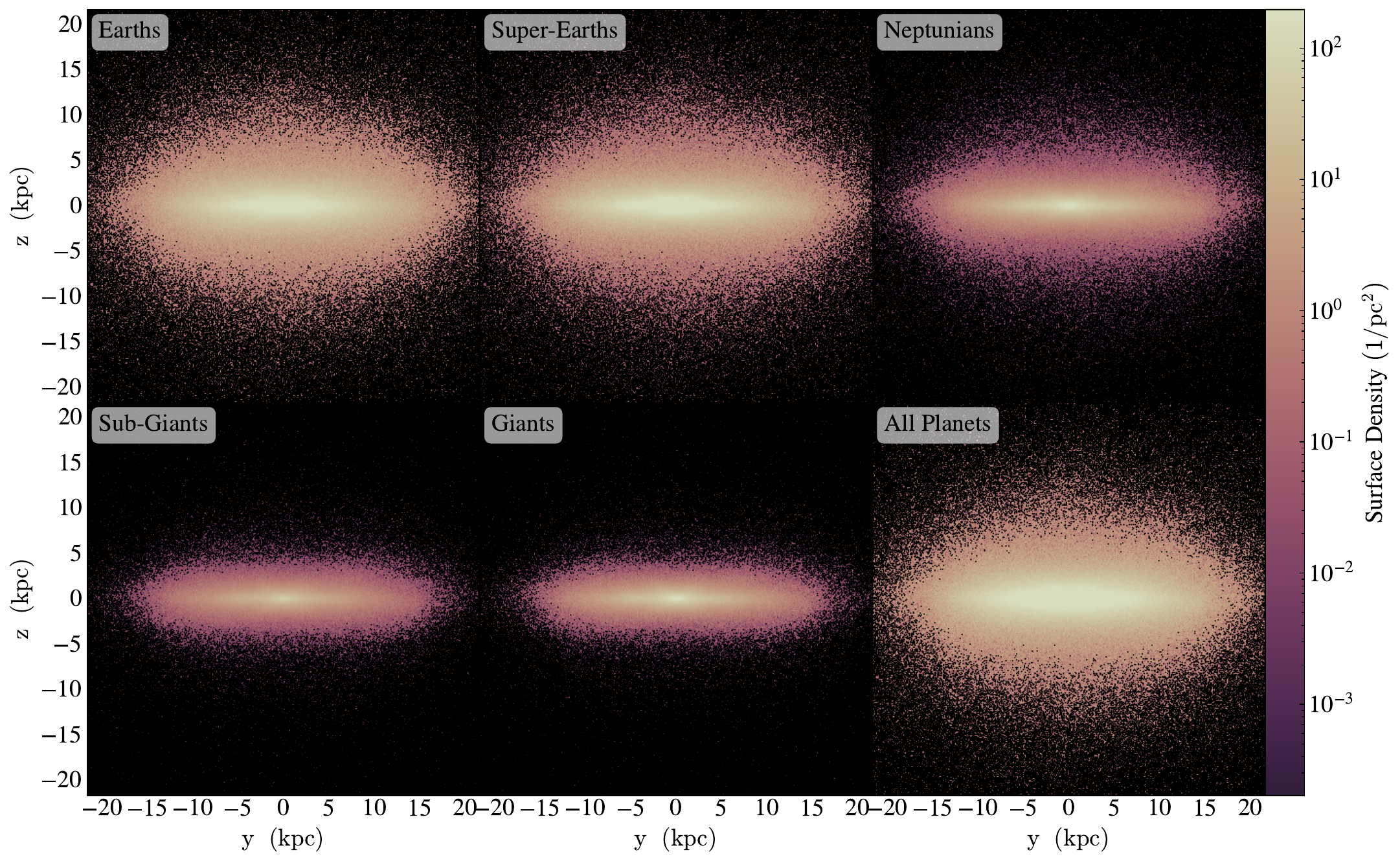}
        \caption{Side-on view.}
        \label{fig:side_map_sunlike}
     \end{subfigure}
     \caption{Maps of distribution of planets for each planet Type in the HESTIA MW analogue for stars with $M_\star = 1 M_\Sun$. The assignment is based on the $N_\mathrm{Embryo} = 50$ NGPPS population.}
     \label{fig:planet_maps_sunlike}
\end{figure*}

Focussing initially on planets orbiting Sun-like stars, which (for our purposes) are stars within ±5\% of the Sun's mass, we observe distinctive patterns in occurrence rates across different planet categories. For the $N_\mathrm{Embryo}=50$ run, the occurrence rates in the different components are shown in \cref{fig:occurrence_rates}. The planet populations in the bulge and thin disc are remarkably similar, the same is true for the thick disc and halo. This similarity is primarily attributed to these components covering similar metallicity regimes, which significantly influences planet occurrence rates in our model. A list of the occurrence rates can be found in \cref{tab:occurrence_rates}.

These occurrence rates of planets in the thin disc stellar population align closely with those produced by \citet{Emsenhuber2021a}, which is consistent with their initial aim to model the nearby (primarily thin disc) planet population. The population in the bulge has similar occurrence rates, with the bulge favouring more massive planets slightly. The most common type of planets in these regions are Earth-like planets and Super-Earths with an occurrence rate of 3.5 -- 4 planets per star. Sub-giants and giants have a frequency of 0.1-0.4, around one order of magnitude lower. However, the average age of a planetary system within these components differ. The average planet in the bulge is 7.6 Gyr old, while the average planet in the thin disc is younger at an age of 5.8 Gyr. Within a component, low-mass planets tend to be younger than more massive planets, due to the time scales of metal enrichment needed for giant planet formation (see \cref{tab:planet_ages}).

The thick disc and halo bear a distinct planet population compared to the more metal-rich regions. The lower metallicity of the stars favour low-mass planet formation over more massive ones. The occurrence rate of Earth-like planets is a 1.5 times larger than for the bulge and thin disc component, at 5.5 planets per star, with a comparable drop in the occurrence rate of super-Earths to $\sim$ 2.5 planets per star. More striking is the difference in occurrence rates for Neptunian and giant planets. Neptunian planets are five times more common in the thin disc and bulge compared to the thick disc and halo, with frequencies around 0.6. This ratio increases with increasing planet mass, the occurrence rates of giant planets with masses $>300 M_\Earth$ are 20 times higher in the metal-rich components compared to the thick disc and halo. The planets in these regions are also older on average, at 7.7 Gyr for the thick disc and 8.2 Gyr for the halo (see also \cref{tab:planet_ages}).

Taking a higher level view, we analysed the distribution of planet types across the whole galaxy. We show radial profiles in \cref{fig:normalized_planet_density}, while 2D distribution maps are shown in \cref{fig:planet_maps_sunlike}. Giant planets have a higher concentration in the centre of the galaxy, correlating with the region's higher metallicity. \cref{fig:normalized_planet_density} shows the normalised planet density as a function of distance from the galactic centre. The spatial density of planets depends on the stellar density and planet occurrence rates. Thus, unsurprisingly, the highest density of planets is found towards the galactic centre due to the large stellar density in this region. Moving away from the galactic centre, the number of giant planets drops faster than for lower mass planets, due to the higher metallicity-dependence of the giant planet occurrence rates. In total, 90\% of giant planets and 84\% of Earth-like planets can be found within 8 kpc from the galactic centre, roughly the distance of the Sun from the MW centre.

\begin{table}[!ht]
    \centering
    \caption{Average age of planets around Sun-like stars in the different galaxy components (in Gyr).}
    \begin{tabularx}{\columnwidth}{l|R|R|R|R}
          & Bulge & Thin disc & Thick disc & Halo\\
        \addlinespace
        \hline\hline
        \addlinespace
        Earth-Like & 8.0 & 6.1 & 7.8 & 8.3 \\
        Super-Earth & 7.5 & 5.7 & 7.5 & 8.0 \\
        Neptunian & 6.8 & 5.1 & 7.2 & 7.3 \\
        Sub-giant & 6.9 & 5.1 & 6.9 & 6.6 \\
        Giant & 6.5 & 4.8 & 6.8 & 6.1 \\
    \end{tabularx}
    \label{tab:planet_ages}
\end{table}

Further, we find that the total number of planets is dependent on the assumed number of initial planet embryos. While the dependence is relatively weak for massive planets -- echoing the conclusions of \citet{Emsenhuber2021} -- a more pronounced dependence is observed for lower mass planets. We calculated the total number of planets within $R = 42$ kpc of the HESTIA galaxy for $N_\mathrm{Embryo}$ = 10, 20, 50, and 100 and tabulate them in \cref{tab:planet_numbers}. Depending on the assumed number of embryos, this yields between $9 \cdot 10^9$ and $25 \cdot 10^9$ planets around Sun-like stars in the HESTIA analogue. It is probable that the number of embryos is not universal for protoplanetary systems and depends, amongst other things on the metallicity (see \cref{sec:initial_embryo_distribution}). The $N_\mathrm{Embryo}$ = 10 and $N_\mathrm{Embryo}$ = 100 cases can however be treated as a lower and upper estimate for the total number of planets.  

\begin{table}[!ht]
    \centering
    \caption{Total number of planets around Sun-like stars within 42 kpc from the galactic centre, for all different values of $N_\mathrm{Embryo}$.}
    \begin{tabularx}{\columnwidth}{l|R|R|R|R}
         $N_\mathrm{Embryo}$ & 10 & 20 & 50 & 100\\
         \addlinespace
         & $\cdot 10^9$ & $\cdot 10^9$ & $\cdot 10^9$ & $\cdot 10^9$\\
         \hline\hline
         \addlinespace
         Earth       &  3.05 &   5.18 &   9.36 &  11.00 \\
         Super-Earth &  3.73 &   5.14 &   9.09 &  10.97 \\
         Neptunian   &  1.25 &   1.37 &   1.31 &   1.42 \\
         Sub-giant   &  0.35 &   0.38 &   0.29 &   0.32 \\
         Giant       &  0.82 &   0.82 &   0.71 &   0.74 \\
         \addlinespace
         \hline
        \addlinespace
         Total       &  9.21 &  12.89 &  20.77 &  24.45 \\
    \end{tabularx}
    \label{tab:planet_numbers}
\end{table}

\subsection{Low-metallicity environments}
\label{sec:metal_poor_regions}
Our analysis in the previous section assumed the occurrence rates are unchanging for metallicities below -0.6, since the original NGPPS sample is constrained to this region of parameter space. However, there are observational and theoretical considerations suggesting the occurrence rates differ in this regime (see \cref{sec:metalpoor_stars}), and the flat extrapolation is likely to overestimate the planet occurrence rates, especially in the thick disc and halo.

\begin{figure}
    \centering
    \includegraphics[width=\columnwidth]{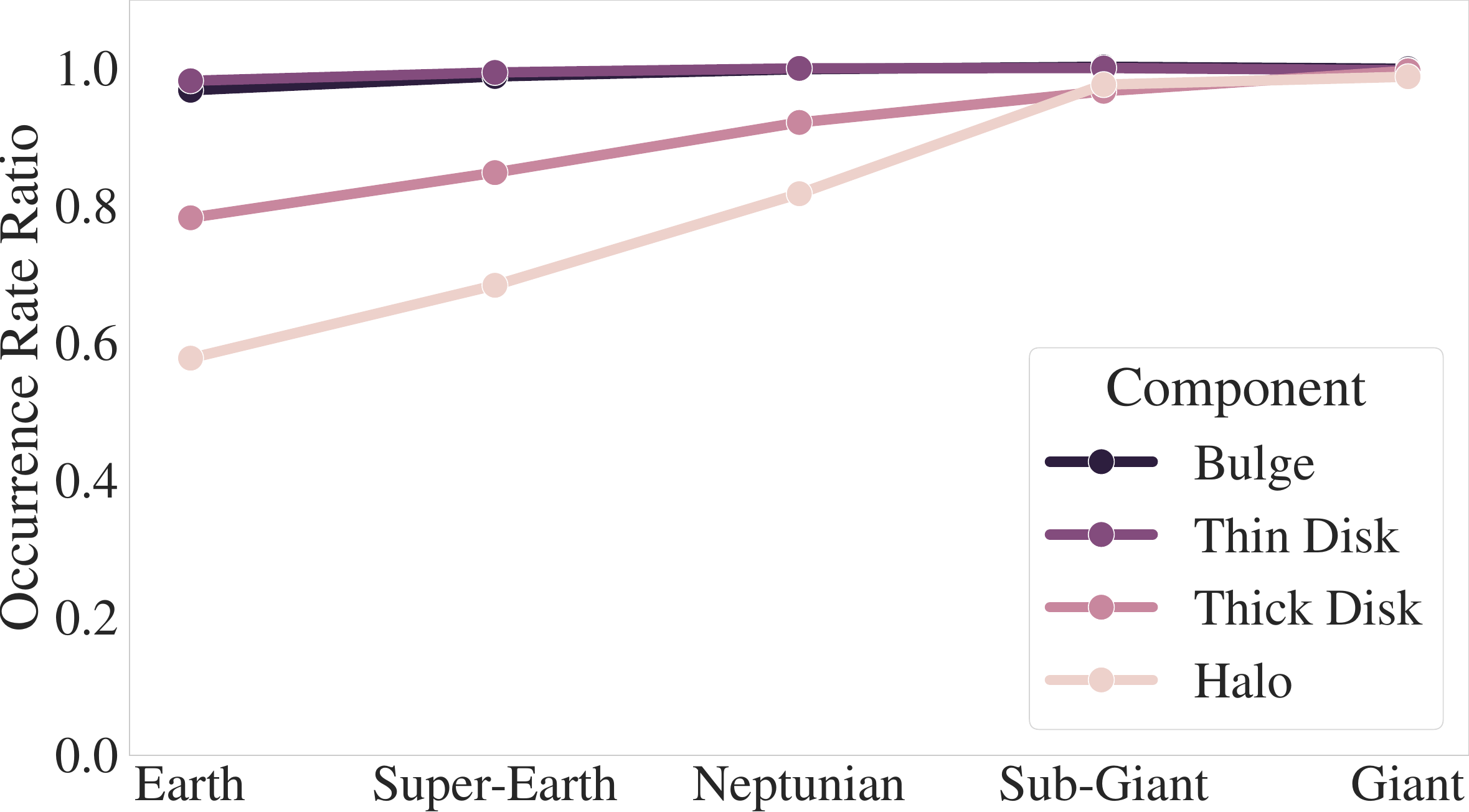}
    \caption{Ratio between the occurrence rates in the fiducial metallicity model (flat extrapolation) and the alternative case where we assign zero planets. The ratios are calculated for $M_\star=1 M_\Sun$ and $N_\mathrm{Embryo}=50$.}
    \label{fig:metallicity_effect}
\end{figure}

To quantify this effect, we considered an alternative metallicity model, where we assigned zero planets to stars below the metallicity threshold [Fe/H] = -0.6. We compared the occurrence rates to the previous prescriptions. \cref{fig:metallicity_effect} shows the ratio between the occurrence rate for the fiducial and this truncated model for the different planet types and galaxy components. The variations are minimal for the bulge and thin disc, since most stars in these regions are above the metallicity threshold. For the low-metallicity environments, the occurrence rates of low-mass planets in the halo is up to 50\% lower when assuming this more restrictive metallicity assumption.

Interestingly, the occurrence rate for giant planets does not significantly differ between the two models, not even in the low-metallicity galaxy components such as the halo, even though the occurrence rates of these planets have the highest metallicity dependence. This stems from the fact that, within our model, giant planets already show near-zero occurrence rates at a metallicity of [Fe/H] = -0.6. Consequently, the variation in the metallicity models does not significantly change their estimated frequency. In contrast, the occurrence rates for Earth-like planets show severe reduction. Given that the occurrence of lower mass planets is only weakly dependent on metallicity, and considering that approximately half of the star particles in the halo have metallicities below our cutoff, the occurrence rate for Earth-like planets is effectively halved in this scenario.

It is important to note that the actual MW halo is more metal-poor than that of the HESTIA analogue. The effect demonstrated here is therefore likely to be even more pronounced in reality.

\subsection{Planets around intermediate- and low-mass dwarfs}
\label{sec:dwarf_stars}
Within the Bern model and NGPPS populations, the average number of planets per star decreases with host star mass. This is at odds with some observations which find higher occurrence rates around colder and less massive stars \citep{Howard2012, Mulders2015, Yang2020, He2021}. Further, while the NGPPS populations created by \citet{Burn2021} span the host mass range from 0.1 to 1 $M_\Sun$, to reduce computing time only $N_\mathrm{Embryo}=50$ initial embryos have been considered. We therefore are unable to analyse the sensitivity of occurrence rates on $N_\mathrm{Embryo}$ for low-mass stars. Consequently, care should be taken when comparing the occurrence rates and absolute numbers for these stars directly to the ones discussed for Sun-like stars in previous section. Nonetheless, the relative frequencies still contain valuable information about the planet populations around cooler stars.

\begin{figure}
     \centering
     \begin{subfigure}{\columnwidth}
        \centering
        \includegraphics[width=\textwidth]{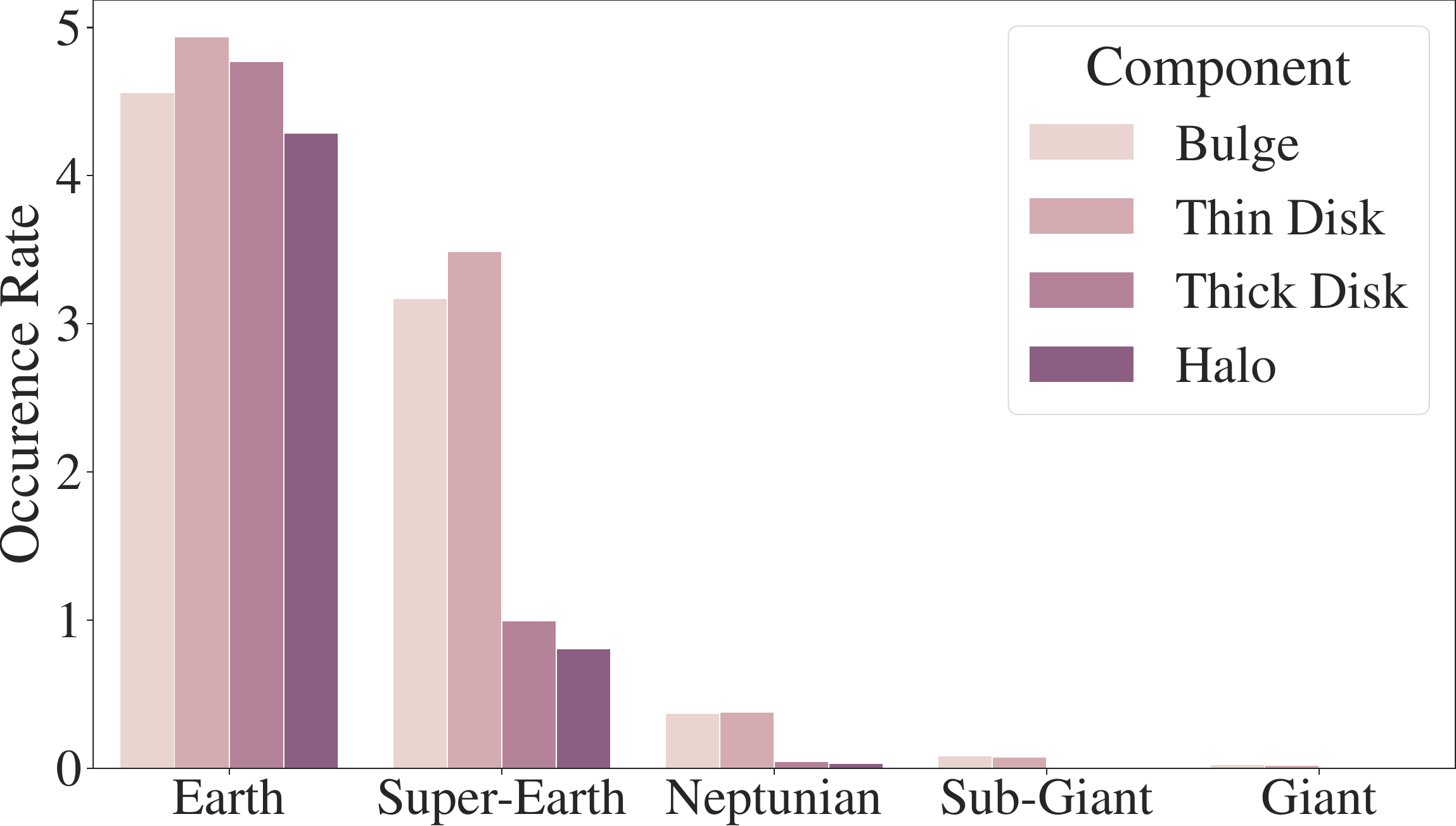}
        \caption{$M_\star = 0.5 M_\Sun$.}
        \label{fig:occurrence_rates_0.5}
     \end{subfigure}
     \begin{subfigure}{\columnwidth}
        \centering
        \includegraphics[width=\textwidth]{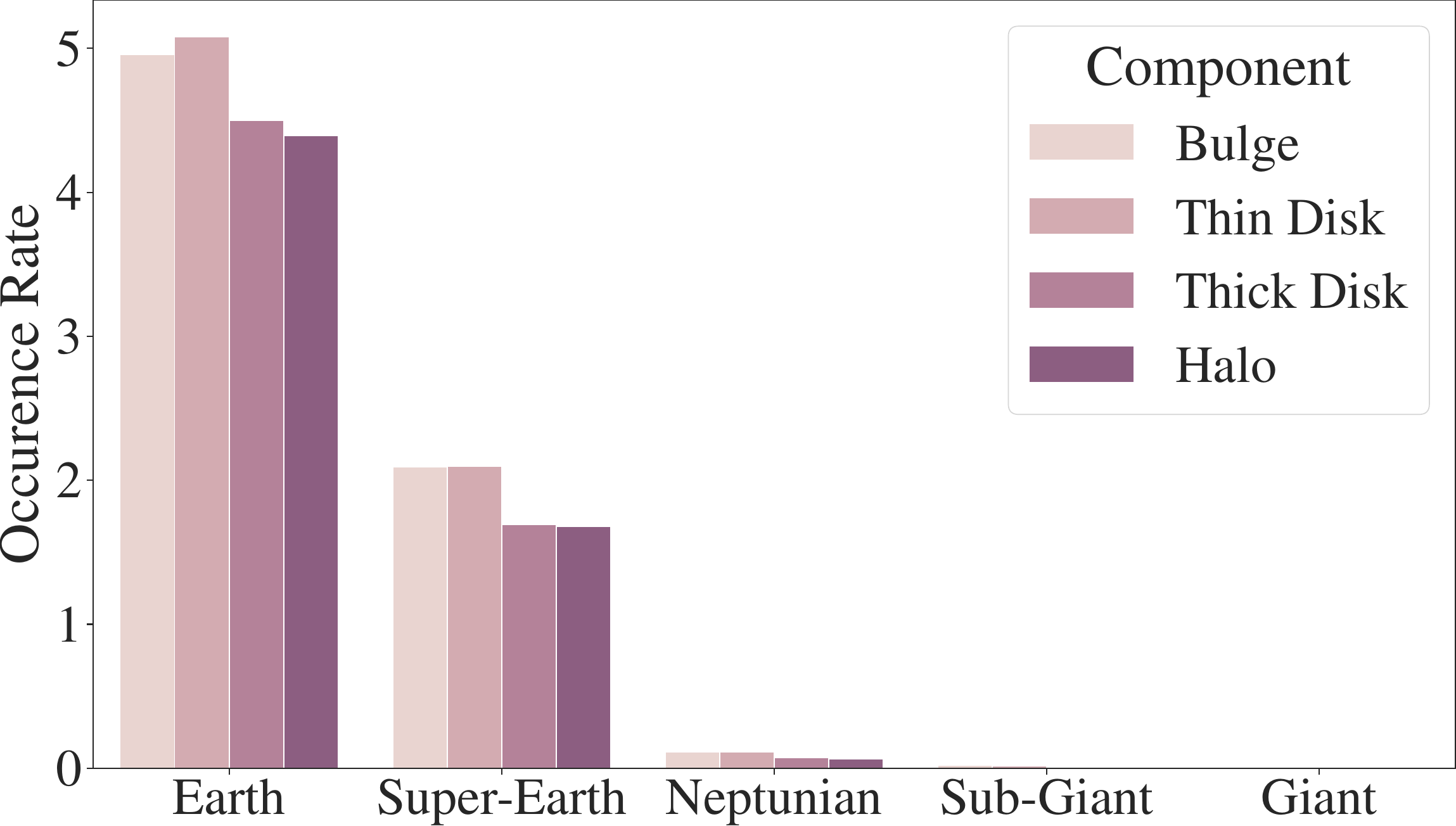}
        \caption{$M_\star = 0.3 M_\Sun$.}
        \label{fig:occurrence_rates_0.1}
     \end{subfigure}
     \caption{Occurrence rates of planet around host star with masses of $M_\star=0.5 M_\Sun$ and $M_\star=0.3 M_\Sun$, for the $N_\mathrm{Embryo}=50$ run.}
     \label{fig:occurrence_rates_dwarfs}
\end{figure}

For lower mass stars, the correlation between planet occurrence rates with metallicity becomes weaker, although the average metallicity to form a given planet type is larger \citep{Burn2021}. Consequently, the planet populations in the different regions of the galaxy are more homogeneous compared to those around Sun-like stars. The occurrence rates are shown in \cref{fig:occurrence_rates_dwarfs} and \cref{tab:occurrence_rates}.

Around intermediate-mass dwarfs ($M_\star = 0.5 M_\Sun$), giant planets are extremely rare in the thick disc and halo with occurrence rates of 0.001 -- 0.002. These values increase by a factor of 10 for the metal-rich regions. The occurrence rate of Earth-like planets for these stars is rather insensitive to the metallicity and and stays between four and five planets per star. Super-Earths on the other hand are more metallicity-dependent with occurrence rates around one planet per star for the metal poor region, three times lower than for the metal rich regions.

For small M-type dwarfs with masses $M_\star = 0.3 M_\Sun$, no giant planets with masses $>300 M_\Earth$ are found in any of the regions of the galaxy. The accretion discs around these stars simply do not provide enough solid material to form massive planetary cores within the lifetimes of the gas disc. Earth-like planets and Super-Earth occurrence rates are more uniform across the different regions, with occurrence rates of 4.5 -- 5 and 1.7 -- 2.1, respectively.  

If the occurrence rates across planet masses are taken at face values, Earth-like planets are most common around intermediate mass stars $M_\star = 0.5 M_\Sun$ for metal-rich regions, while they are most common around Sun-like stars in the thick disc and halo. Neptunians and giants are most common around more massive stars, irrespective of galactic environment.

The weaker metallicity dependence also affects the trends in the occurrence rates with stellar mass (see \cref{tab:occurrence_rates}). For massive planets (Neptunians, sub-giants, giants), the occurrence rates consistently increase with stellar mass in all components. In contrast, the occurrence rate of Earth-like planets decreases with stellar mass in the metal-rich regions (bulge, thin disc) but increases in the metal-poor regions (thick disc, halo). For Super-Earths, the occurrence rates have a minimum at $M_\star = 0.5 M_\Sun$ in the metal-poor regions.

\section{Discussion}
\label{sec:discussion}
\subsection{Planet populations in the galactic context}
Our results show that galactic chemical evolution and variations in stellar populations profoundly affect planet demographics throughout the galaxy. For Sun-like stars, we observe that metal-rich regions, specifically the bulge and thin disc of the MW, harbour the largest number of planets with masses $>M_\Earth$, in both absolute numbers and relative frequencies. In contrast, the metal-poor regions, namely, the thick disc and halo, show a significant suppression in giant planet formation, which is attributed to the lack of solid material building blocks, which are crucial for forming the massive planet cores before the dispersal of the protoplanetary gas discs. Specifically, the occurrence rate of giant planets around Sun-like stars is an order of magnitude higher in the metal-rich regions. Earth-like planets and Super-Earths are more insensitive to the metallicity variations, but still affected. Earth-like planets around Sun-like stars are actually most frequent in the low metallicity environments of the thick disc and halo, being around twice as common as in the thin disc. Super-Earths show the opposite trend and are about half as frequent in the low metallicity regions.

For lower mass stars, with $M_\star = 0.5 M_\Sun$ and $0.3 M_\Sun$, we observe that planet populations are less dependent on metallicity, leading to a more uniform distribution across the galaxy. The lower amounts of solid material available in the protoplanetary discs around these stars generally results in host stars having a higher average metallicity compared to the Sun-like case for the same planet type. Thus, giant and sub-giant planets remain significantly rarer in low-metallicity regions. No giant planets with masses greater $>300 M_\Earth$ are found around any stars of masses $0.3 M_\Sun$, regardless of the galactic environment.

Occasionally, the concept of a galactic habitable zone is discussed in the literature \citep{Gonzalez2001a, Lineweaver2001, Prantzos2008a, Gowanlock2011a}, suggesting that life-supporting conditions are more likely in certain galactic regions, including some of our previous work involving hydrodynamical galaxy simulations \citep{Forgan2017a}. The key idea is that the galactic centre, with its extreme environmental conditions, is less conducive to life due to higher radiation levels and more frequent stellar encounters. Conversely, the outer regions are thought to be too metal-depleted for habitable planet formation. While we have not explored the radiation-related arguments in the inner region, we find no evidence of an outer limit. Low-mass, rocky planets remain abundant throughout the galaxy far into the stellar halo. Moreover, some of these models suggest that giant planets might be detrimental to habitability or the formation of Earth-like planets \citep{Lineweaver2001, Forgan2017a}. Under this assumption, stars in the far-out metal-poor regions could be considered favourable, since giant planet formation is suppressed in these regions. 

\subsection{Empirical occurrence rates}
The robustness of our results applied to the real MW depends on the accuracy of the NGPPS occurrence rates compared to true occurrence rates. \citet{Matuszewski2023} conducted a planet yield analysis for the PLATO mission \citep{Rauer2014}, and compared the occurrence rate estimates around FGK stars from various sources. This comparison includes the NGPPS $N_\mathrm{Embryo}=100$ run, and two empirical models based on Kepler data - one by \citet{Hsu2019} and the other by \citet{Kunimoto2020}. Unlike the mass-based categorization in NGPPS, these models focus on planet radius, aligning with the observational constraints of transit missions.

The NGPPS model predicts an overall occurrence rate of 5.62 planets per star within an orbital period of 500 days, slightly higher than the 5.02 planets per star estimated by \citet{Hsu2019}. In contrast, \citet{Kunimoto2020} report a significantly lower rate of approximately 1 planet per star, though this model is limited to orbital periods shorter than 400 days.

\begin{figure}
    \centering
    \includegraphics[width=\columnwidth]{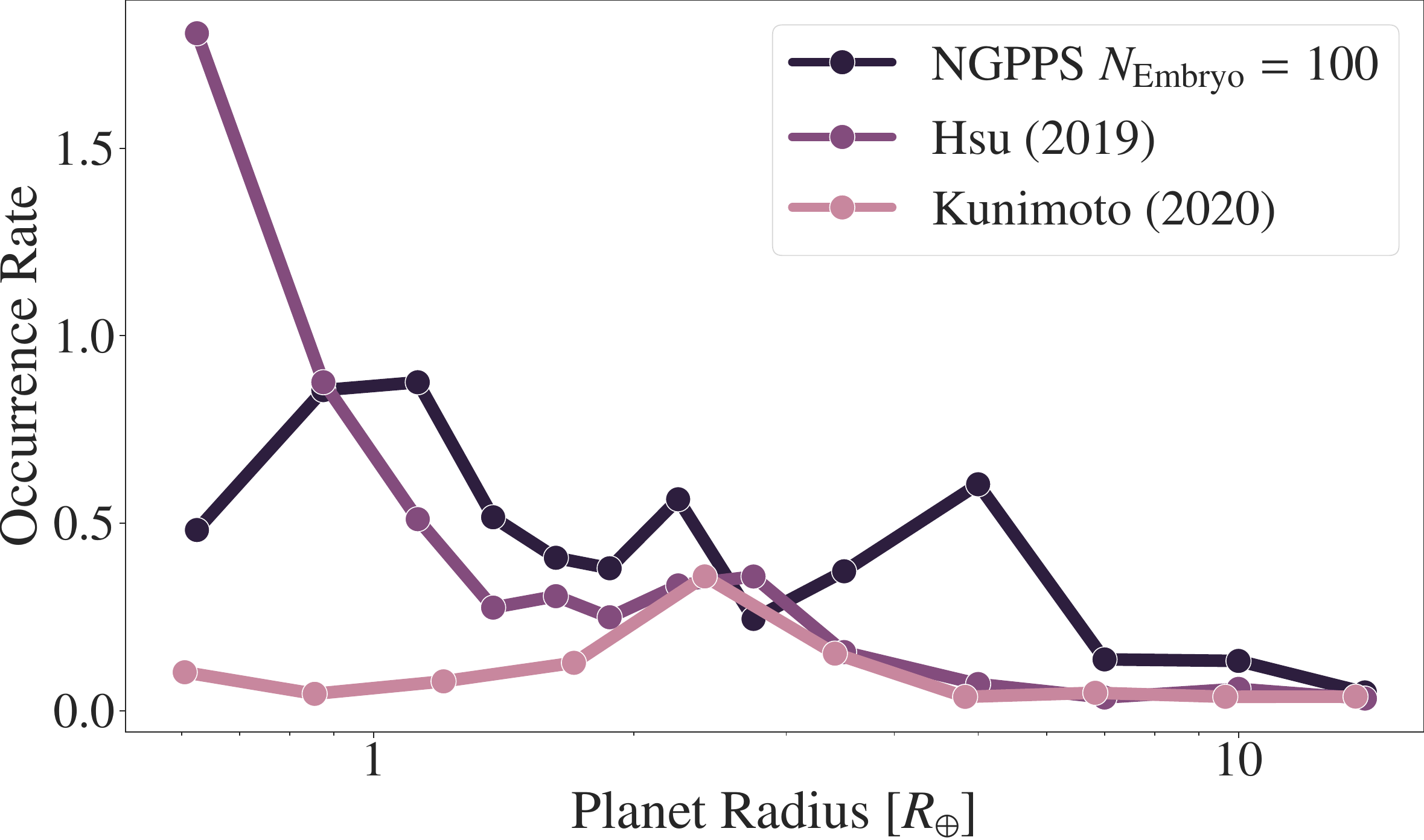}
    \caption{Occurrence rates of planets around FGK stars within an orbital period of 500 days, as estimated by \citet{Matuszewski2023}, based on the works of \citet{Hsu2019}, \citet{Kunimoto2020} and \citet{Emsenhuber2021a}, as a function of planet radius.}
    \label{fig:empirical_occurence_rate}
\end{figure}

As shown in \cref{fig:empirical_occurence_rate}, the distribution of planet sizes predicted by these models show some notable differences. The \citet{Hsu2019} model predicts a substantially higher fraction of planets with $R_\mathrm{p}<1 R_\Earth$ compared to NGPPS and \citet{Kunimoto2020} models, especially for periods between 60 and 500 days. On the other hand, NGPPS has a larger fraction of planets above $R_\mathrm{p} > 1R_\Earth$ , with a marked peak around $R_\mathrm{p} \sim 5 R_\Earth$. This primarily stems from NGPPS predicting a large number of cold Neptune-sized planets with orbital periods >100 days, which remain largely unobserved to date.

The \citet{Kunimoto2020} model aligns with the \citet{Hsu2019} predictions for planets $ R_\mathrm{p} > 3 R_\Earth$ but shows far fewer low-mass planets. The disagreement between the three models
for smaller planets stems from the fact that observational constraints on Earth-sized planet occurrence rates are still limited. The upcoming PLATO mission aims to address this gap, and will improve these estimates.

Translating these findings to the galactic context, the true nature of low-mass planet occurrence rates and their dependence on metallicity is an important factor in understanding exoplanet demographics throughout the galaxy. Massive planets are  more comprehensively understood and our findings more robust. However, other systematic effects have the potential to alter the giant planet population that we have not addressed in this study. We discuss some of these limitations in the next section.

\subsection{Model assumptions and limitations}
\label{sec:limitations}
\subsubsection{Disc lifetimes}
The lifetime of protoplanetary discs, primarily governed by external photoevaporation, plays a crucial role in planet formation. We adopt the range of photoevaporation rates from the \citet{Emsenhuber2021a} NGPPS sample, which are chosen to reproduce typical disc lifetimes in the solar neighbourhood. Efficient massive planet formation ($> 10 M_\Earth$) in the Bern model require disc lifetimes around 3 -- 6 Myr, indicating that early disc dispersal is not a dominant mechanism with respect to halting giant formation \citep{Schlecker2021}. This is reflected in the weak correlation coefficient between photoevaporation rate and planet type (\cref{fig:original_sample_correlation_matrix}).

However, it is important to note that most studies of disc lifetimes, including those by \citet{Emsenhuber2021} and \citet{Ansdell2017}, have focussed on regions with relatively low stellar densities. In contrast, the galactic centre presents a drastically different environment, where higher stellar densities lead to more intense photoevaporation and, consequently, shorter disc lifetimes \citep{Winter2022}. The Central Molecular Zone (CMZ) in particular, exhibits gas surface density of  $\sim 1000 M_{\odot}/\mathrm{pc}^2$ \citep{Henshaw2016}, exceeding that of the solar neighbourhood by a factor of 100 \citep{Kruijssen2012}, which is connected to larger and denser stellar clusters \citep{Farias2023}. This heightened density can significantly amplify photoevaporation, and thus reduce disc lifetime. \citet{Winter2020} constructed an analytical model for the disc lifetime as a function of galactic environment, and find that virtually all discs in the CMZ have lifetimes <3 Myr, severely hindering giant planet formation.

While the CMZ is an exceptional region within the MW, with a diameter of approximately 0.6 kpc, it is not sufficiently resolved in our simulations. Nonetheless, it is plausible that the shortened disc lifetimes in high-density regions towards the galactic centre could lower the occurrence rates of giant planets.

\subsubsection{Initial embryo distribution}
\label{sec:initial_embryo_distribution}
The formation of planet embryos is not explicitly modeled in the Bern model. Instead, a fixed number of embryos are distributed logarithmically at the beginning of each run. However, an analysis of the connection between initial parameter and resulting planet population for the NGPPS population performed by \citet{Schlecker2021} found the location of a planet embryo is the single strongest predictor for the final planetary type and this distribution might vary significantly across different galactic environments.

\citet{Voelkel2021} have studied the connection between planetesimal formation and planet embryo formation using a one-dimensional evolution model coupled to the N-body code LIPAD. They find that the mean orbital separation of planet embryos tends towards a constant value when expressed in terms of the embryo Hill radii. This value ($\sim$ 10--20 $R_\mathrm{Hill}$) is largely insensitive to planetesimal disc mass and density profile, and leads directly to a logarithmic distribution of embryos. This is consistent with previous studies \citep{Kokubo1998}, and with assumptions made for the NGPPS runs.

In terms of the number of planet embryos, they find that the total number of embryos increases with increasing planetesimal disc mass and density slope. For low solid disc masses, specifically $6 M_\Earth$ between 0.5 and 5 AU, around 20--40 embryos are formed, depending on the assumed density profile. At a higher planetesimal disc mass of $27 M_\Earth$, around 60 embryos are formed, irrespective of the assumed profile. Translated into the context of our planet population model, which assumes universal distribution of gas disc masses the galaxy, these findings imply that stars in metal-rich regions (boasting higher solid disc masses) would, on average, form a greater number of planet embryos. Since the occurrence rates of low-mass planets in the model increases with the initial number of embryos (\cref{tab:planet_numbers}), this would indicate that Earth-like planets and Super-Earths are more common in metal-rich regions. The frequencies of giant planets on the other hand are insensitive to $N_\mathrm{Embryo}$ and are not strongly affected.

\subsection{Relation to previous works}
Few studies have been dedicated to planet populations in the galactic context, primarily due to the difficulty in observing distant planets. Despite these challenges, some statistical studies have been performed to study the variations of these populations.

\citet{Bashi2022} analysed a sample of 506 planet candidates orbiting around 369 FGK host stars, categorising them into thin- and thick-disc components. They found that the occurrence rate of close-in super-Earths in the thin disc is approximately 1.3-3.6 larger than in the thick disc, albeit with low number statistics as only 8 of their planets are associated with thick disc stars. This is consistent with our findings, where the lower average [Fe/H] in the thick disc leads to an increased occurrence rate of Earth-like planets, with a corresponding drop in the frequency of super-Earths by a factor of 1.5 . \citet{Bashi2022}, however, attribute this variation primarily to the dynamical history of the stars, including potential disruptive encounters, and gravitational perturbations among planets in typically older systems, rather than differences in metallicity.

\citet{Koshimoto2021} estimated the planet-hosting probability of stars as a function of galactic radius by analysing gravitational microlensing events. Unlike transit and radial velocity detections, which are typically limited to within 1 kpc from the Sun, microlensing can probe much larger distances. Their analysis of 28 planetary microlensing events has shown no strong dependence of occurrence rates on galactic radius,  consistent with our findings. However, larger microlensing surveys are needed to study statistical differences between the bulge and disc populations. The Nancy Grace Roman Space Telescope \citep{Spergel2015} is expected to detect more than 50000 microlensing events and is expected to find around 1400 planets in the galactic bulge \citep{Penny2019}, allowing for a robust estimation of planet occurrence rates around bulge stars.  

\citet{Nielsen2023} constructed an analytical planet formation model to study the occurrence rates of planets around alpha-poor (broadly representing the thin disc), alpha-rich (broadly representing the thick disc), and halo stars. Their findings show minimal dependence of Earth-like planets on the galactic component, a slight increase in the occurrence of Super-Earths, and a strong dependence of giant planets on these galactic components. Similarly to our finding, this is primarily attributed to the metallicity-dependence of the occurrence rates. Earth-like planet frequency in their model is largely independent of metallicity, while giant planets strongly favour higher metallicities. Their overall findings are therefore similar to ours.

Furthermore, the works of \citet{Bashi2019}, \citet{Bashi2020}, \citet{Chen2022}, and \citet{Nielsen2023} have expanded the scope of study to include the [$\alpha$/Fe] fraction in addition to the [Fe/H] fraction of the host star and its implications for planet formation. This aspect is potentially important, as thick disc stars are $\alpha$-enhanced compared to thin disc. We have not included this quantity, as the NGPPS populations have no direct $\alpha$ element dependence. The findings of \citet{Nielsen2023} however suggest that the $\alpha$-enhancement of the host stars plays a secondary role compared to the overall metallicity, as traced by [Fe/H]. Yet, \citet{Swastik2022} find that planet masses increase marginally with decreasing [$\alpha$/Fe].

\section{Summary and conclusions}
\label{sec:summary}
In this study, we combined a simulated  MW analogue for HESTIA suite of high-resolution galaxy formation simulations \citep{Libeskind2020} and the NGPPS dataset from the Bern model for planet formation and evolution \citep{Emsenhuber2021} to study the planet demographics throughout the galaxy. Using the stellar ages and metallicities provided by HESTIA, we were able to map the occurrence rate of various planet types as a function of galactic environment and quantify variations in the frequency of Earth-like planets, Super-Earths, Neptunians, and giant planets (defined through their planetary masses). Our key findings are as follows:
\begin{enumerate}
    \item The occurrence rates of specific planet types within the framework of the NGPPS population model depend primarily on the available solid mass in the planetesimal disc, determined by the initial gas disc mass and host star metallicity. For a fixed initial gas mass, planet masses increase with increasing [Fe/H]. (\cref{sec:assignment_model})
    \item Based on this, the overall planet populations in the metal-rich galactic bulge and thin disc differ significantly from the populations found in the metal-poor thick disc and stellar halo. Planets with masses $> 300 M_\Earth$ are 10-20 times more common around stars in the thick disc than compared to the thin disc. Earth-mass planets around Sun-like stars are most abundant in the thick disc, with occurrence rates 1.4 times higher than in the thin disc. (\cref{sec:sunlike_stars})
    \item The planet populations around Sun-like stars are youngest in the thin disc, with Earth-like planets having an average age of 6.1 Gyr in the simulated MW analogue. The oldest Earth-like planets are found in the halo, at an average age of 8.3 Gyr. The average age decreases with increasing planet mass. (\cref{sec:sunlike_stars})
    \item The planet populations in different galactic environments vary more strongly around Sun-like stars than those around lower mass stars ($M_\star = 0.3$ -- $0.5 M_\Sun$). This is caused by stronger correlation between metallicity and occurrence rate for massive stars in the NGPPS sample. Earth-like planets are most common around intermediate mass dwarfs with masses 0.5 -- 0.7 $M_\Sun$.   (\cref{sec:assignment_model}, \cref{sec:dwarf_stars}).
\end{enumerate}
The study of planet formation as a function of galactic environment is still in its infancy. Currently, transit and RV surveys are restricted mostly to stars within 1 kpc of the Sun. Future exoplanet surveys will however increase our understanding of planet formation and the effects of the environment. The upcoming PLATO mission \citep{Rauer2014} will observe around 200 000 stars covering a large area of the sky and is likely to detect a sample of exoplanets around thick-disc stars. In addition, the Nancy Grace Roman Space Telescope \citep{Spergel2015}  is expected to find a statistical sample of around 1400 planets in the galactic bulge \citep{Penny2019} via microlensing. Understanding the exoplanet demographics in different environments will yield new insights into processes involved in planet formation and evolution, while helping improve planet formation models.

\section*{Software}
The analysis has been performed using Python, and the scientific data analysis libraries \texttt{numpy} \citep{Harris2020}, \texttt{pandas} \citep{Thepandasdevelopmentteam2023}, \texttt{statsmodels} \citep{Seabold2010} and \texttt{scikit-learn} \citep{Pedregosa2011}. The HESTIA simulation data was accessed and processed using \texttt{yt} \citep{Smith2015}. Visualisation were made using \texttt{matplotlib} \citep{ThomasACaswell2023}, \texttt{seaborn} \citep{Waskom2021} and \texttt{yt}. The galaxy decomposition was performed using \texttt{MORDOR} \citep{Zana2022}. 

\begin{acknowledgements}
CB thanks the Young Academy Groningen for their generous support through an interdisciplinary PhD fellowship. PD and MT acknowledge support from the NWO grant 016.VIDI.189.162 (``ODIN"). PD warmly thanks the European Commission's and University of Groningen's CO-FUND Rosalind Franklin program. 

K.R. is grateful for support from the UK STFC via grant ST/V000594/1.

We want to thank Alexandre Emsenhuber for providing us with the Monte Carlo variable dataset used for the original NGPPS sample, and the Data \& Analysis Center for Exoplanets (\url{dace.unige.ch}) for a seamless access to the NGPPS and observational exoplanet data. Access to the HESTIA data \citep{Libeskind2020} was kindly provided by the CLUES collaboration (\url{clues-project.org}), and can be requested from them. The tables and further data presented in this work are available upon request to \url{boettner@astro.rug.nl}. 

We also would like to thank the referee for insightful feedback, which helped to improve this work.
\end{acknowledgements}

%%%%%%%%%%%%%%%%%%%%%%%%%%%%%%%%%%%%%%%%
% The Bibliography
\bibliographystyle{aa} % style aa.bst
\bibliography{main.bib} % your references Yourfile.bib

\begin{thebibliography}{124}
\expandafter\ifx\csname natexlab\endcsname\relax\def\natexlab#1{#1}\fi

\bibitem[{Abadi {et~al.}(2003)Abadi, Navarro, Steinmetz, \& Eke}]{Abadi2003}
Abadi, M.~G., Navarro, J.~F., Steinmetz, M., \& Eke, V.~R. 2003, The
  Astrophysical Journal, 597, 21

\bibitem[{Alessi \& Pudritz(2018)}]{Alessi2018a}
Alessi, M. \& Pudritz, R.~E. 2018, Monthly Notices of the Royal Astronomical
  Society, 478, 2599

\bibitem[{Alessi {et~al.}(2020)Alessi, Pudritz, \& Cridland}]{Alessi2020}
Alessi, M., Pudritz, R.~E., \& Cridland, A.~J. 2020, Monthly Notices of the
  Royal Astronomical Society, 493, 1013

\bibitem[{Ansdell {et~al.}(2017)Ansdell, Williams, Manara, Miotello, Facchini,
  {van der Marel}, Testi, \& {van Dishoeck}}]{Ansdell2017}
Ansdell, M., Williams, J.~P., Manara, C.~F., {et~al.} 2017, The Astronomical
  Journal, 153, 240

\bibitem[{Bashi \& Zucker(2019)}]{Bashi2019}
Bashi, D. \& Zucker, S. 2019, The Astronomical Journal, 158, 61

\bibitem[{Bashi \& Zucker(2022)}]{Bashi2022}
Bashi, D. \& Zucker, S. 2022, Monthly Notices of the Royal Astronomical
  Society, 510, 3449

\bibitem[{Bashi {et~al.}(2020)Bashi, Zucker, Adibekyan, Santos, {Tal-Or},
  Trifonov, \& Mazeh}]{Bashi2020}
Bashi, D., Zucker, S., Adibekyan, V., {et~al.} 2020, Astronomy and
  Astrophysics, 643, A106

\bibitem[{Beers \& Christlieb(2005)}]{Beers2005}
Beers, T.~C. \& Christlieb, N. 2005, Annual Review of Astronomy and
  Astrophysics, 43, 531

\bibitem[{Bensby {et~al.}(2014)Bensby, Feltzing, \& Oey}]{Bensby2014}
Bensby, T., Feltzing, S., \& Oey, M.~S. 2014, Astronomy and Astrophysics, 562,
  A71

\bibitem[{Bitsch {et~al.}(2019)Bitsch, Izidoro, Johansen, Raymond, Morbidelli,
  Lambrechts, \& Jacobson}]{Bitsch2019}
Bitsch, B., Izidoro, A., Johansen, A., {et~al.} 2019, Astronomy and
  Astrophysics, 623, A88

\bibitem[{{Bland-Hawthorn} \& Gerhard(2016{\natexlab{a}})}]{Bland-Hawthorn2016}
{Bland-Hawthorn}, J. \& Gerhard, O. 2016{\natexlab{a}}, Annual Review of
  Astronomy and Astrophysics, 54, 529

\bibitem[{{Bland-Hawthorn} \&
  Gerhard(2016{\natexlab{b}})}]{Bland-Hawthorn2016a}
{Bland-Hawthorn}, J. \& Gerhard, O. 2016{\natexlab{b}}, Annual Review of
  Astronomy and Astrophysics, 54, 529

\bibitem[{Bond {et~al.}(2010)Bond, Ivezi{\'c}, Sesar, Juri{\'c}, Munn,
  Kowalski, Loebman, Ro{\v s}kar, Beers, Dalcanton, Rockosi, Yanny, Newberg,
  Allende~Prieto, Wilhelm, Lee, Sivarani, Majewski, Norris, {Bailer-Jones},
  Re~Fiorentin, Schlegel, Uomoto, Lupton, Knapp, Gunn, Covey, Allyn~Smith,
  Miknaitis, Doi, Tanaka, Fukugita, Kent, Finkbeiner, Quinn, Hawley, Anderson,
  Kiuchi, Chen, Bushong, Sohi, Haggard, Kimball, McGurk, Barentine, Brewington,
  Harvanek, Kleinman, Krzesinski, Long, Nitta, Snedden, Lee, Pier, Harris,
  Brinkmann, \& Schneider}]{Bond2010}
Bond, N.~A., Ivezi{\'c}, {\v Z}., Sesar, B., {et~al.} 2010, The Astrophysical
  Journal, 716, 1

\bibitem[{Boss(1997)}]{Boss1997}
Boss, A.~P. 1997, Science, 276, 1836

\bibitem[{Buchhave {et~al.}(2012)Buchhave, Latham, Johansen, Bizzarro, Torres,
  Rowe, Batalha, Borucki, Brugamyer, Caldwell, Bryson, Ciardi, Cochran, Endl,
  Esquerdo, Ford, Geary, Gilliland, Hansen, Isaacson, Laird, Lucas, Marcy,
  Morse, Robertson, Shporer, Stefanik, Still, \& Quinn}]{Buchhave2012}
Buchhave, L.~A., Latham, D.~W., Johansen, A., {et~al.} 2012, Nature, 486, 375

\bibitem[{Burn {et~al.}(2021)Burn, Schlecker, Mordasini, Emsenhuber, Alibert,
  Henning, Klahr, \& Benz}]{Burn2021}
Burn, R., Schlecker, M., Mordasini, C., {et~al.} 2021, Astronomy and
  Astrophysics, 656, A72

\bibitem[{Caswell {et~al.}(2023)Caswell, {Elliott Sales de Andrade}, Lee,
  Droettboom, Hoffmann, Klymak, Hunter, Firing, Stansby, Varoquaux, Nielsen,
  Gustafsson, Sunden, Root, May, Elson, Sepp{\"a}nen, {hannah}, {Jae-Joon Lee},
  Dale, McDougall, Straw, Hobson, Lucas, Comer, Gohlke, Vincent, Yu, Ma, \&
  Silvester}]{ThomasACaswell2023}
Caswell, T.~A., {Elliott Sales de Andrade}, Lee, A., {et~al.} 2023,
  Matplotlib/Matplotlib: {{REL}}: V3.7.4, Zenodo

\bibitem[{Chabrier(2003)}]{Chabrier2003a}
Chabrier, G. 2003, Publications of the Astronomical Society of the Pacific,
  115, 763

\bibitem[{Chen {et~al.}(2022)Chen, Xie, Zhou, Yang, Dong, Zhu, Zheng, Liu,
  Zong, \& Luo}]{Chen2022}
Chen, D.-C., Xie, J.-W., Zhou, J.-L., {et~al.} 2022, The Astronomical Journal,
  163, 249

\bibitem[{Chevance {et~al.}(2022)Chevance, Krumholz, McLeod, Ostriker,
  Rosolowsky, \& Sternberg}]{Chevance2022b}
Chevance, M., Krumholz, M.~R., McLeod, A.~F., {et~al.} 2022, The {{Life}} and
  {{Times}} of {{Giant Molecular Clouds}}

\bibitem[{Courteau {et~al.}(2011)Courteau, Widrow, McDonald, Guhathakurta,
  Gilbert, Zhu, Beaton, \& Majewski}]{Courteau2011}
Courteau, S., Widrow, L.~M., McDonald, M., {et~al.} 2011, The Astrophysical
  Journal, 739, 20

\bibitem[{Crain \& {van de Voort}(2023)}]{Crain2023}
Crain, R.~A. \& {van de Voort}, F. 2023, Annual Review of Astronomy and
  Astrophysics, 61, 473

\bibitem[{Deason {et~al.}(2019)Deason, Belokurov, \& Sanders}]{Deason2019}
Deason, A.~J., Belokurov, V., \& Sanders, J.~L. 2019, Monthly Notices of the
  Royal Astronomical Society, 490, 3426

\bibitem[{Dr{\k a}{\.z}kowska {et~al.}(2023)Dr{\k a}{\.z}kowska, Bitsch,
  Lambrechts, Mulders, Harsono, Vazan, Liu, Ormel, Kretke, \&
  Morbidelli}]{Drazkowska2023}
Dr{\k a}{\.z}kowska, J., Bitsch, B., Lambrechts, M., {et~al.} 2023, arXiv, 534,
  717

\bibitem[{Dubois {et~al.}(2016)Dubois, Peirani, Pichon, Devriendt, Gavazzi,
  Welker, \& Volonteri}]{Dubois2016}
Dubois, Y., Peirani, S., Pichon, C., {et~al.} 2016, Monthly Notices of the
  Royal Astronomical Society, 463, 3948

\bibitem[{Emsenhuber {et~al.}(2021{\natexlab{a}})Emsenhuber, Mordasini, Burn,
  Alibert, Benz, \& Asphaug}]{Emsenhuber2021}
Emsenhuber, A., Mordasini, C., Burn, R., {et~al.} 2021{\natexlab{a}}, Astronomy
  and Astrophysics, 656, A69

\bibitem[{Emsenhuber {et~al.}(2021{\natexlab{b}})Emsenhuber, Mordasini, Burn,
  Alibert, Benz, \& Asphaug}]{Emsenhuber2021a}
Emsenhuber, A., Mordasini, C., Burn, R., {et~al.} 2021{\natexlab{b}}, Astronomy
  and Astrophysics, 656, A70

\bibitem[{Farias {et~al.}(2023)Farias, Offner, Grudi{\'c}, Guszejnov, \&
  Rosen}]{Farias2023}
Farias, J.~P., Offner, S. S.~R., Grudi{\'c}, M.~Y., Guszejnov, D., \& Rosen,
  A.~L. 2023, Stellar {{Populations}} in {{STARFORGE}}: {{The Origin}} and
  {{Evolution}} of {{Star Clusters}} and {{Associations}}

\bibitem[{Fischer \& Valenti(2005)}]{Fischer2005}
Fischer, D.~A. \& Valenti, J. 2005, The Astrophysical Journal, 622, 1102

\bibitem[{Forgan {et~al.}(2017)Forgan, Dayal, Cockell, \&
  Libeskind}]{Forgan2017a}
Forgan, D., Dayal, P., Cockell, C., \& Libeskind, N. 2017, International
  Journal of Astrobiology, 16, 60

\bibitem[{Frebel \& Norris(2015)}]{Frebel2015}
Frebel, A. \& Norris, J.~E. 2015, Annual Review of Astronomy and Astrophysics,
  53, 631

\bibitem[{Genovali {et~al.}(2014)Genovali, Lemasle, Bono, Romaniello, Fabrizio,
  Ferraro, Iannicola, Laney, Nonino, Bergemann, Buonanno, Fran{\c c}ois, Inno,
  Kudritzki, Matsunaga, Pedicelli, Primas, \& Th{\'e}venin}]{Genovali2014}
Genovali, K., Lemasle, B., Bono, G., {et~al.} 2014, Astronomy and Astrophysics,
  566, A37

\bibitem[{Gilmore \& Reid(1983)}]{Gilmore1983}
Gilmore, G. \& Reid, N. 1983, Monthly Notices of the Royal Astronomical
  Society, 202, 1025

\bibitem[{Gonzalez(1997)}]{Gonzalez1997}
Gonzalez, G. 1997, Monthly Notices of the Royal Astronomical Society, 285, 403

\bibitem[{Gonzalez {et~al.}(2001)Gonzalez, Brownlee, \& Ward}]{Gonzalez2001a}
Gonzalez, G., Brownlee, D., \& Ward, P. 2001, Icarus, 152, 185

\bibitem[{Gottloeber {et~al.}(2010)Gottloeber, Hoffman, \&
  Yepes}]{Gottloeber2010}
Gottloeber, S., Hoffman, Y., \& Yepes, G. 2010, Constrained {{Local UniversE
  Simulations}} ({{CLUES}})

\bibitem[{Gowanlock {et~al.}(2011)Gowanlock, Patton, \&
  McConnell}]{Gowanlock2011a}
Gowanlock, M.~G., Patton, D.~R., \& McConnell, S.~M. 2011, Astrobiology, 11,
  855

\bibitem[{Grand {et~al.}(2017)Grand, G{\'o}mez, Marinacci, Pakmor, Springel,
  Campbell, Frenk, Jenkins, \& White}]{Grand2017}
Grand, R. J.~J., G{\'o}mez, F.~A., Marinacci, F., {et~al.} 2017, Monthly
  Notices of the Royal Astronomical Society, 467, 179

\bibitem[{Harris {et~al.}(2020)Harris, Millman, {van der Walt}, Gommers,
  Virtanen, Cournapeau, Wieser, Taylor, Berg, Smith, Kern, Picus, Hoyer, {van
  Kerkwijk}, Brett, Haldane, {del R{\'i}o}, Wiebe, Peterson,
  {G{\'e}rard-Marchant}, Sheppard, Reddy, Weckesser, Abbasi, Gohlke, \&
  Oliphant}]{Harris2020}
Harris, C.~R., Millman, K.~J., {van der Walt}, S.~J., {et~al.} 2020, Nature,
  585, 357

\bibitem[{Hattori {et~al.}(2018)Hattori, Valluri, Bell, \&
  Roederer}]{Hattori2018}
Hattori, K., Valluri, M., Bell, E.~F., \& Roederer, I.~U. 2018, The
  Astrophysical Journal, 866, 121

\bibitem[{Hawkins {et~al.}(2015)Hawkins, Jofr{\'e}, Masseron, \&
  Gilmore}]{Hawkins2015}
Hawkins, K., Jofr{\'e}, P., Masseron, T., \& Gilmore, G. 2015, Monthly Notices
  of the Royal Astronomical Society, 453, 758

\bibitem[{Haywood(2001)}]{Haywood2001}
Haywood, M. 2001, Monthly Notices of the Royal Astronomical Society, 325, 1365

\bibitem[{He {et~al.}(2021)He, Ford, \& Ragozzine}]{He2021}
He, M.~Y., Ford, E.~B., \& Ragozzine, D. 2021, The Astronomical Journal, 161,
  16

\bibitem[{Henshaw {et~al.}(2016)Henshaw, Longmore, \& Kruijssen}]{Henshaw2016}
Henshaw, J.~D., Longmore, S.~N., \& Kruijssen, J. M.~D. 2016, Monthly Notices
  of the Royal Astronomical Society, 463, L122

\bibitem[{Hoffman(2009)}]{Hoffman2009}
Hoffman, Y. 2009, Data Analysis in Cosmology, 665, 565

\bibitem[{Howard {et~al.}(2012)Howard, Marcy, Bryson, Jenkins, Rowe, Batalha,
  Borucki, Koch, Dunham, Gautier, Van~Cleve, Cochran, Latham, Lissauer, Torres,
  Brown, Gilliland, Buchhave, Caldwell, {Christensen-Dalsgaard}, Ciardi,
  Fressin, Haas, Howell, Kjeldsen, Seager, Rogers, Sasselov, Steffen, Basri,
  Charbonneau, Christiansen, Clarke, Dupree, Fabrycky, Fischer, Ford, Fortney,
  Tarter, Girouard, Holman, Johnson, Klaus, Machalek, Moorhead, Morehead,
  Ragozzine, Tenenbaum, Twicken, Quinn, Isaacson, Shporer, Lucas, Walkowicz,
  Welsh, Boss, Devore, Gould, Smith, Morris, Prsa, Morton, Still, Thompson,
  Mullally, Endl, \& MacQueen}]{Howard2012}
Howard, A.~W., Marcy, G.~W., Bryson, S.~T., {et~al.} 2012, The Astrophysical
  Journal Supplement Series, 201, 15

\bibitem[{Hsu {et~al.}(2019)Hsu, Ford, Ragozzine, \& Ashby}]{Hsu2019}
Hsu, D.~C., Ford, E.~B., Ragozzine, D., \& Ashby, K. 2019, The Astronomical
  Journal, 158, 109

\bibitem[{Ida \& Lin(2004{\natexlab{a}})}]{Ida2004}
Ida, S. \& Lin, D. N.~C. 2004{\natexlab{a}}, The Astrophysical Journal, 604,
  388

\bibitem[{Ida \& Lin(2004{\natexlab{b}})}]{Ida2004a}
Ida, S. \& Lin, D. N.~C. 2004{\natexlab{b}}, The Astrophysical Journal, 616,
  567

\bibitem[{Johansen {et~al.}(2019)Johansen, Ida, \& Brasser}]{Johansen2019}
Johansen, A., Ida, S., \& Brasser, R. 2019, Astronomy and Astrophysics, 622,
  A202

\bibitem[{Johnson {et~al.}(2010)Johnson, Aller, Howard, \& Crepp}]{Johnson2010}
Johnson, J.~A., Aller, K.~M., Howard, A.~W., \& Crepp, J.~R. 2010, Publications
  of the Astronomical Society of the Pacific, 122, 905

\bibitem[{Kafle {et~al.}(2014)Kafle, Sharma, Lewis, \&
  {Bland-Hawthorn}}]{Kafle2014}
Kafle, P.~R., Sharma, S., Lewis, G.~F., \& {Bland-Hawthorn}, J. 2014, The
  Astrophysical Journal, 794, 59

\bibitem[{Karachentsev \& Nasonova(2010)}]{Karachentsev2010}
Karachentsev, I.~D. \& Nasonova, O.~G. 2010, Monthly Notices of the Royal
  Astronomical Society, 405, 1075

\bibitem[{Kilic {et~al.}(2017)Kilic, Munn, Harris, von Hippel, Liebert,
  Williams, Jeffery, \& DeGennaro}]{Kilic2017}
Kilic, M., Munn, J.~A., Harris, H.~C., {et~al.} 2017, The Astrophysical
  Journal, 837, 162

\bibitem[{Kokubo \& Ida(1998)}]{Kokubo1998}
Kokubo, E. \& Ida, S. 1998, Icarus, 131, 171

\bibitem[{Koshimoto {et~al.}(2021)Koshimoto, Bennett, Suzuki, \&
  Bond}]{Koshimoto2021}
Koshimoto, N., Bennett, D.~P., Suzuki, D., \& Bond, I.~A. 2021, The
  Astrophysical Journal, 918, L8

\bibitem[{Kruijssen(2012)}]{Kruijssen2012}
Kruijssen, J. M.~D. 2012, Monthly Notices of the Royal Astronomical Society,
  426, 3008

\bibitem[{Kunimoto \& Matthews(2020)}]{Kunimoto2020}
Kunimoto, M. \& Matthews, J.~M. 2020, The Astronomical Journal, 159, 248

\bibitem[{Lemasle {et~al.}(2007)Lemasle, Fran{\c c}ois, Bono, Mottini, Primas,
  \& Romaniello}]{Lemasle2007}
Lemasle, B., Fran{\c c}ois, P., Bono, G., {et~al.} 2007, Astronomy and
  Astrophysics, 467, 283

\bibitem[{Lemasle {et~al.}(2018)Lemasle, Hajdu, Kovtyukh, Inno, Grebel,
  Catelan, Bono, Fran{\c c}ois, Kniazev, da~Silva, \& Storm}]{Lemasle2018}
Lemasle, B., Hajdu, G., Kovtyukh, V., {et~al.} 2018, Astronomy \& Astrophysics,
  618, A160

\bibitem[{Libeskind {et~al.}(2020)Libeskind, Carlesi, Grand, Khalatyan, Knebe,
  Pakmor, Pilipenko, Pawlowski, Sparre, Tempel, Wang, Courtois, Gottl{\"o}ber,
  Hoffman, Minchev, Pfrommer, Sorce, Springel, Steinmetz, Tully, Vogelsberger,
  \& Yepes}]{Libeskind2020}
Libeskind, N.~I., Carlesi, E., Grand, R. J.~J., {et~al.} 2020, Monthly Notices
  of the Royal Astronomical Society, 498, 2968

\bibitem[{Licquia \& Newman(2015)}]{Licquia2015a}
Licquia, T.~C. \& Newman, J.~A. 2015, The Astrophysical Journal, 806, 96

\bibitem[{Lineweaver(2001)}]{Lineweaver2001}
Lineweaver, C.~H. 2001, Icarus, 151, 307

\bibitem[{Lu {et~al.}(2020)Lu, Schlaufman, \& Cheng}]{Lu2020}
Lu, C.~X., Schlaufman, K.~C., \& Cheng, S. 2020, The Astronomical Journal, 160,
  253

\bibitem[{Luck \& Lambert(2011)}]{Luck2011}
Luck, R.~E. \& Lambert, D.~L. 2011, The Astronomical Journal, 142, 136

\bibitem[{Masseron \& Gilmore(2015)}]{Masseron2015}
Masseron, T. \& Gilmore, G. 2015, Monthly Notices of the Royal Astronomical
  Society, 453, 1855

\bibitem[{Matsuyama {et~al.}(2003)Matsuyama, Johnstone, \&
  Hartmann}]{Matsuyama2003a}
Matsuyama, I., Johnstone, D., \& Hartmann, L. 2003, The Astrophysical Journal,
  582, 893

\bibitem[{Matuszewski {et~al.}(2023)Matuszewski, Nettelmann, Cabrera,
  B{\"o}rner, \& Rauer}]{Matuszewski2023}
Matuszewski, F., Nettelmann, N., Cabrera, J., B{\"o}rner, A., \& Rauer, H.
  2023, Astronomy \& Astrophysics, 677, A133

\bibitem[{McBride {et~al.}(2009)McBride, Fakhouri, \& Ma}]{McBride2009}
McBride, J., Fakhouri, O., \& Ma, C.-P. 2009, Monthly Notices of the Royal
  Astronomical Society, 398, 1858

\bibitem[{McMillan(2017)}]{McMillan2017}
McMillan, P.~J. 2017, Monthly Notices of the Royal Astronomical Society, 465,
  76

\bibitem[{McWilliam(1997)}]{McWilliam1997}
McWilliam, A. 1997, Annual Review of Astronomy and Astrophysics, 35, 503

\bibitem[{McWilliam(2016)}]{McWilliam2016}
McWilliam, A. 2016, Publications of the Astronomical Society of Australia, 33,
  e040

\bibitem[{Monari {et~al.}(2018)Monari, Famaey, Carrillo, Piffl, Steinmetz,
  Wyse, Anders, Chiappini, \& Jan{\ss}en}]{Monari2018}
Monari, G., Famaey, B., Carrillo, I., {et~al.} 2018, Astronomy and
  Astrophysics, 616, L9

\bibitem[{Mordasini {et~al.}(2009{\natexlab{a}})Mordasini, Alibert, \&
  Benz}]{Mordasini2009a}
Mordasini, C., Alibert, Y., \& Benz, W. 2009{\natexlab{a}}, Astronomy and
  Astrophysics, 501, 1139

\bibitem[{Mordasini {et~al.}(2009{\natexlab{b}})Mordasini, Alibert, Benz, \&
  Naef}]{Mordasini2009}
Mordasini, C., Alibert, Y., Benz, W., \& Naef, D. 2009{\natexlab{b}}, Astronomy
  and Astrophysics, 501, 1161

\bibitem[{Mulders {et~al.}(2015)Mulders, Pascucci, \& Apai}]{Mulders2015}
Mulders, G.~D., Pascucci, I., \& Apai, D. 2015, The Astrophysical Journal, 798,
  112

\bibitem[{Mulders {et~al.}(2016)Mulders, Pascucci, Apai, Frasca, \&
  {Molenda-{\.Z}akowicz}}]{Mulders2016}
Mulders, G.~D., Pascucci, I., Apai, D., Frasca, A., \& {Molenda-{\.Z}akowicz},
  J. 2016, The Astronomical Journal, 152, 187

\bibitem[{Narang {et~al.}(2018)Narang, Manoj, Furlan, Mordasini, Henning,
  Mathew, Banyal, \& Sivarani}]{Narang2018}
Narang, M., Manoj, P., Furlan, E., {et~al.} 2018, The Astronomical Journal,
  156, 221

\bibitem[{Navarro \& White(1994)}]{Navarro1994}
Navarro, J.~F. \& White, S. D.~M. 1994, Monthly Notices of the Royal
  Astronomical Society, 267, 401

\bibitem[{Nielsen {et~al.}(2023)Nielsen, Gent, Bergemann, Eitner, \&
  Johansen}]{Nielsen2023}
Nielsen, J., Gent, M.~R., Bergemann, M., Eitner, P., \& Johansen, A. 2023,
  Planet Formation throughout the {{Milky Way}}: {{Planet}} Populations in the
  Context of {{Galactic}} Chemical Evolution

\bibitem[{Pedregosa {et~al.}(2011)Pedregosa, Varoquaux, Gramfort, Michel,
  Thirion, Grisel, Blondel, Prettenhofer, Weiss, Dubourg, Vanderplas, Passos,
  Cournapeau, Brucher, Perrot, \& Duchesnay}]{Pedregosa2011}
Pedregosa, F., Varoquaux, G., Gramfort, A., {et~al.} 2011, Journal of Machine
  Learning Research, 12, 2825

\bibitem[{Penny {et~al.}(2019)Penny, Gaudi, Kerins, Rattenbury, Mao, Robin, \&
  Novati}]{Penny2019}
Penny, M.~T., Gaudi, B.~S., Kerins, E., {et~al.} 2019, The Astrophysical
  Journal Supplement Series, 241, 3

\bibitem[{Petigura {et~al.}(2018)Petigura, Marcy, Winn, Weiss, Fulton, Howard,
  Sinukoff, Isaacson, Morton, \& Johnson}]{Petigura2018}
Petigura, E.~A., Marcy, G.~W., Winn, J.~N., {et~al.} 2018, The Astronomical
  Journal, 155, 89

\bibitem[{Piffl {et~al.}(2014)Piffl, Scannapieco, Binney, Steinmetz, Scholz,
  Williams, {de Jong}, Kordopatis, Matijevi{\v c}, Bienaym{\'e},
  {Bland-Hawthorn}, Boeche, Freeman, Gibson, Gilmore, Grebel, Helmi, Munari,
  Navarro, Parker, Reid, Seabroke, Watson, Wyse, \& Zwitter}]{Piffl2014}
Piffl, T., Scannapieco, C., Binney, J., {et~al.} 2014, Astronomy and
  Astrophysics, 562, A91

\bibitem[{Pollack {et~al.}(1996)Pollack, Hubickyj, Bodenheimer, Lissauer,
  Podolak, \& Greenzweig}]{Pollack1996}
Pollack, J.~B., Hubickyj, O., Bodenheimer, P., {et~al.} 1996, Icarus, 124, 62

\bibitem[{Portail {et~al.}(2015)Portail, Wegg, Gerhard, \&
  {Martinez-Valpuesta}}]{Portail2015}
Portail, M., Wegg, C., Gerhard, O., \& {Martinez-Valpuesta}, I. 2015, Monthly
  Notices of the Royal Astronomical Society, 448, 713

\bibitem[{Posti \& Helmi(2019)}]{Posti2019a}
Posti, L. \& Helmi, A. 2019, Astronomy \& Astrophysics, 621, A56

\bibitem[{Prantzos(2008)}]{Prantzos2008a}
Prantzos, N. 2008, Space Science Reviews, 135, 313

\bibitem[{Prochaska {et~al.}(2000)Prochaska, Naumov, Carney, McWilliam, \&
  Wolfe}]{Prochaska2000}
Prochaska, J.~X., Naumov, S.~O., Carney, B.~W., McWilliam, A., \& Wolfe, A.~M.
  2000, The Astronomical Journal, 120, 2513

\bibitem[{Rafikov(2005)}]{Rafikov2005}
Rafikov, R.~R. 2005, The Astrophysical Journal, 621, L69

\bibitem[{Rauer {et~al.}(2014)Rauer, Catala, Aerts, Appourchaux, Benz,
  Brandeker, {Christensen-Dalsgaard}, Deleuil, Gizon, Goupil, G{\"u}del,
  {Janot-Pacheco}, {Mas-Hesse}, Pagano, Piotto, Pollacco, Santos, Smith,
  Su{\'a}rez, Szab{\'o}, Udry, Adibekyan, Alibert, Almenara, {Amaro-Seoane},
  Eiff, Asplund, Antonello, Barnes, Baudin, Belkacem, Bergemann, Bihain, Birch,
  Bonfils, Boisse, Bonomo, Borsa, Brand{\~a}o, Brocato, Brun, Burleigh,
  Burston, Cabrera, Cassisi, Chaplin, Charpinet, Chiappini, Church, Csizmadia,
  Cunha, Damasso, Davies, Deeg, D{\'i}az, Dreizler, Dreyer, Eggenberger,
  Ehrenreich, Eigm{\"u}ller, Erikson, Farmer, Feltzing, {de Oliveira Fialho},
  Figueira, Forveille, Fridlund, Garc{\'i}a, Giommi, Giuffrida, Godolt, {Gomes
  da Silva}, Granzer, Grenfell, {Grotsch-Noels}, G{\"u}nther, Haswell, Hatzes,
  H{\'e}brard, Hekker, Helled, Heng, Jenkins, Johansen, Khodachenko,
  Kislyakova, Kley, Kolb, Krivova, Kupka, Lammer, Lanza, Lebreton, Magrin,
  {Marcos-Arenal}, Marrese, Marques, Martins, Mathis, Mathur, Messina, Miglio,
  Montalban, Montalto, Monteiro, Moradi, Moravveji, Mordasini, Morel, Mortier,
  Nascimbeni, Nelson, Nielsen, Noack, Norton, Ofir, Oshagh, Ouazzani,
  P{\'a}pics, Parro, Petit, Plez, Poretti, Quirrenbach, Ragazzoni, Raimondo,
  Rainer, Reese, Redmer, Reffert, {Rojas-Ayala}, Roxburgh, Salmon, Santerne,
  Schneider, Schou, Schuh, Schunker, {Silva-Valio}, Silvotti, Skillen, Snellen,
  Sohl, Sousa, Sozzetti, Stello, Strassmeier, {\v S}vanda, Szab{\'o},
  Tkachenko, Valencia, Van~Grootel, Vauclair, Ventura, Wagner, Walton,
  Weingrill, Werner, Wheatley, \& Zwintz}]{Rauer2014}
Rauer, H., Catala, C., Aerts, C., {et~al.} 2014, Experimental Astronomy, 38,
  249

\bibitem[{Reddy {et~al.}(2006)Reddy, Lambert, \& Allende~Prieto}]{Reddy2006}
Reddy, B.~E., Lambert, D.~L., \& Allende~Prieto, C. 2006, Monthly Notices of
  the Royal Astronomical Society, 367, 1329

\bibitem[{Reddy {et~al.}(2003)Reddy, Tomkin, Lambert, \&
  Allende~Prieto}]{Reddy2003}
Reddy, B.~E., Tomkin, J., Lambert, D.~L., \& Allende~Prieto, C. 2003, Monthly
  Notices of the Royal Astronomical Society, 340, 304

\bibitem[{Sanders \& Binney(2015)}]{Sanders2015}
Sanders, J.~L. \& Binney, J. 2015, Monthly Notices of the Royal Astronomical
  Society, 449, 3479

\bibitem[{Santos {et~al.}(2001)Santos, Israelian, \& Mayor}]{Santos2001}
Santos, N.~C., Israelian, G., \& Mayor, M. 2001, Astronomy and Astrophysics,
  373, 1019

\bibitem[{Schaye {et~al.}(2015)Schaye, Crain, Bower, Furlong, Schaller, Theuns,
  Dalla~Vecchia, Frenk, McCarthy, Helly, Jenkins, {Rosas-Guevara}, White, Baes,
  Booth, Camps, Navarro, Qu, Rahmati, Sawala, Thomas, \& Trayford}]{Schaye2015}
Schaye, J., Crain, R.~A., Bower, R.~G., {et~al.} 2015, Monthly Notices of the
  Royal Astronomical Society, 446, 521

\bibitem[{Schib {et~al.}(2021)Schib, Mordasini, Wenger, Marleau, \&
  Helled}]{Schib2021}
Schib, O., Mordasini, C., Wenger, N., Marleau, G.~D., \& Helled, R. 2021,
  Astronomy and Astrophysics, 645, A43

\bibitem[{Schlaufman(2018)}]{Schlaufman2018}
Schlaufman, K.~C. 2018, The Astrophysical Journal, 853, 37

\bibitem[{Schlecker {et~al.}(2021)Schlecker, Pham, Burn, Alibert, Mordasini,
  Emsenhuber, Klahr, Henning, \& Mishra}]{Schlecker2021}
Schlecker, M., Pham, D., Burn, R., {et~al.} 2021, Astronomy and Astrophysics,
  656, A73

\bibitem[{Schuster {et~al.}(2006)Schuster, Moitinho, M{\'a}rquez, Parrao, \&
  Covarrubias}]{Schuster2006}
Schuster, W.~J., Moitinho, A., M{\'a}rquez, A., Parrao, L., \& Covarrubias, E.
  2006, Astronomy and Astrophysics, 445, 939

\bibitem[{Seabold \& Perktold(2010)}]{Seabold2010}
Seabold, S. \& Perktold, J. 2010, in Python in {{Science Conference}}, {Austin,
  Texas}, 92--96

\bibitem[{Sit \& Ness(2020)}]{Sit2020}
Sit, T. \& Ness, M.~K. 2020, The Astrophysical Journal, 900, 4

\bibitem[{Smith(2015)}]{Smith2015}
Smith, B. 2015, Software, 11661446 Bytes

\bibitem[{Spergel {et~al.}(2015)Spergel, Gehrels, Baltay, Bennett,
  Breckinridge, Donahue, Dressler, Gaudi, Greene, Guyon, Hirata, Kalirai,
  Kasdin, Macintosh, Moos, Perlmutter, Postman, Rauscher, Rhodes, Wang,
  Weinberg, Benford, Hudson, Jeong, Mellier, Traub, Yamada, Capak, Colbert,
  Masters, Penny, Savransky, Stern, Zimmerman, Barry, Bartusek, Carpenter,
  Cheng, Content, Dekens, Demers, Grady, Jackson, Kuan, Kruk, Melton, Nemati,
  Parvin, Poberezhskiy, Peddie, Ruffa, Wallace, Whipple, Wollack, \&
  Zhao}]{Spergel2015}
Spergel, D., Gehrels, N., Baltay, C., {et~al.} 2015, Wide-{{Field InfrarRed
  Survey Telescope-Astrophysics Focused Telescope Assets WFIRST-AFTA}} 2015
  {{Report}}

\bibitem[{Swastik {et~al.}(2022)Swastik, Banyal, Narang, Manoj, Sivarani,
  Rajaguru, Unni, \& Banerjee}]{Swastik2022}
Swastik, C., Banyal, R.~K., Narang, M., {et~al.} 2022, The Astronomical
  Journal, 164, 60

\bibitem[{{The pandas development team}(2023)}]{Thepandasdevelopmentteam2023}
{The pandas development team}. 2023, Pandas-Dev/Pandas: {{Pandas}}, Zenodo

\bibitem[{Thorngren {et~al.}(2016)Thorngren, Fortney, {Murray-Clay}, \&
  Lopez}]{Thorngren2016}
Thorngren, D.~P., Fortney, J.~J., {Murray-Clay}, R.~A., \& Lopez, E.~D. 2016,
  The Astrophysical Journal, 831, 64

\bibitem[{Tully {et~al.}(2009)Tully, Rizzi, Shaya, Courtois, Makarov, \&
  Jacobs}]{Tully2009}
Tully, R.~B., Rizzi, L., Shaya, E.~J., {et~al.} 2009, The Astronomical Journal,
  138, 323

\bibitem[{Tychoniec {et~al.}(2018)Tychoniec, Tobin, Karska, Chandler, Dunham,
  Harris, Kratter, Li, Looney, Melis, P{\'e}rez, Sadavoy, {Segura-Cox}, \& van
  Dishoeck}]{Tychoniec2018}
Tychoniec, {\L}., Tobin, J.~J., Karska, A., {et~al.} 2018, The Astrophysical
  Journal Supplement Series, 238, 19

\bibitem[{{Van der Marel} \& Mulders(2021)}]{vanderMarel2021}
{Van der Marel}, N. \& Mulders, G.~D. 2021, The Astronomical Journal, 162, 28

\bibitem[{Voelkel {et~al.}(2021)Voelkel, Deienno, Kretke, \&
  Klahr}]{Voelkel2021}
Voelkel, O., Deienno, R., Kretke, K., \& Klahr, H. 2021, Astronomy and
  Astrophysics, 645, A132

\bibitem[{Vogelsberger {et~al.}(2014)Vogelsberger, Genel, Springel, Torrey,
  Sijacki, Xu, Snyder, Nelson, \& Hernquist}]{Vogelsberger2014}
Vogelsberger, M., Genel, S., Springel, V., {et~al.} 2014, Monthly Notices of
  the Royal Astronomical Society, 444, 1518

\bibitem[{Wallerstein(1962)}]{Wallerstein1962}
Wallerstein, G. 1962, The Astrophysical Journal Supplement Series, 6, 407

\bibitem[{Waskom(2021)}]{Waskom2021}
Waskom, M. 2021, Journal of Open Source Software, 6, 3021

\bibitem[{Watkins {et~al.}(2019)Watkins, {van der Marel}, Sohn, \&
  Evans}]{Watkins2019}
Watkins, L.~L., {van der Marel}, R.~P., Sohn, S.~T., \& Evans, N.~W. 2019, The
  Astrophysical Journal, 873, 118

\bibitem[{Winter \& Haworth(2022)}]{Winter2022}
Winter, A.~J. \& Haworth, T.~J. 2022, European Physical Journal Plus, 137, 1132

\bibitem[{Winter {et~al.}(2020)Winter, Kruijssen, Chevance, Keller, \&
  Longmore}]{Winter2020}
Winter, A.~J., Kruijssen, J. M.~D., Chevance, M., Keller, B.~W., \& Longmore,
  S.~N. 2020, Monthly Notices of the Royal Astronomical Society, 491, 903

\bibitem[{Yang {et~al.}(2020)Yang, Xie, \& Zhou}]{Yang2020}
Yang, J.-Y., Xie, J.-W., \& Zhou, J.-L. 2020, The Astronomical Journal, 159,
  164

\bibitem[{Yasui(2021)}]{Yasui2021}
Yasui, C. 2021, The Astrophysical Journal, 914, 115

\bibitem[{Yasui {et~al.}(2009)Yasui, Kobayashi, Tokunaga, Saito, \&
  Tokoku}]{Yasui2009}
Yasui, C., Kobayashi, N., Tokunaga, A.~T., Saito, M., \& Tokoku, C. 2009, The
  Astrophysical Journal, 705, 54

\bibitem[{Yu {et~al.}(2023)Yu, Bullock, Gurvich, Hafen, Stern,
  {Boylan-Kolchin}, {Faucher-Gigu{\`e}re}, Wetzel, Hopkins, \& Moreno}]{Yu2023}
Yu, S., Bullock, J.~S., Gurvich, A.~B., {et~al.} 2023, Monthly Notices of the
  Royal Astronomical Society, 523, 6220

\bibitem[{Zana {et~al.}(2022)Zana, Lupi, Bonetti, Dotti, {Rosas-Guevara},
  {Izquierdo-Villalba}, Bonoli, Hernquist, \& Nelson}]{Zana2022}
Zana, T., Lupi, A., Bonetti, M., {et~al.} 2022, Monthly Notices of the Royal
  Astronomical Society, 515, 1524

\bibitem[{Zaritsky \& Courtois(2017)}]{Zaritsky2017}
Zaritsky, D. \& Courtois, H. 2017, Monthly Notices of the Royal Astronomical
  Society, 465, 3724

\bibitem[{Zoccali {et~al.}(2008)Zoccali, Hill, Lecureur, Barbuy, Renzini,
  Minniti, G{\'o}mez, \& Ortolani}]{Zoccali2008}
Zoccali, M., Hill, V., Lecureur, A., {et~al.} 2008, Astronomy and Astrophysics,
  486, 177

\end{thebibliography}

%%%%%%%%%%%%%%%%%%%%%%%%%%%%%%%%%%%%%%%%
% The Appendix

\begin{appendix}
\renewcommand\thefigure{A.\arabic{figure}}  
\setcounter{figure}{0}  % Reset figure counter
\begin{figure*}[p]
     \centering
     \begin{subfigure}[b]{0.40\textwidth}
        \centering
        \includegraphics[width=\textwidth]{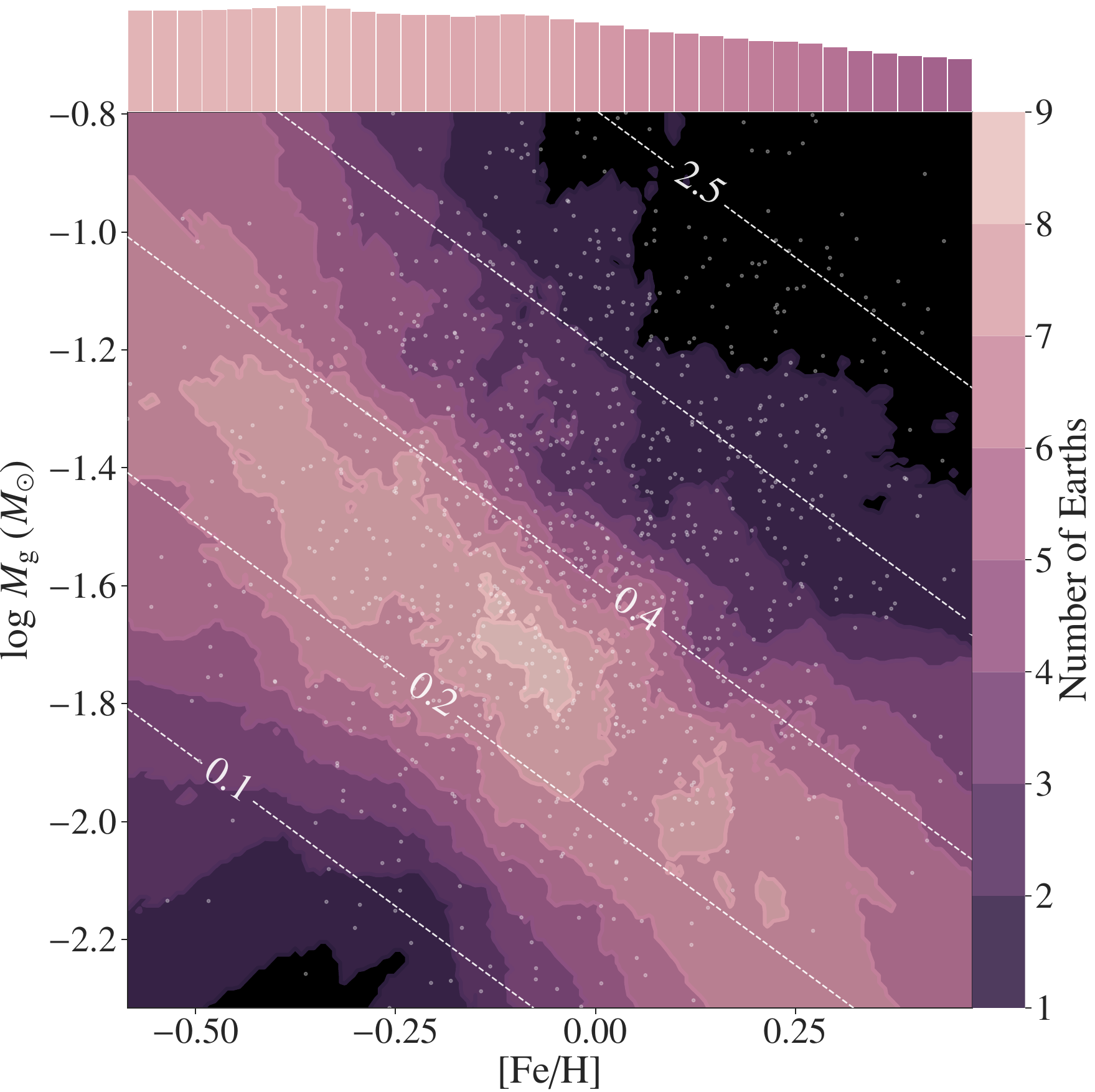}
        \caption{Earth-like.}
        \label{fig:contourplot_Earth}
     \end{subfigure}
     \hspace{40pt}
     \begin{subfigure}[b]{0.39\textwidth}
        \centering
        \includegraphics[width=\textwidth]{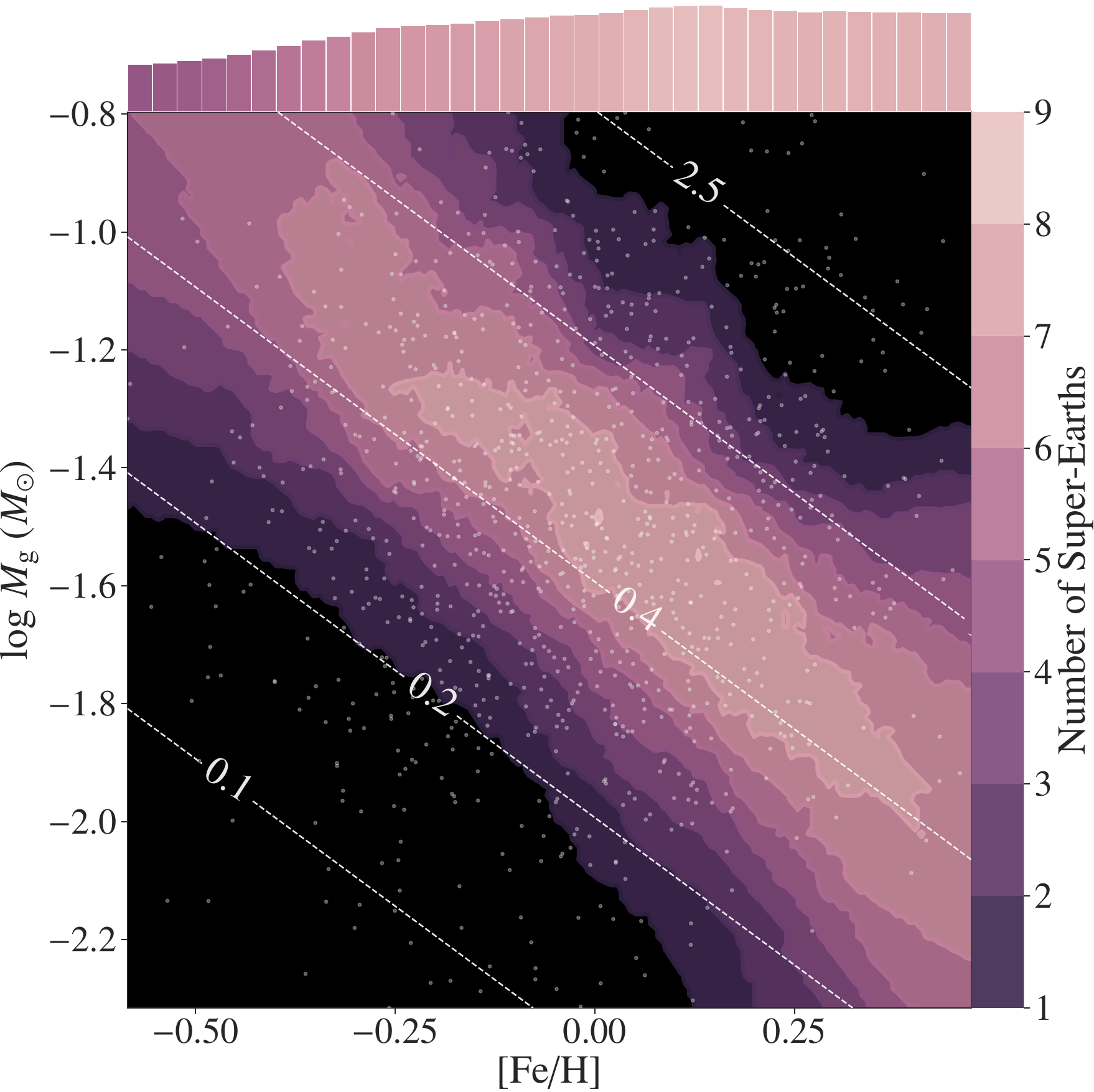}
        \caption{Super-Earth.}
        \label{fig:contourplot_Super}
     \end{subfigure}
     \hspace{40pt}
     \begin{subfigure}[b]{0.39\textwidth}
        \centering
        \includegraphics[width=\textwidth]{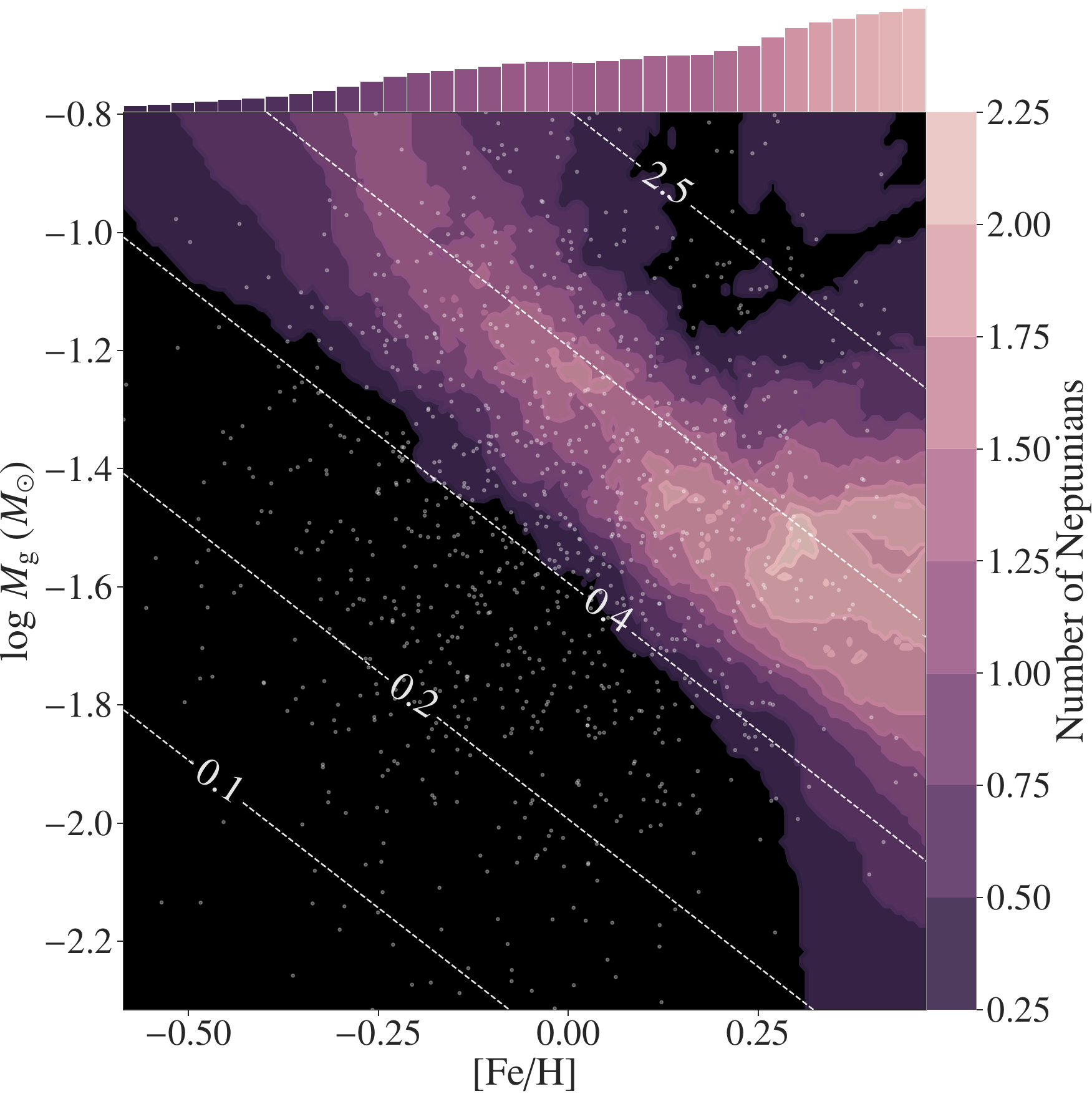}
        \caption{Neptunian.}
        \label{fig:contourplot_Neptunian}
     \end{subfigure}
     \hspace{40pt}
     \begin{subfigure}[b]{0.39\textwidth}
        \centering
        \includegraphics[width=\textwidth]{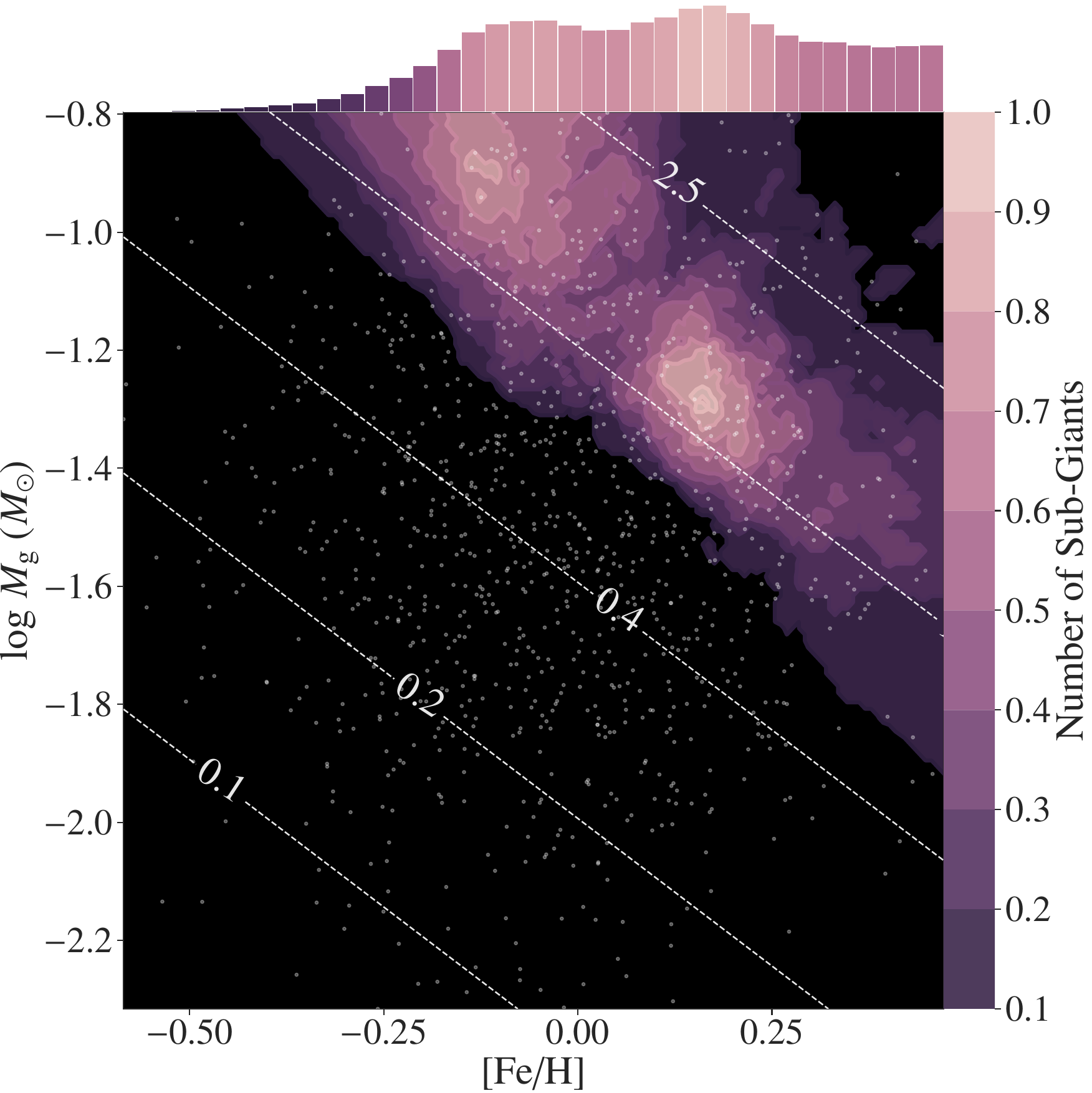}
        \caption{Sub-giant.}
        \label{fig:contourplot_Sub}
     \end{subfigure}
     \begin{subfigure}[b]{0.39\textwidth}
        \centering
        \includegraphics[width=\textwidth]{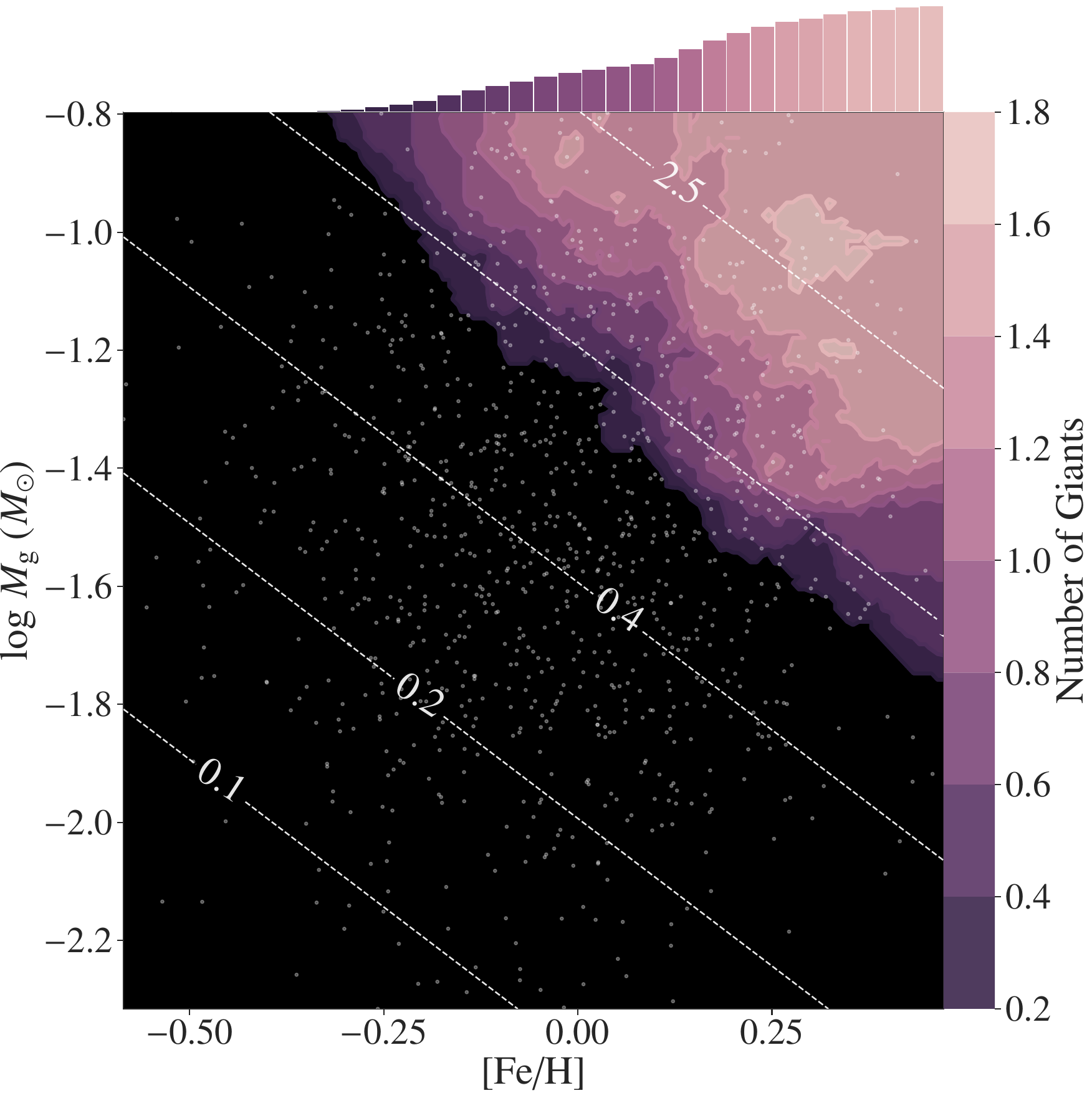}
        \caption{Giant.}
        \label{fig:contourplot_Giant}
     \end{subfigure}
     \hspace{40pt}
     \begin{subfigure}[b]{0.39\textwidth}
        \centering
        \includegraphics[width=\textwidth]{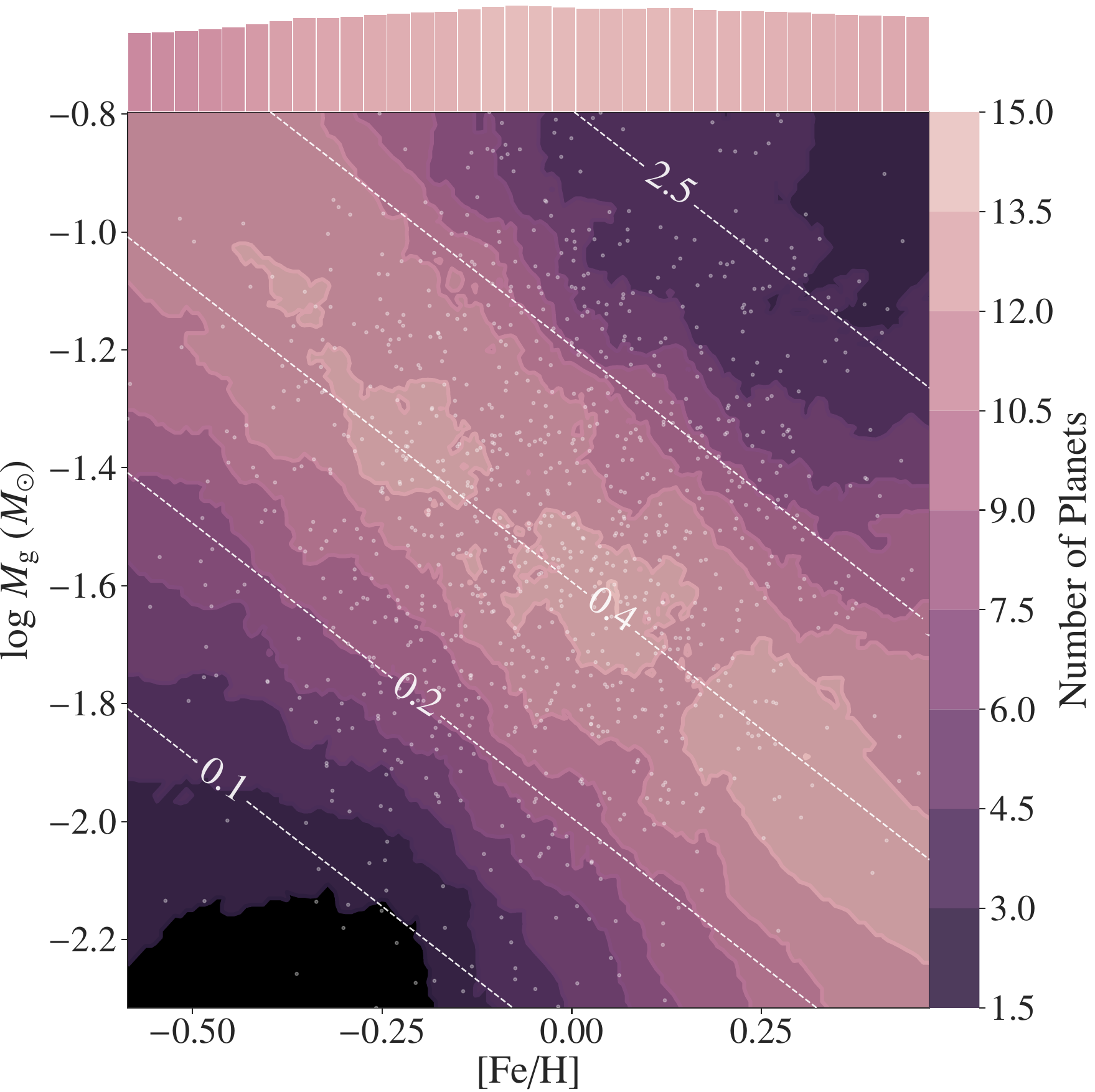}
        \caption{All.}
        \label{fig:contourplot_Planet}
     \end{subfigure}
     \caption{Predicted number of planets by type from the planet model ($M_\star=1 M_\Sun$, $N_\mathrm{Embryo}=50$), based on initial gas disc mass and [Fe/H]. The white diagonal lines represent contours of equal solid disc mass, measured in Jupiter masses, showing a pronounced correlation between solid disc mass and planet count. The histograms on top illustrate the marginal distribution for [Fe/H], while dots mark the original Monte Carlo variable sample from \citet{Emsenhuber2021a}.}
    \label{fig:contourplots_planet_model}
\end{figure*}

\renewcommand\thetable{A.\arabic{table}}  
\setcounter{table}{0}  % Reset figure counter
\begin{table*}[b]
    \centering
    \caption{Planet occurrence rates in the different galactic components and for host star masses $M_\star=$ 0.3, 0.5, and 1. All occurrence rates are calculated for the NGPPS populations with $N_\mathrm{Embryo}=50$ initial embryos.}
    \begin{tabularx}{\textwidth}{r|L|R|R|R|R}
        $M_\star$ & & Bulge & Thin Disc & Thick Disc & Halo \\
        \addlinespace
        \hline\hline
        \addlinespace
        \multirow[t]{5}{*}{1} 
        & Earth & 3.6 & 3.9 & 5.5 & 5.4 \\
        & Super-Earth & 3.9 & 4.1 & 2.7 & 2.4 \\
        & Neptunian & 0.68 & 0.60 & 0.13 & 0.11 \\
        & Sub-giant & 0.146 & 0.135 & 0.019 & 0.014 \\
        & Giant & 0.407 & 0.313 & 0.021 & 0.018 \\
        \addlinespace
        \hline
        \addlinespace
        \multirow[t]{5}{*}{0.5} 
        & Earth & 4.6 & 5.0 & 4.8 & 4.3 \\
        & Super-Earth & 3.2 & 3.5 & 1.0 & 0.8 \\
        & Neptunian & 0.34 & 0.38 & 0.04 & 0.03 \\
        & Sub-giant & 0.084 & 0.075 & 0.002 & 0.003 \\
        & Giant & 0.022 & 0.020 & 0.002 & 0.001 \\
        \addlinespace
        \hline
        \addlinespace
        \multirow[t]{5}{*}{0.3} 
        & Earth & 5.0 & 5.1 & 4.5 & 4.4 \\
        & Super-Earth & 2.1 & 2.1 & 1.7 & 1.7 \\
        & Neptunian & 0.11 & 0.12 & 0.08 & 0.07 \\
        & Sub-giant & 0.019 & 0.016 & 0.004 & 0.002 \\
        & Giant & 0.000 & 0.000 & 0.000 & 0.000 \\
        \addlinespace
        \hline
    \end{tabularx}
    \label{tab:occurrence_rates}
\end{table*}

\begin{figure*}[b]
     \centering
     \begin{subfigure}{0.7\textwidth}
        \centering
        \includegraphics[width=\textwidth]{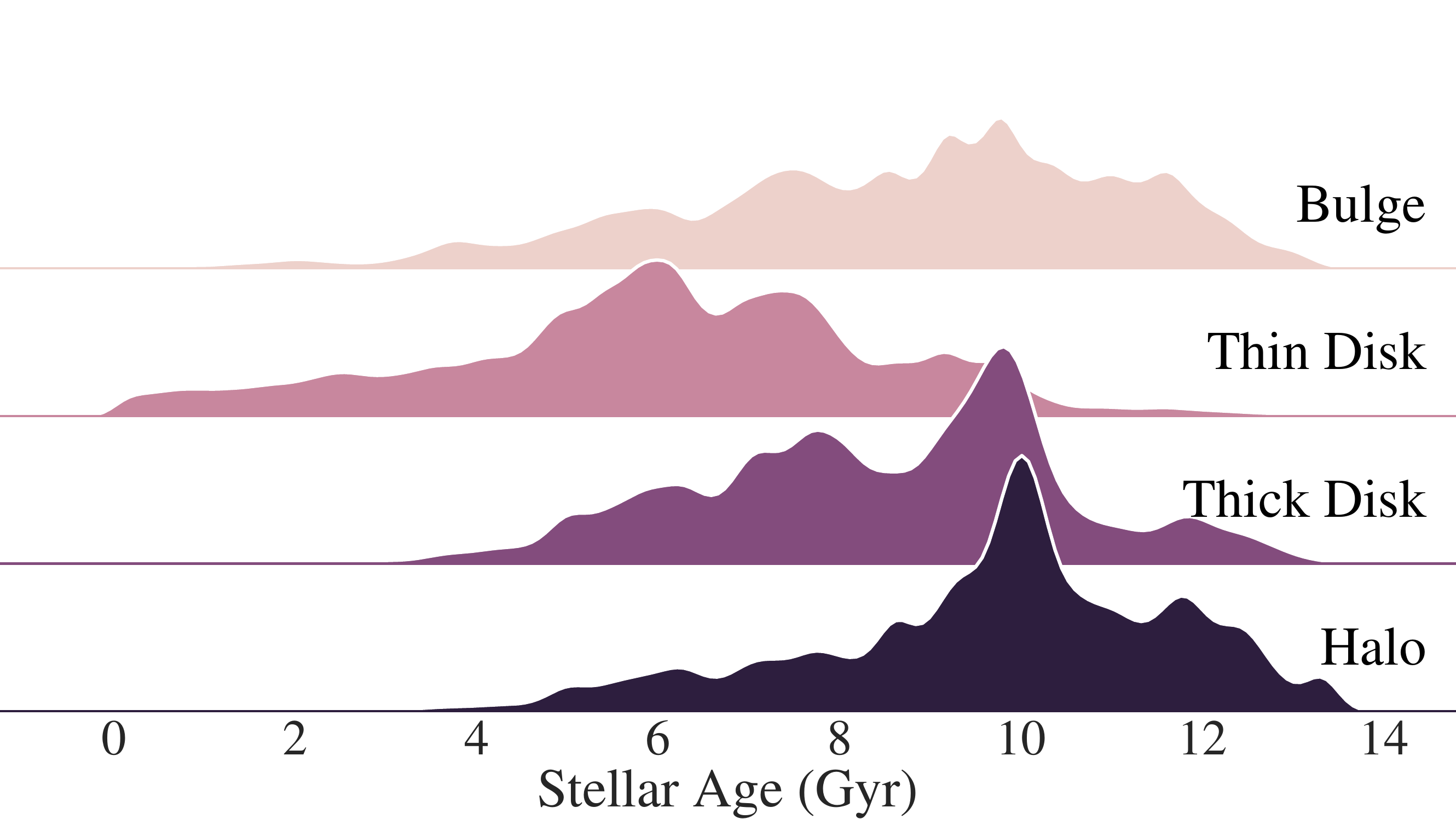}
        \label{fig:ridgeplot_age}
     \end{subfigure}
     
     \begin{subfigure}{0.7\textwidth}
        \centering
        \includegraphics[width=\textwidth]{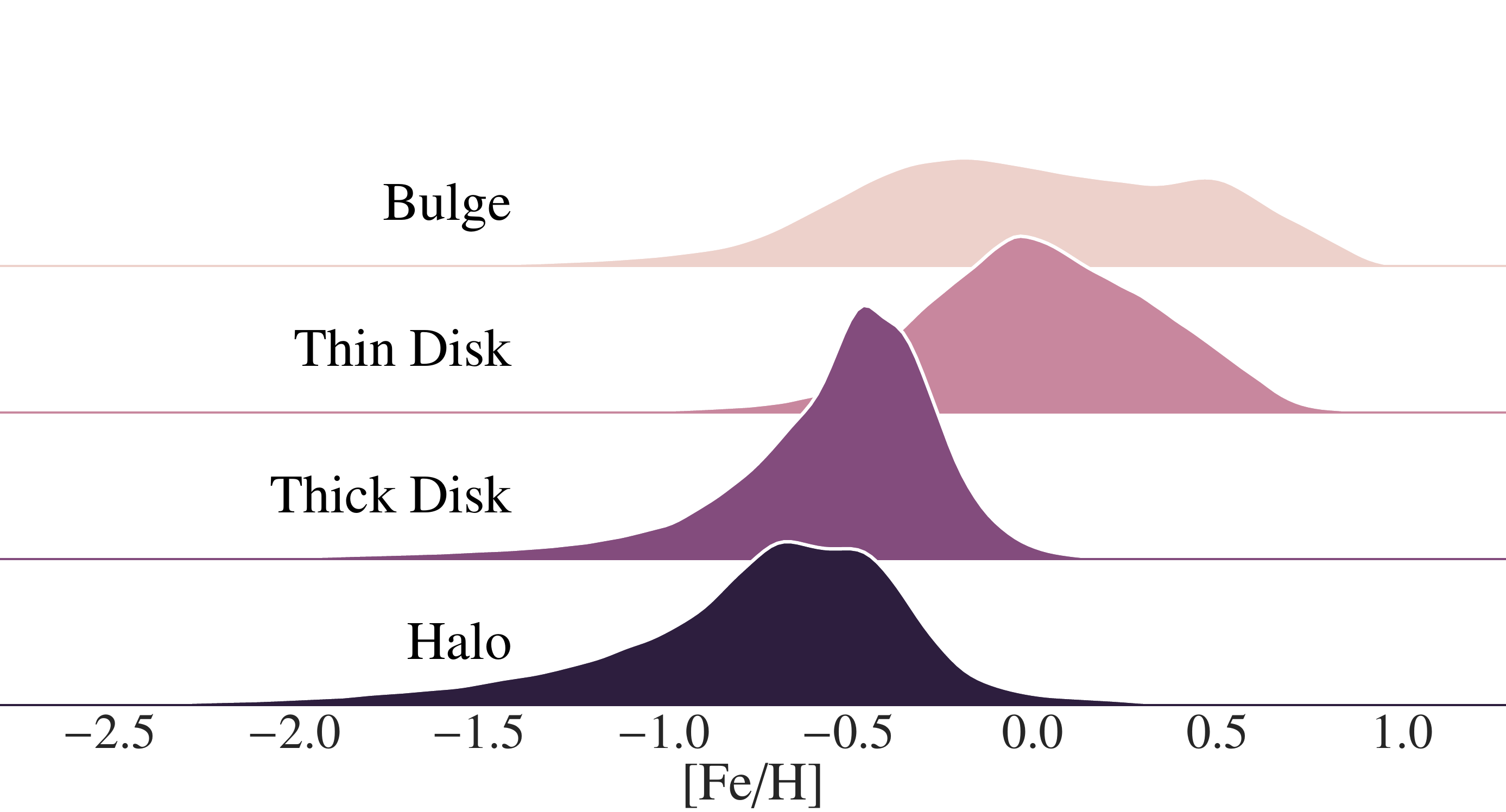}
        \label{fig:ridgeplot_metallicity}
     \end{subfigure}
     \caption{Normalised distributions of age and metallicity distribution of the galaxy components.}
     \label{fig:ridgeplots_galaxy_components}
\end{figure*}
\end{appendix}

\end{document}